%% file: sample-acmsmall.tex
\documentclass[acmsmall]{acmart}
\AtBeginDocument{%
  \providecommand\BibTeX{{%
    \normalfont B\kern-0.5em{\scshape i\kern-0.25em b}\kern-0.8em\TeX}}}

\usepackage{graphicx}

\usepackage{algorithm}
\usepackage{algorithmic}
\usepackage{colortbl}
\usepackage{tikz}
\usetikzlibrary{calc}
\usetikzlibrary{arrows}
\usepackage{pgfplots}
\usepackage{pgf-pie}
\usepackage{pdflscape}
\usepackage{amsmath,amsfonts}
\usepackage{subcaption}
\usetikzlibrary{patterns}
\usepackage{multicol}
\usepackage{multirow}
\usepackage{makecell}
\usepackage[capitalize]{cleveref} 

\usepackage{adjustbox}

\setcopyright{acmcopyright}
\copyrightyear{2018}
\acmYear{2018}
\acmDOI{XXXXXXX.XXXXXXX}

\acmJournal{JACM}
\acmVolume{37}
\acmNumber{4}
\acmArticle{111}
\acmMonth{8}

\begin{document}

\title{Exploring Automated Code Evaluation Systems and Resources for Code Analysis: A Comprehensive Survey}

\author{Md. Mostafizer Rahman}
\email{mostafiz@duet.ac.bd}
\orcid{0000-0001-9368-7638}

\affiliation{%
  \institution{Dhaka University of Engineering \& Technology, Gazipur}
  \country{Bangladesh}  
  \postcode{1707}
}
\additionalaffiliation{
\institution{The University of Aizu, Japan} 
  \city{Aizuwakamatsu}
  \country{Japan}
  }

\author{Yutaka Watanobe}
\affiliation{%
  \institution{The University of Aizu} 
  \city{Aizuwakamatsu}
  \country{Japan}}
\email{yutaka@u-aizu.ac.jp}

\author{Atsushi Shirafuji}
\affiliation{%
  \institution{The University of Aizu} 
  \city{Aizuwakamatsu}
  \country{Japan}}
\email{atsushirafuji@gmail.com}

\author{Mohamed Hamada}
\affiliation{%
  \institution{The University of Aizu} 
  \city{Aizuwakamatsu}
  \country{Japan}}
\email{hamada@u-aizu.ac.jp}

\renewcommand{\shortauthors}{M. M. Rahman et al.}

\begin{abstract}
 The automated code evaluation system (AES) is mainly designed to reliably assess user-submitted code. The code is compiled and then tested in a unified environment with predefined input and output test cases. Due to their extensive range of applications and the accumulation of valuable resources, AESs are becoming increasingly popular. Research on the application of AES and their real-world resource exploration for diverse coding tasks is still lacking. In this study, we conducted a comprehensive survey on AESs and their resources. This survey explores the application areas of AESs, available resources, and resource utilization for coding tasks. AESs are categorized into programming contests, programming learning and education, recruitment, online compilers, and additional modules, depending on their application. We explore the available datasets and other resources of these systems for research, analysis, and coding tasks. The success of machine learning models for inference procedures depends primarily on the purity of the data, where the accumulated real-life data (e.g., codes and submission logs) from AESs can be a valuable treasure. Moreover, we provide an overview of machine learning-driven coding tasks, such as bug detection, code review, comprehension, refactoring, search, representation, and repair. These tasks are performed using real-life datasets. In addition, we briefly discuss the Aizu Online Judge platform as a real example of an AES from the perspectives of system design (hardware and software), operation (competition and education), and research. This is due to the scalability of the AOJ platform (programming education, competitions, and practice), open internal features (hardware and software), attention from the research community, open source data (e.g., solution codes and submission documents), and transparency. We also analyze the overall performance of this system and the perceived challenges over the years.
\end{abstract}




\maketitle

\section{Introduction} \label{sec:intro}

Competitive programming contests (CPCs) are usually held online or offline (local area network), where participants try to solve programming problems according to given specifications. CPCs, also called programming sports, are recognized worldwide. Many educational institutions and multinational tech giants (e.g., Google, Facebook, etc.) foster these programming events. Generally, CPC organizers provide a set of mathematical or logical problems for participants. Typically, problems are related to categories such as data structures, string analysis, graph theory, combinatorics, computational geometry, number theory, and game theory. In addition, artificial intelligence (AI) and constraint programming are also included in certain competitions. Participants write code to solve those problems. The problem-solving process can be divided into two main steps: ($i$) development of an efficient algorithm and ($ii$) implementation of the algorithm in an appropriate programming language. The correctness of the submitted solution code is judged in a special environment called an online judge (OJ) or automated code evaluation system (AES). Numerous factors are considered in the judging, including output quality, program size, memory usage, and CPU time. In most contests, the submitted solution code is automatically evaluated by the host computer (judge server), where each solution code is judged against a set of test cases. A solution code is \emph{accepted} only if it passes all test cases, otherwise, it is \emph{rejected}. However, in some contests, partial judging may occur depending on the quality of the results, the number of test cases passed, or other criteria.

In the last 30 years, we have observed a significant rise in the popularity of programming computing events, exemplified by the International Collegiate Programming Contest (ICPC). This competition holds the distinction of being the largest, oldest, and most fiercely CPC for university students worldwide. The ICPC serves as a platform where students can engage with one another, enhance their programming abilities, develop algorithmic thinking, teamwork, and problem-solving skills. It not only provides a valuable opportunity for academic, community, and industry collaboration but also holds the overarching goal of nurturing the next generation of skilled computing professionals. The inaugural edition of the ICPC took place at Texas A\&M University in 1970 \cite{s_ref1}. Over the years, the ICPC has evolved into one of the most esteemed programming competitions worldwide. A testament to its widespread appeal can be observed through the participation numbers. For instance, the event attracted over 52,709 students from 3,233 universities across 110 countries in 2018, a substantial increase compared to the participation of 840 teams from 560 universities in 1997. Following the success of the first ICPC final, numerous algorithmic competitions subsequently adopted similar AES. The International Olympiad in Informatics (IOI) is one of them, which integrated AES into its evaluation process starting in 1989. Contrasting the ICPC, the IOI targets specifically to high school students, with participants tackling problems individually instead of working in teams. Notably, participants in the IOI receive scores that can be full or partial based on the merit of their submitted solutions, rather than solely binary (correct or incorrect) scores like ICPC. Furthermore, big companies such as Google and Facebook regularly organize CPCs \cite{li2022alphacode}. The popular Codeforces CPC platform hosts weekly contests that attract a substantial number of participants, reaching tens of thousands \cite{nref62}. Impressively, this platform has over 500,000 active users. In a study conducted by Cormack et al. \cite{s_ref2}, the fundamental regulations, evaluation procedures, and scoring mechanisms employed by prominent CPCs such as the ICPC and IOI were elucidated.

The CPC environment encompasses a distinct system designed to automatically verify the correctness of submitted solution code relying on predefined input/output test cases. Moreover, the system calculates various resources such as time and memory usage limits during the evaluation process of the submitted solution code. Such a specialized system is commonly referred to as an AES\footnote{The terms automated code evaluation system (AES), automated code assessment system (AAS), and online judge system (OJ) are employed interchangeably to refer to the same concept.} or OJ system. The concept of an OJ system was first introduced by Kurnia and collaborators \cite{s_ref3} in 2001, enabling the automated and real-time evaluation of solution code. However, the development, implementation, and maintenance of an OJ system is a complex endeavor, as various essential factors must be carefully considered to ensure its security and seamless operation. Typically, an AES assesses submitted codes using either local or cloud-based infrastructures and should be well-equipped to handle various threats that may arise during the evaluation process. For instance, submitted solution codes can take the form of executable files (\emph{.exe}) or malicious source code that have the potential to alter the test environment, prolong compilation time, or manipulate restricted resources such as memory and time. A comprehensive exploration of potential threat types and corresponding countermeasures for CPCs can be found in a study \cite{s_ref4}. To mitigate such threats and ensure secure execution of the submitted solution code, recent advancements have introduced specialized sandboxes \cite{s_ref5}, virtualization techniques, containerization \cite{s_ref6}, and the utilization of Docker frameworks \cite{s_ref7}. These approaches hold promise in enhancing the security and reliability of AES.

Apart from CPC, AESs offer a range of auxiliary functionalities that find application in diverse domains such as programming education, online compilers, data mining, recruitment, and software development. Furthermore, the resources generated by AESs, including evaluation results, solution codes, and logs, are highly regarded as valuable assets for both educational and industrial research purposes. Over the years, numerous academic research endeavors have yielded significant findings by leveraging the resources provided by AESs, which are extensively utilized for programming tutoring across various academic institutions \cite{s_ref8,s_ref9}. These results have shed light on submission patterns, verdict statistics, problem-solving capabilities for constraint-oriented tasks, and prevalent solution patterns among different student groups. Educators are capitalizing on the insights gained from such analytical research to enhance their lesson plans and teaching methodologies.  Moreover, the amassed data resources within AESs are recognized as one of the largest real-world code repositories. This has paved the way for software engineering research in areas such as software code analysis, vulnerability prediction, source code mining, suggestions for class and method names, and software refactoring \cite{s_ref10,s_ref11,s_ref12,s_ref13,s_ref14,s_ref15,s_ref16,s_ref17}.

In recent times, the field of AI, particularly machine learning (ML) and deep learning (DL), has witnessed remarkable advancements in text \cite{s_ref18,s_ref19}, speech \cite{s_ref22,s_ref23,s_ref24}, and image \cite{s_ref20,s_ref21} processing. These breakthroughs in ML and DL have been facilitated by the proliferation of open-source codes and the rapid development of computational hardware. This progress has inspired both practitioners and researchers to explore source code and software engineering challenges \cite{s_ref25,s_ref26,s_ref27,s_ref28,s_ref29}. Within the domain of source code analysis, researchers and professionals have embraced the utilization of DL models for various code-related tasks. These tasks encompass code representation \cite{s_ref25,s_ref30}, program synthesis \cite{s_ref27,s_ref28,s_ref29,s_ref30,s_ref31,s_ref32,s_ref33,s_ref34}, code completion \cite{s_ref14,s_ref35_ref101}, code testing  \cite{s_ref31,s_ref32,s_ref33}, code summarization \cite{s_ref16,s_ref35_ref101,s_ref36}, code refactoring \cite{s_ref15},  code repair \cite{s_ref12}, and analysis of source code quality and vulnerability \cite{s_ref26,s_ref37,s_ref38,s_ref39,s_ref40}. The adoption of DL for code analysis is on the rise, accompanied by a growing array of methods, techniques, resources, tools, and datasets.  Consequently, researchers face the challenge of comprehending the expansive landscape of available resources and research directions in this domain. However, numerous endeavors have been made to consolidate application-specific research through survey publications \cite{s_ref1,nref60,nref61}. In addition to the state-of-the-art (SOTA) surveys, our survey offers several advantages to researchers and professionals engaged in code analysis tasks. Firstly, it encompasses a comprehensive summary of a substantial number of SOTA surveys. Secondly, it provides a literature review categorized by specific areas such as quality assessment, testing, code repair, and program synthesis within the field of software engineering. Lastly,  an exploration of the existing tools and datasets available for code analysis.

In the past few years, the resources provided by AESs have emerged as valuable datasets for a variety of code analysis tasks.  One notable dataset is CodeNet \cite{s_ref41}, a comprehensive collection encompassing 14 million real-world solution codes and approximately 500 million lines of code spanning 55 programming languages. These codes have been sourced from the Aizu Online Judge (AOJ) \cite{s_ref42} and AtCoder  \cite{s_ref43} OJ systems. CodeNet is a meticulously curated dataset specifically designed for code analysis applications, including code classification and code similarity measurement. Another large dataset, CodeContests \cite{li2022alphacode}, is tailored for ML applications and is extensively employed in the training of models like Alphacode. CodeContests incorporates datasets sourced from diverse CPC platforms, including AOJ, Atcoder, CodeChef, Codeforces, and HackerEarth. After implementing the necessary filtering procedures, the CodeContests dataset contains a total of 715.1 GB of code. POJ-104 \cite{li2022alphacode} and GCJ \cite{s_ref45} are extensively utilized benchmark datasets that originated from a pedagogical AES and the Google Code Jam competition held between 2008 and 2020, respectively. Notably, GCJ-297 \cite{s_ref46} stands as a benchmark dataset comprising 297 problems and approximately 208,000 solution codes. For an in-depth exploration of code analysis tasks and their associated benchmark datasets, a comprehensive discussion can be found in CodeXGLUE \cite{s_ref47}.

The objective of this paper is as follows: To the best of our knowledge, there is a lack of existing surveys on AESs and their available resources for code analysis tasks, our aim is to bridge this gap. Initially, we provide an overview of the current landscape of AESs and their potential application domains. Considering the diverse nature of these systems, a classification based on their specific applications holds significant value for both researchers and professionals. We summarized the resources, including datasets, generated by AESs and other similar platforms. These accumulated datasets can serve as valuable assets for practitioners and researchers engaging in further research and analysis. Moreover, we provide a comprehensive review of recent studies focusing on code analysis tasks employing ML techniques. We also present an in-depth investigation of a real-world AES, exploring aspects such as system design (software and hardware), operation, maintenance, and research opportunities.

The remainder of this paper is organized as follows: \cref{sec:aes} introduces the evaluation methodology and scoring of AESs, \cref{sec:classification:aes} classifies AES based on their use, \cref{sec:resources} examines the available resources for AESs, \cref{sec:code-analysis:ml} summarizes the ML-based coding analysis tasks, \cref{sec:aes-example:aoj} explores aspects of a real-world AES, including system design (software and hardware), operation, maintenance and research opportunities, and finally, \cref{sec:conclusion} concludes the study.

\section{Automated Code Evaluation Systems} \label{sec:aes}

\subsection{Evaluation Method} \label{sec:aes:eval-method}
AES is a reliable, secure, and continuous online evaluation system for algorithms submitted by users around the world. For better understanding, the AES evaluation method can be defined as follows:

\paragraph{Evaluation Method:} The evaluation method consists of three main steps. ($i$) code submission, ($ii$) code evaluation with test datasets, and ($iii$) evaluation score.

In the code submission phase, the submitted code is compiled and verified whether the code is executable in a homogeneous evaluation environment or not. If the verification is successful, each solution code is reliably evaluated on a coherent evaluation infrastructure using problem-specific test cases. The evaluation of the test cases determines for each submission: ($i$) the code executes without errors, ($ii$) the resource constraints (time and memory) have not been exceeded for a given problem, ($iii$) the obtained result satisfies all problem definitions. 

Code evaluation is basically a complex process in which the submitted code must be compiled considering each test instance. We can formally define the test instance as follows.

\paragraph{Test Instance:} Let, $\Omega$ be an alphabet containing input and output data. A test instance $\tau_i \in \mathcal{T}$, where $\mathcal{T}$ is the set of all test instances of a particular problem. Here, $\tau_i$ can be defined as a triple $\tau_i = (g_i, h_i, p_i)$, where $g_i=\Omega^*$ and $h_i=\Omega^*$ are the input data and reference output, respectively, and $p_i=\Omega^*$  is the set of parameters.

The parameter set $p_i$ represents the resource limits, such as the maximum usage of primary memory (RAM), CPU time, etc., that cannot be exceeded during evaluation for a given instance. The  parameter set $p_i$ can be empty ($p_i=\emptyset$) if the configuration of the evaluation engine is set to default resource limits. In code evaluation, the solution is compiled as an output of the submission phase. Here, the solution can be defined as follows.

\paragraph{Solution:} A solution is a function, denoted $b(g_i, p_i^\prime) \longrightarrow z_i^\prime$, that computes output data $z_i^\prime$ based on input data $g_i$  and  parameters $p_i^\prime$ given by the evaluation engine. It represents the binary form of submission. 

Here, the execution parameter set $p_i^\prime$ provided by the evaluation engine can be the same as the original parameter set $p_i$ defined as part of the respective test instance (i.e., $p_i^\prime = p_i$). However, some variations may also occur, e.g., $p_i^\prime$ can be a subset of $p_i$ ($p_i^\prime \subset p_i$), $p_i^\prime$ can be completely different from $p_i$ (i.e., $p_i^\prime \not\subset p_i$), or $p_i^\prime$ can be an empty set (i.e., $p_i^\prime= \emptyset$). In contrast to the parameter set $p_i$, $p_i^\prime$ is usually set to empty because the parameters affecting the evaluation are hidden from the solution.

The evaluation engine can be formally defined as follows.

\paragraph{Evaluation Engine:} The evaluation engine is a function that executes a binary file ($b$) with a test instance $\tau_i$, denoted $e(b, \tau_i)\longrightarrow (v_i, s_i, l_i)$. This function returns the evaluation verdict $v_i$, the evaluation score $s_i$ ($s_i \in \mathbb{R}$), and a list of statistics $l_i$ for the entire execution process.

The evaluation verdict $v_i$ can be one of the following: Accepted (AC), Time Limit Exceeded (TLE), Memory Limit Exceeded (MLE), Wrong Answer (WA), Runtime Error (RE), Presentation Error (PE), and Output Limit Exceeded (OLE). Details of these evaluation verdicts can be found in \cite{s_ref135}. Many OJ systems evaluate submissions according to the ICPC rule, which evaluates the submissions on each test instance $\tau_i$ in a binary way, i.e., correct or incorrect. 
In this case, the evaluation score $s_i$ is always 0 (i.e., $s_i=0$) because $s_i$ is not used in evaluation according to ICPC rules \cite{s_ref1}.
In addition, $s_i$ characterizes the evaluation even if users have received the verdict "Accepted" (i.e. $v_i=AC$) for a particular instance. In general, the test instances $\tau_i$ can be characterized by the different values of $s_i$, but there are also more complex scoring methods that can be applied. Some evaluation scoring methods are presented in \cref{sec:aes:eval-score}. Finally, the statistics $l_i$ collects information about the maximum utilization of resources, such as memory and time during the execution. If OJ does not share this information with the user, $l_i$ can be empty, $l_i=\emptyset$.

\subsection{Evaluation Score} \label{sec:aes:eval-score}

Evaluation score is basically the aggregated evaluation of $s_i$ and $v_i$ to calculate the final score $s$ and verdict $v$ of each user submission. The final score of a submission is used to rank the solutions to the problem.

A solution receives an AC verdict if and only if it passes all test instances (\cref{verd1}). Otherwise, the solution receives another verdict that is different from AC (e.g., WA, TLE) (\cref{verd2}).

\begin{equation} \label{verd1}
      v=AC \Longleftrightarrow \forall_i v_i=AC 
\end{equation}

\begin{equation}\label{verd2}
v=v_j \Longleftrightarrow (\forall_{i<j}v_i = AC) \wedge (v_j \neq AC)    
\end{equation}

If the problem is not an optimization problem, the total evaluation score $s$ is calculated by the following two ways. (1) According to the ICPC evaluation rules, it is a binary evaluation of the submission, and thus, $s = 0$. (2) Otherwise, it is the sum of the evaluation scores for all correctly solved instances, denoted as \cref{sum_score_calculation}.

\begin{equation} \label{sum_score_calculation}
    s=\sum_{i=1}^{|\mathcal{T}|} 
\begin{cases}
s_i, &\text{$if$ $v_i=AC$}\\
0, &\text{$otherwise$}
\end{cases}
\end{equation}

When calculating the score for the optimization problems, the best solution among all the submissions submitted by the participants is taken into account. In most cases, \cref{optim_score_calculation} is used to calculate the evaluation score $s$ for the optimization problems where the objective function is maximized \cite{s_ref1}. Here, $b_i$ denotes the score of the best solution at the $i^{th}$ instance. Usually, \cref{optim_score_calculation} is used to rank all submitted solutions to the problems.

\begin{equation} \label{optim_score_calculation}
    s=\frac{100}{|\mathcal{T}|}\sum_{i=1}^{|\mathcal{T}|} 
\begin{cases}
\frac{s_i}{b_i}, &\text{$if$ $v_i=AC$}\\
0, &\text{$otherwise$}
\end{cases}
\end{equation}

In addition, more customized and complex evaluation procedures can be used for scoring. For example, the Polish Olympiad in Informatics uses \cref{poi_calculation}, which penalizes the solution if it takes more than half of the time limit \cite{s_ref1}, where $\Upsilon_i$ denotes the maximum point that can be assigned to a test instance $\tau_i$, $\Lambda_i$ is the maximum time limit for a solution, and $\lambda_i$ is the processing time required for the solution to produce output for an instance $\tau_i$.

\begin{equation} \label{poi_calculation}
    s=\Upsilon_i ~.~ min\Biggl(1.0,2.0.\frac{\Lambda_i - \lambda_i}{\Lambda_i}\Biggl)
\end{equation}

\section{Classification of Automated Code Evaluation Systems} \label{sec:classification:aes}
In this section, we present the classification of AESs on the basis of application domains. Considering the programming contests, there are few literature reviews on the classification of contests organized using AES. Pohl \cite{s_ref48} was the first to propose a classification of programming contests based on criteria such as contest style, duration, submission and evaluation methods, scoring, and entrance. Furthermore, classifications based on programming contests, types of programming exercises, and characteristics of the AES system have also been discussed in studies \cite{s_ref49,s_ref50}. However, most of these classifications are limited to a single application such as education or programming contests. There is no clear or explicit classification of AESs based on their potential applications that can be useful for users. Therefore, we decided to classify AESs based on their applications as follows.

\subsection{Competitive Programming Contest}
OJ systems have a wide range of applications for competitive programming. Many educational institutions use this platform to prepare their students to participate in competitive programming contests. Competitive programming contests are also held by various organizations and have gained popularity. The first OJ is UVa \cite{s_ref51}, which received great attention worldwide. It was founded in 1995 at the University of Valladolid in Spain. Based on the collected UVa dataset, Skiena and Revilla \cite{s_ref52} wrote the book "Programming Challenges: The Programming Contest Training Manual" to help students in programming contests. Judge0 \cite{nref59} is an open-source\footnote{\url{https://github.com/judge0/judge0}}, scalable, and powerful online code execution tool that can be used in a wide range of programming-related applications such as programming competitions, e-learning platforms, recruitment, and online code editors and IDEs. A partial list of OJ systems is given in \cref{OJ_competitive_programming}.

\begin{table*} [!t]
\centering
\caption{OJ systems used for the competitive programming contests\label{OJ_competitive_programming}}
\begin{tabular}{|p{3.5cm}|p{2.2cm}|p{1.8cm}|p{2.1cm}|p{1.4cm}|p{1.5cm}|}
\hline
\textbf{Name} & \textbf{In Operation} & \textbf{Language} & \textbf{\# Problems} & \textbf{\# Users} & \textbf{Founded}\\
\hline
UVa Online Judge & Yes & Eng & 4,300 & 100,000 & 1995  \\ \hline
Aizu Online Judge (AOJ) &	Yes &	Eng, Jpn &	3,000 &	130,000 &	2004 \\ \hline
National Tsing Hua University Online Judge &	Yes &	Eng &	10,000 &	- &	2015 \\ \hline
National Taiwan University Online Judge &	Yes &	Chi &	2600 &	600 &	2016 \\ \hline
Sphere Online Judge (SPOJ) &	Yes &	Eng, Pol, Por, Vie &	20,000 &	315,000 &	2004 \\ \hline
PKU Judge Online (POJ) &	Yes &	Eng, Chi &	3,000 &	250,000 &	2003 \\ \hline

Topcoder &	Yes &	Eng &	5,200 &	4,000 &	2001 \\ \hline
Codeforces &	Yes &	Eng, Rus &	3,000 &	600,000 &	2010 \\ \hline
AtCoder & Yes & Eng, Jpn & 5,900 & 400,000 & 2012 \\ \hline

Google Code Jam (GCJ) &	Yes &	Eng &	450 &	670,000 &	2003 \\ \hline
Facebook Hacker Cup &	Yes &	Eng &	- &	80,000 &	2011 \\ \hline

HUSTOJ &	Yes &	Eng, Chi &	650 &	26,000 &	2014 \\ \hline

Timus Online Judge &	Yes &	Eng &	1,000 &	110,000 &	2000 \\ \hline
IEEEXtreme &	Yes &	Eng &	- &	-&	2006 \\ \hline
ICPC &	Yes &	Eng &	- &	-&	1970 \\ \hline
IOI &	Yes &	Eng &	- &	-&	1989 \\ \hline
Judge0 & Yes &	Eng &	- &	-&	2017 \\ \hline
LeetCode & Yes &	Eng &	- &	-&	2015 \\ \hline

\end{tabular}

\end{table*}

Furthermore, data science, AI and ML experts compete against each other on various online platforms, adding a new dynamic to competitive programming. In particular, AI game competitions are known as bot programming competitions or AI programming competitions. Many online platforms are used for AI, ML, and data science competitions, including the Kaggle platform, which is mainly used for data science and ML competitions. Google AI Challenge is a biennial competition for students; Battlecode is one of the longest-running AI programming competitions hosted by the Massachusetts Institute of Technology; Lux AI is an AI programming competition held on the Kaggle platform. There are also numerous platforms and AI/ML challenges such as Halite, CodinGame, AI Coliseum, Russian AI Cup, AWS DeepRacer, SamurAI, CodeCup, Coder One, Terminal by Correlation One, SIGNATE, and so on.

\subsection{Academic Tool}

Recently, OJ systems have emerged as academic tools for programming learning, programming assignments, and assessment. Teachers/instructors at many educational institutions automatically grade students' assignments using OJ. The benefits of using OJ system in education are innumerable. For example, the submitted solution codes are checked with higher accuracy, no wrong solutions are accepted, students can get their result immediately, and the teacher can take action to improve students' programming skills based on the result. A successful application of the OJ system for algorithms and data structures and analytical investigation based on the collected data is presented in studies \cite{s_ref8,s_ref9,s_ref53}. Peking University has integrated POJ \cite{wen2005peking} as an essential tool for programming education. POJ is being utilized in various programming courses, such as Introduction to Computing, Problem Solving and Programming, and Data Structures and Algorithms. Ala-Mutka \cite{s_ref54} gave a detailed review of the application of OJ systems in education. A review \cite{s_ref55} addresses the available software for automatic assessment of programming tasks that may be of interest for learning programming. Fonte et al. \cite{s_ref56} presented the advanced version of OJ system that can be used to provide valuable feedback to programmers to help them understand where the problem lies and how to improve the solution code. Additionally, literature \cite{nref2,nref3,nref4,nref5} explores topics such as code plagiarism detection, automated code correction, and personalized feedback through OJ systems.

In contrast, the limited scope of black box testing and oversimplified feedback (correct or incorrect) restricts its widespread application in programming education. To overcome this constraint, Zhou et al. \cite{nref6} introduced a novel OJ system framework that enables effective programming education and provides comprehensive assessment of submitted solutions. The new OJ system comprises four modules, including code quality checking, similarity checking, personalized feedback for students, and advising on teaching adjustments. This OJ system has been implemented in a programming in C course for over ten years and has demonstrated substantial improvements in students' programming abilities. 
Lu et al. \cite{s_ref58} showed the positive influence of the OJ system, which increases the performance level and also arouses students' interest in programming throughout the year. Teachers are increasingly incorporating OJ systems into their daily programming teaching activities. Although the performance of OJ systems is impressive, the contributions and effects of teachers are invaluable and irreplaceable. OJ systems can serve as excellent tools to assist teachers in programming education. By providing reliable, high-efficiency, and objective evaluations, as well as fair and accurate scoring, OJ systems can greatly reduce teachers' workload \cite{nref6}. Teachers can utilize OJ systems to assign programming tasks, analyze student performance through reports and statistics, identify code similarities, and more. For example, in a recursive algorithm lesson, students may be given a problem that can be solved using either brute force or a recursive algorithm. If the teacher finds that most students used the brute force algorithm (which is relatively easy than recursive algorithm), they may reinforce the use of the recursive algorithm to improve students' algorithmic skills. \cref{OJ_education} shows a partial list of OJ systems that have been incorporated into programming education.

\begin{table} []
\centering
\caption{OJ systems used for the Education\label{OJ_education}}
\begin{tabular}{|p{3cm}|p{2.3cm}|p{2.5cm}|p{2.3cm}|p{1.5cm}|}
\hline
\textbf{Name} & \textbf{In Operation} & \textbf{Language} & \textbf{\# Problems} &  \textbf{Founded}\\
\hline

UVa Online Judge &	Yes &	Eng &	4,300 &	1995 \\ \hline
Aizu Online Judge &	Yes &	Eng, Jpn &	3,000 &	2004 \\ \hline
Jutge.org &	Yes &	Eng, Esp, Fre, Cat, Dut &	4,843 &	2006 \\ \hline
POJ &  Yes & Eng, Chi & 3,000 & 2003 \\ \hline 
CodeChef &	Yes &	Eng &	3,000 &	2009 \\ \hline
CodeHunt &	Yes &	Eng &	8,300 &	2014 \\ \hline
Codecademy &	Yes &	Eng &	- &	2011 \\ \hline
CodeWars &	Yes &	Eng &	1,200 &	2012 \\ \hline
URI Online Judge &	Yes &	Eng, Spa, Por &	1,100 &	2011 \\ \hline

Hanghou Dianzi University (HDU) OJ & Yes & Eng, Chi & 6,000 & - \\ \hline

Judge0 & Yes & Eng & - & 2017 \\ \hline
LeetCode & Yes &	Eng &	- &	2015 \\ \hline

\end{tabular}

\end{table}

\subsection{Recruitment and Selection}

In recent years, the use of technologies such as big data analytics, AI, and OJ in the recruitment process  is becoming more common among companies and organizations \cite{nref7}. These technologies provide rigorous, effective, and cost-efficient ways to find suitable candidates. One study \cite{nref7} explores how prescreening applications, practical skills testing, and candidate communication can be handled using AI. Furthermore, technology in the different phases of the recruitment and selection process is presented in the study \cite{nref8}. The phases are attracting, screening, selection, and on-boarding and socialization. 
There are numerous platforms that use OJ systems to support the recruiting process. These platforms are mainly commercial and automatically evaluate the submitted codes and rank the programmers. We present some OJ systems that are used for recruitment. For example, \emph{HackerEarth}, \emph{HackerRank}, \emph{Qualified}, \emph{CodeEval}, \emph{Codility}, \emph{Track Test}, and so on. \emph{HackerEarth} is an online platform dedicated to hiring talented developers, hosting crowdsourcing-based ideas, and organizing hackathons. \emph{Codility} helps hiring managers to find the best developers from a large pool of skilled programmers in the shortest possible time. A list of OJ systems applicable to the recruitment and selection process can be found in \cite{s_ref1}.   

\subsection{Online Compilers}

Another category of OJ systems are online compilers, where the codes developed in different languages by user solutions can be compiled and executed remotely through a browser. \emph{Codeanywhere} is one of the most feature-rich online compilers that offers a dedicated custom development environment and real-time collaboration. It allows users to automatically connect to GitHub, Amazon Cloud, FTP servers and Bitbucket. \emph{Coding Ground} offers a full-featured IDE that allows users to edit, compile and run their projects in the cloud. Codio is a cloud-based IDE that supports a large number of programming languages, can be integrated into e-learning platforms, and can be used to detect plagiarism. Furthermore, there are many online compilers whose goal is to provide support for code compilation without any physical configuration. CodeSkulptor, for example, is an online compiler for the Python programming language and also supports the learning process of programmers. The Codepad compiler allows users to share their code with other collaborators via URLs. In addition, C++ Shell, Web compiler, OnlineGDB C, Tutorialspoint, Replit, Rextester, myCompiler, OneCompiler, Online Compiler CodeChef,  and Techiedelight are compilers used for online code compilation.

\subsection{Development Platforms}
In contrast, many OJ development platforms are available to host programming competitions or educational activities in local infrastructure. \emph{DOMjudge} is a well-known OJ development platform that can be easily installed to host programming contests. It allows users to prepare and run programming contests according to ACM ICPC rules. Mooshak \cite{s_ref57} is an automatic judge and full-fledged contest manager. It is an open system that can be installed on a Linux operating system with Apache HTTP server. The architecture of Mooshak is suitable not only for small competitions with one server, but also for competitions with multiple sites. CloudCoder, a web-based platform inspired by CodingBat \cite{nref9}, is designed to assess students' submitted code assigned by the instructor. At Tsinghua University, an online assessment system was developed for university-level code evaluation \cite{nref10}, allowing users to add new problems. However, the code resources of this system are not open source. Xavier and collaborators described an OJ platform developed at the University of Porto, which drew inspiration from DOMjudge and Moodle \cite{nref11}. They reviewed the OJ development platforms. Furthermore, various plug-ins for the Moodle platform have been created for code evaluation, such as Code Runner, Virtual Programming Lab, and Online Judge plug-in for Moodle \cite{nref12,nref13}. \cref{OJ_compilers_platforms} provides a list of online compilers and development platforms.

\begin{table*} [!t]
\centering
\caption{OJ systems are used for online compilers and development platforms\label{OJ_compilers_platforms}}
\begin{tabular}{|p{3.5cm}|p{2.3cm}|p{1.5cm}|p{1.8cm}|p{2.3cm}|}
\hline
\textbf{Name} & \textbf{In Operation} & \textbf{Founded} & \textbf{Compiler} &  \textbf{Development Platform}\\
\hline

Codeanywhere &	Yes & 2013 & $\checkmark$ &	$\times$ \\ \hline
Coding Ground &	Yes & 2006 & $\checkmark$ &	$\times$ \\ \hline	
DOMJudge &	Yes &	2004 &	$\times$  &	$\checkmark$ \\ \hline
C ++ Shell & Yes & 2014 & $\checkmark$ & $\times$ \\ \hline
Mooshak & 2015 & 2005 &	$\times$  &	$\checkmark$ \\ \hline
SIO2 &	Yes &	2012 &	$\times$  &	$\checkmark$ \\ \hline
Ideone & Yes & 2009 &  $\checkmark$ &	$\times$ \\ \hline	
CodeChef & Yes & - & $\checkmark$ & $\times$ \\ \hline

Virtual Programming Lab & Yes & 2012 &	$\times$  &	$\checkmark$ \\ \hline

Online Compiler & Yes & 2009 & $\checkmark$ & $\times$ \\ \hline
A+ & Yes & 2017 &	$\times$  &	$\checkmark$ \\ \hline
Codio &	Yes &	2013 & 	$\checkmark$ &	$\times$ \\ \hline	
TestMyCode & Yes & 2013 &	$\times$  &	$\checkmark$ \\ \hline
Programiz &	Yes &	- &	 $\checkmark$ &	$\times$ \\ \hline	
xLx & Yes & 2001 &	$\times$  &	$\checkmark$ \\ \hline
Web-CAT & Yes & 2003 &	$\times$  &	$\checkmark$ \\ \hline
Web Compiler & Yes & 2014 & $\checkmark$ & $\times$ \\ \hline
Codepad & Yes & 2008 & $\checkmark$ & $\times$ \\ \hline
BOSS & Yes & 2012 &	$\times$  &	$\checkmark$ \\ \hline
CodeSkulptor &	Yes & 2012 & $\checkmark$ &	$\times$ \\ \hline	
CloudCoder & Yes &	2012 &	$\times$  &	$\checkmark$ \\ \hline
Tsinghua Online Judge &	Yes &	2012 &	$\times$  &	$\checkmark$ \\ \hline
paiza.io & Yes & 2014 &	$\checkmark$ & $\times$ \\ \hline

Judge0 & Yes & 2017 & $\checkmark$  & $\checkmark$  \\ \hline

\end{tabular}

\end{table*}

\section{Resources of Automated Code Evaluation Systems} \label{sec:resources}

In this section, we present a consolidate overview of AESs datasets and tools. Since AESs are regularly used for a variety of purposes, including recruitment, programming education and contests, data mining, and research, a huge number of data resources (codes, problems, analysis results, scores, submission logs, etc.) are generated.  Kaggle is an online platform for data scientists and ML programmers \cite{nref14}. It is also used to organize data mining and ML competitions. Kaggle allows users to collaborate with others, publish datasets, find datasets of others, access GPU-integrated notebooks, and compete against each other to solve data science problems. There are many data science and ML codes and datasets (e.g., sports, health, travel, software, and food) that are available (open source) and mostly reliable. DREAM Challenges is another platform for solving biological problems, especially scientific challenges, and is therefore targeted at researchers only \cite{nref16,nref17,nref18}. OpenML \cite{nref15} is an open platform for sharing datasets, code, algorithms, and experimental results. It has a rich and diverse data repository that stores more than 5,000 datasets and supports Python, Java, REST API and many other scientific programming languages. Recently, Hugging Face~\footnote{\url{https://huggingface.co/}.} has gained attention as the open-source platform for sharing ML models, datasets, evaluation metrics, and demo applications for open science. Since the platform is a Git-based code repository similar to GitHub, researchers can easily share their own models and datasets. In addition, users can easily try the public models and datasets through Python libraries.

Furthermore, numerous solution codes and logs are available on various platforms (AOJ, AtCoder, etc.) and can be used for education and programming research. These real-world and rich data resources have become attractive for coding, education, learning analytics, ML, and data mining research \cite{s_ref9,s_ref59}. In a study \cite{s_ref8}, a comprehensive data-driven analysis based on OJ data was conducted. The experimental results show the shortcomings of students' programming and the scope for possible improvements.  Hsiao et al. \cite{s_ref60} leveraged an educational learning analysis tool called "Web Programming Grading Assistant (WPGA)" to study the effectiveness of students' programming learning. Rahman et al. \cite{s_ref9} conducted educational data mining using OJ data to support programming learning. However, benchmark datasets have had a significant impact on the growth of coding-related research using ML. Here we present some benchmark datasets for code intelligence research. Code representation is a fundamental activity for preparing source code that is compatible with ML models. It involves converting code into a numerical representation to feed ML models to solve specific tasks (e.g., pattern recognition, method name prediction, and comment classification). code2vec~\cite{nref19} and GHTorrent \cite{nref20} can be useful datasets for code representation tasks. Code quality assessment is a broad category of coding tasks with subcategories such as clone detection, smell detection, quality assessment and prediction. BigCloneBench, Multi-label Smells, DL Smells, QScored, and ML for software refactoring datasets can be useful for code quality assessment \cite{s_ref15,nref22,nref23,nref24,nref25}.  Vulnerability analysis is another important area of code analysis that is used to determine whether the code is vulnerable or not.  Datasets such as Draper VDISC, Genome, TRL, and Project-KB \cite{nref26,nref27,nref28,nref29} are used to identify vulnerabilities in code. Code testing datasets such as Defect4j, BugDetection, DeepBugs, and DTLDP \cite{nref30,nref31,nref32,nref33} can be useful for the testing task.

In addition, Code search \cite{s_ref61,s_ref62,s_ref63} is important when programmers want to use other codes. This system automatically finds semantically similar codes based on a natural language query. The code completion system \cite{s_ref14,s_ref64,s_ref65,s_ref66,s_ref67} can help programmers automatically complete their code. Also, the code-to-code translation system \cite{s_ref68,s_ref69,s_ref70} assists programmers translate their code from one language to another (e.g., Python to Java and Java to Python). As the use of real datasets for coding-related research increases, therefore, we present a list of available datasets from platforms such as OJ, contest platforms, Stack Overflow, and GitHub in \cref{OJ_benchmark_dataset}.

\begin{table*} []
\centering
\caption{A list of datasets and their application in coding analysis \label{OJ_benchmark_dataset}}
\begin{tabular}{|p{.5cm}|p{3cm}|p{5.5cm}|p{4cm}|}
\hline
\textbf{Sl.} & \textbf{Dataset Name} & \textbf{Coding Tasks} & \textbf{Size} \\
\hline

1 &	CodeNet \cite{s_ref41} &	code classification, code similarity &	14 million \\ \hline

2&	Aizu \cite{s_ref42} &	code classification, code-to-code translation, code completion, refactoring, summarization	& 8 million \\ \hline
3 &	AtCoder \cite{s_ref43} & code classification, code-to-code translation, code completion, refactoring, summarization &	7.5 million \\ \hline
4 &	BigCloneBench \cite{s_ref71} & code clone	 & 6 million (true clone pair) and 260,000 (false clone pair) \\ \hline
5 &	POJ-104 \cite{s_ref44} & code classification, code similarity, code clone & 52,000 \\ \hline
6 & GCJ~\cite{ullah2019gcj} & code plagiarism detection & 2.4 million \\ \hline
7 &	PY150 \cite{s_ref72} & code completion &	150,000 \\ \hline
8 &	Devign \cite{s_ref73} & fault detection &	27,318 \\ \hline
9 & QuixBugs~\cite{lin2017quixbugs} & code repair & 40 \\ \hline
10 &	Bugs2Fix \cite{s_ref74} &	code repair & 122,000 \\ \hline
11 &	CodeSearchNet \cite{s_ref62} & code summarization & 	1.1 million \\ \hline

12 & CodeContests~\cite{li2022alphacode} & text-to-code generation & 4.5 million \\ \hline 
13 & APPS~\cite{hendrycks2021apps} & text-to-code generation & 10,000 \\ \hline 
14 & CONCODE~\cite{iyer2018concode} & text-to-code generation & 104,000 \\ \hline
15 & MBPP~\cite{austin2021program} & text-to-code generation & 974 \\ \hline 
16 & MathQA-Python~\cite{austin2021program} & text-to-code generation & 23,914 \\ \hline 
17 & HumanEval~\cite{chen2021codex} & text-to-code generation & 164 \\ \hline
18 & HumanEval-X\footnote{\url{https://huggingface.co/datasets/THUDM/humaneval-x}.} & text-to-code generation & 820 \\ \hline
19 & HumanEval Infilling~\cite{bavarian2022efficient} & code infilling & 8,488 \\ \hline
20 & code-docstring-corpus~\cite{miceli2017parallel} & text-to-code generation, code summarization & 150,370 \\ \hline
21 & CoNaLa~\cite{yin2018conala} & text-to-code generation, code summarization & 2,879 (annotated) and 598,237 (not annotated) \\ \hline
22 & MCoNaLa~\cite{wang2023mconala} & text-to-code generation, code summarization & 896 \\ \hline
23 & MBXP~\cite{athiwaratkun2022multilingual} & text-to-code generation, code summarization, code translation, code infilling, prompt robustness & --- \\ \hline
24 & CodeXGLUE \cite{s_ref47} & clone detection, defect detection, cloze test, code completion, code translation, code search, code repair, code summarization, text-to-code generation, document translation &	--- \\ \hline
25 & StaQC \cite{nref34} & code repair &  148K Python and 120K SQL question-code pairs from StackOverflow \\ \hline
26 & IntroClass \cite{nref35} & code repair &  --- \\ \hline
\end{tabular}

\end{table*}

\section{Code Analysis Using Machine Learning} \label{sec:code-analysis:ml}
According to Evans Data Corporation\footnote{\url{https://evansdata.com/press/viewRelease.php?pressID=278}}, there were approximately 23.9 million professional developers in 2019, and that number is expected to reach approximately 28.7 million by the end of 2024 \cite{s_ref47}. The ML-based code intelligence can be used to improve the productivity of a growing number of professional programmers. At the same time, benchmark datasets have a significant impact on the growth of applied ML research. Recently, researchers have begun to utilize statistical models such as neural networks for code analysis tasks. Also, the application of pre-trained models that learn from large programming data (e.g., code), such as BERT~\cite{devlin2019bert}-based models~\cite{liu2019roberta,feng2020codebert,kanade2020cubert,wang2021syncobert,jiang2021treebert,guo2021graphcodebert} and GPT~\cite{radford2018gpt,radford2019gpt2,brown2020gpt3}-based models~\cite{chen2021codex,ouyang2022instructgpt}, have achieved great success in a wide range of coding tasks. With the growth of resources, datasets, tools, and methods, code analysis research is also expanding. Therefore, \cref{code_analysis_tasks_ML} summarizes recent attempts of code analysis tasks using ML.

\begin{table*} []
\centering
\caption{Machine learning approaches for coding analysis tasks \label{code_analysis_tasks_ML}}
\begin{tabular}{|p{.5cm}|p{5.5cm}|p{5.5cm}|}
\hline
\textbf{Sl.} & \textbf{Code Analysis Task} & \textbf{Article}\\
\hline

1 & Defect Detection &	 \cite{s_ref73,s_ref75,s_ref76,s_ref77,s_ref78,s_ref79,s_ref80,s_ref81,s_ref82}  \\ \hline
2 & Clone Detection & \cite{s_ref71,s_ref88,s_ref89,s_ref90,s_ref92,s_ref93,s_ref94,s_ref95,s_ref96,s_ref97} \\ \hline
3 & Code Completion &	\cite{s_ref14,s_ref35_ref101,s_ref64,s_ref65,s_ref66,s_ref67,s_ref100,s_ref102} \\ \hline
4 & Code Repair &	\cite{s_ref12,s_ref74,s_ref103,s_ref104,s_ref106,s_ref108,s_ref109,s_ref110,s_ref111}\\ \hline
5 & Code Search &	\cite{s_ref61,s_ref62,s_ref112,s_ref113,s_ref115,s_ref117} \\ \hline
6 & Code Summarization &	\cite{s_ref118,s_ref119,s_ref120,s_ref121,s_ref122,s_ref123,s_ref124} \\ \hline
7 & Text-to-Code Generation &	\cite{s_ref119,s_ref124,s_ref126,s_ref127,s_ref129,s_ref130}\\ \hline
8 & Code Classification &	\cite{s_ref12,s_ref14,s_ref17,s_ref131,s_ref132,s_ref133,s_ref134} \\ \hline

\end{tabular}

\end{table*}

A brief description of the coding tasks with ML is as follows. The purpose of the task \emph{defect detection} is to identify errors/defects within the body of the source code. Classification of codes as buggy or not is based on the identification of errors in the code. Compiling the code is not enough to detect all errors in the code, because logic errors may exist in the code after compilation. Sometimes, the entire code needs to be manually inspected to detect logic errors, which is time-consuming and costly. ML models are able to detect these logic errors in the code based on a large number of correct code corpus. Ghassan \cite{nref63} proposed a practical approach to detect logic errors in code for object-oriented environments such as C\# .Net Framework. To avoid logic errors, the proposed environment includes some predefined behaviors using Xceed, Alsing, and Mind Fusion Components. Al-Ashwal et al. \cite{nref65} presented a tool for identifying logic errors in Java code using both static and dynamic methods. Song et al. \cite{nref64} proposed an automatic logic error detection method for functional programming tasks using correct reference solutions for each counter assignment. This method was very efficient than manually identifying logic errors in code. The BiLSTM model is used to detect both logical and syntactic errors in code \cite{s_ref12}.

\emph{Clone detection} is used to identify semantically similar code. It has two subtasks: searching for similar codes and classification. Code cloning is a well-known approach of reusing source code in software development.  Ain et al. \cite{nref66} provided a comprehensive review of the latest tools, trends, and techniques in the field of code clone detection. Code clones pose a challenge to software maintenance because verification is required for similar segments when a bug is identified in one segment. Feng and collaborators \cite{nref67} proposed a token-based approach to clone detection using a DL model called CCLEARNER. A classifier is trained using known method-level code clones and non-clones and then used the trained classifier to detect code clones of a given code.  Fang et al. \cite{nref70} proposed a supervised DL model for function-level code clone detection. They proposed a joint code representation that uses fusion embedding techniques to learn the unseen semantic and syntactic features from the source code. The proposed fusion learning approach outperformed state-of-the-art models in detecting function-level code clones on a large C++ program dataset.  White et al. \cite{nref68} introduced a learning-based code clone detection technique using DL.  Moreover, a systematic review of DL applications for code clone detection has been conducted in the study \cite{nref69}.

\emph{Code completion} is another task that helps programmer complete code correctly. Programmers sometimes get confused what to write next, and in such cases, code completion helps them to complete the code. Code completion can be done in two ways ($i$) token-level completion and ($ii$) line-level completion.  \emph{Code repair} task identifies bugs and automatically fix them. Typically, they identify bugs in the code and fix the code according to the context of the code. Rahman et al. \cite{s_ref12} proposed a BiLSTM la nguage model for code evaluation and repair. The model is trained on a large number of correct codes and used for code evaluation and repair. The model also proposes correct words as replacements for error words in the code Li et al. \cite{nref71} proposed a DL model called DEAR that can repair/fix common errors that require dependent changes in the subsequent one or multiple segments in the code. They leveraged a tree-based LSTM model and a divide-and-conquer strategy to understand the actual context of the code.

\emph{Code search} is a task that measures the semantic relevance between the code and natural language. This is the activity of searching the code based on a natural language query. Gu et al. \cite{s_ref61} proposed a novel DL model called CODEnn for code search. The model embeds code and its natural description together in a high-dimensional vector space. Code snippets can be retrieved for a natural language query based on their high-dimensional vector space, which ensures a more relevant code search. Ling et al. \cite{nref72} proposed an adaptive model for deep code search (AdaCS), which is trained once and applied to different code bases. AdaCS divides the learning procedure into two main steps: embedding domain-specific words and identifying common syntactic patterns for matching. Experimental results show that AdaCS exhibits improved resilience and outperforms existing state-of-the-art deep code search techniques when applied to unseen software projects that are not included in the training dataset. 

\emph{Code summarization} provides abstracted summary of the code and helps to better understand the program code and software maintenance. Therefore, producing high-quality annotations is a major challenge for researchers.  Liu et al. \cite{nref73} proposed a novel code summarization method based on neural networks using call dependency information of the source code. All call dependencies were extracted and converted into a sequence of tokens for the Seq2Seq model for code summarization. The experimental results demonstrated that the model understands the call dependency during the code summarization task. \emph{Text-to-code} task is used to generate codes based on the input of the natural language description. \emph{Code classification} can be done in many ways, such as classifying source code based on programming language, application, and error.

To perform coding tasks, researchers have developed various ML models and tools and have had great success in the tasks of code representation, code search, code understanding, code summarization, quality assessment, vulnerability assessment, testing, and code synthesis. Here we present a list of tools and ML models used to analyze code, as shown in \cref{tools_ML_models}. Each model/tool is listed with the actual reference, citation, year of first appearance online, and a short description. All metadata for these models/tools was collected manually by searching author websites, electronic libraries, and publisher websites.

\begin{table} []
\centering
\caption{A list of ML models and tools used to analyze code \label{tools_ML_models}}
\begin{tabular}{|p{2.2cm}|p{3.2cm}|c|p{1.2cm}|p{4.5cm}|}
\hline
\textbf{Tasks} & \textbf{Tools/ML Models} & \textbf{\#Citations} & \textbf{Year of Pub.} &  \textbf{Description}\\
\hline

Code Search &	Deep Code Search \cite{s_ref61} &	418 &	2018 &	Code embedding technique is used to search code \\ \hline

\multirow{4}{*}{\adjustbox{stack=ll}{Code Representation}} &	Graph-based code modeling \cite{nref36} &	702 & 2018 & Code modeling with graph \\ \cline{2-5}

& Code2vec \cite{nref19} & 929 & 2019 & Generate distributed representation of the code \\ \cline{2-5}

& Vocabulary learning on code \cite{nref37} & 40 & 2019 & Generate AST from Java code \\ \cline{2-5}

& CC2vec \cite{nref38} & 110 & 2020 &  Use the representation of code changes\\ \hline

\multirow{5}{*}{\adjustbox{stack=ll}{Quality Assessment}} & SVF \cite{nref39} & 269 & 2016 & Interprocedural static value-flow analysis in LLVM \\ \cline{2-5}

&  ML for software refactoring \cite{s_ref15} & 47 & 2020 & Explores the effectiveness of ML in predicting software refactoring \\ \cline{2-5}

&  CREC \cite{nref40} & 32 & 2018 & Automated code clone recommendation for refactoring\\ \cline{2-5}

&  SMAD \cite{nref41} & 30 & 2020 & SMAD is used to detect two renowned anti-patterns: God Class and Feature Envy\\ \cline{2-5}
&  DL smells \cite{nref23} & 32 & 2021 & Detects code smells by using DL and transfer learning\\ \hline

\multirow{3}{*}{\adjustbox{stack=ll}{Vulnerability Assessment}} & VCCFinder \cite{nref42} & 219 & 2015 & Detects vulnerable codes in repositories\\ \cline{2-5}

 & SWAN \cite{nref43} & 15 & 2019 & Detects vulnerable codes\\ \cline{2-5}
  & WAP \cite{nref44} & 9 & 2013 & Detects and corrects vulnerable codes\\ \hline

\multirow{6}{*}{\adjustbox{stack=ll}{Code Understanding}} & CodeNN \cite{s_ref122} & 607 & 2016 & Neural attention based model is used to summarize the code\\ \cline{2-5}

& ChangeScribe \cite{nref45} & 171 & 2014 & Automatically generate commit messages based on code changes\\ \cline{2-5}

& DeepSim \cite{nref46} & 192 & 2018 & DL is used to find code similarity\\ \cline{2-5}

& NeuralCodeSum \cite{nref47} & 226 & 2020 & Source code summarization using Transformer\\ \cline{2-5}

& CODES \cite{nref48} & 118 & 2012 & Source descriptions are mined from programmers' communication\\ \cline{2-5}

& Rencos \cite{nref49} & 132 & 2020 & Code summarization based on neural network\\ \hline

\multirow{2}{*}{Testing} & AppFlow \cite{nref50} & 61 & 2018 & Using ML to synthesize robust reusable UI tests\\ \cline{2-5}

& DeepFuzz \cite{nref51} & 57 & 2019 & Automatic generation of syntax for Fuzz testing\\ \hline

\end{tabular}

\end{table}

\section{The AOJ: an example of an Automated Code Evaluation System} \label{sec:aes-example:aoj}

The Aizu Online Judge (AOJ)~\cite{s_ref42} is an AES that has been in operation for more than 11 years. It has a wide range of applications such as academic, programming contests, regular practice and research. AOJ has over 130,000 registered users who perform programming activities regularly. It has already evaluated more than eight million program codes. The University of Aizu, Japan, officially uses the AOJ platform for several programming courses as an active academic tool. In addition, AOJ has API services to access its resources for research purposes\footnote{\url{http://developers.u-aizu.ac.jp/}.}, which has led to a large number of research in education, coding analysis, data mining, and datasets based on its data \cite{s_ref8,s_ref9,s_ref12,s_ref14,s_ref17,s_ref41,s_ref131}. Considering the scalability and impact of the AOJ system on practical applications, we present the AOJ system as an example of AES in this study.

\subsection{Hardware Architecture}

\begin{figure}
    \centering
    \includegraphics[width=1\linewidth]{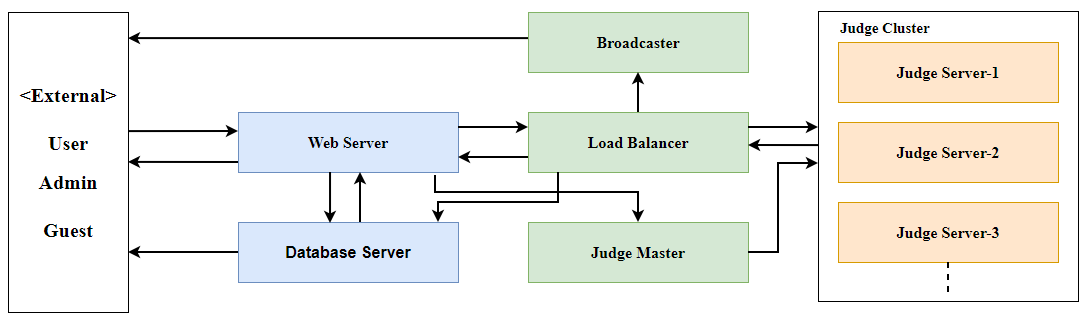}
    \caption{Component diagram of the AOJ system}
    \label{hardware_architecture}
\end{figure}

The AOJ system has a complex component architecture with multiple dedicated components that work together to provide a smooth and uninterrupted evaluation. \cref{hardware_architecture} shows the component diagram of the AOJ system, in which the web server, database server, judge server, broadcaster, and load balancer are interconnected. The web server is the interface used to communicate between the external system and the OJ core system. Basic tasks such as browsing, registration, authentication, and solution submission are implemented in the web server. It also communicates with the database server and load balancer to obtain the evaluation/judgment results of each submission. The database server may consist of one or more servers that store all information related to the evaluation, such as problem set, evaluation results, solution codes, statistical data, and user information. The database servers communicate with the web server and load balancer via private APIs. The judge master is a key server responsible for managing the judge data for the judge cluster. It provides refined judge data for all judge servers and builds the necessary \emph{checker} and \emph{reactive} programs for all judge servers. It also contains a non-stop integration mechanism and is considered one of the special components of the OJ system. Similarly, the load balancer is a core component of the OJ system to interact between the web server and the judge servers. It also communicates with the broadcaster and the database server as well.

The load balancer is primarily responsible for scheduling submissions for judging. It receives the evaluation results from the judge servers and the notification is sent to the web server, database server and broadcaster. The broadcaster is basically used to inform the users about the status of the submissions asynchronously. Finally, the main role of the judge servers is to perform the assigned tasks based on the judge data. The judge servers should be isolated and only allow authorized processes to connect. It is also important to avoid data loss due to malicious or unexpected operations. More details about these hardware components of the AOJ system can be found in \cite{s_ref135}. We have presented the system information of AOJ in \cref{hardware_settings_AOJ}. \cref{hardware_settings_AOJ} shows the AOJ servers\footnote{\url{https://onlinejudge.u-aizu.ac.jp/system\_info}.}, the configurations of each server, and the application of the servers for programming languages. This is a list of servers, including spare/unavailable servers, for quick replacement/update for seamless operation.

\begin{table*} [!t]
\centering
\caption{System information of judge servers and corresponding application for programming languages \label{hardware_settings_AOJ}}
\begin{tabular}{|p{.9cm}|p{2.5cm}|p{5.5cm}|p{2.5cm}|}
\hline
\textbf{Judge Server} & \textbf{Machine} & \textbf{Processor and RAM} & \textbf{Using for} \\
\hline

\#0 &	IBM ThinkPad &	Pentium (R) M 1.30 GHz and 1 GB	 & C, C++ \\ \hline
\#1 &	Dell PowerEdge R300 &	Dual Core Intel(R) Xeon(R) E3113 3.0 GHz and 1 GB &	C, C++ \\ \hline
\#2 &	Dell PowerEdge R300 &	Dual Core Intel(R) Xeon(R) E3113 3.0 GHz and 1 GB &	JAVA \\ \hline
\#3 &	Dell PowerEdge R210 &	Intel(R) Xeon(R) Processor E3-1270 3.40 GHz and 2 GB &	Ruby, Python, Python3, PHP, JavaScript \\ \hline
\#4 &	Dell PowerEdge R210 II & Intel(R) Xeon(R) Processor E3-1270 3.40 GHz and 4 GB &	C, C++ \\ \hline
\#5 & Dell PowerEdge R210 II & Intel(R) Xeon(R) Processor E3-1240v2 3.40 GHz and 8 GB &	C++11, JAVA, C\#, D \\ \hline
\#6 &	Dell PowerEdge R210 II & Intel(R) Xeon(R) Processor E3-1280 v2 3.6 GHz, Turbo 4C/8T and 8 GB & Ruby, Python, Python3, PHP, JavaScript \\ \hline
\#7 &	Dell PowerEdge R220 &	Intel(R) Xeon(R) Processor E3-1286 v3 3.7 GHz, Turbo 4C/8T and 8 GB & C++11, JAVA, Scala, Haskell, OCaml, C\#, D \\ \hline
\#8 &	Dell PowerEdge R220 & Intel(R) Xeon(R) Processor E3-1286 v3 3.7 GHz, Turbo 4C/8T and 8 GB & C, C++, C++14 \\ \hline
\#9 & Dell PowerEdge R210 II & Intel(R) Xeon(R) Processor E3-1280 v2 3.6 GHz, Turbo 4C/8T and 8 GB & Ruby, Python, Python3, PHP, JavaScript, Rust, Go, Kotlin \\ \hline
\#10 &	Dell PowerEdge R240 & Intel(R) Xeon(R) Processor E-2286G 4.0 GHz, Turbo 6C/12T and 8 GB & C++14, C++17, Rust \\ \hline
\#11 &	Dell PowerEdge R240 & Intel(R) Xeon(R) Processor E-2286G 4.0 GHz, Turbo 6C/12T and 8 GB & C++11, JAVA, C\#, PHP, Kotlin \\ \hline
\#12 & Dell PowerEdge R250 & Intel(R) Xeon(R) Processor E-2378G 2.8 GHz, Turbo 8C/16T and 8 GB & PyPy3 \\ \hline
\#13 & Dell PowerEdge R250 &	Intel(R) Xeon(R) Processor E-2378G 2.8 GHz, Turbo 8C/16T and 8 GB &	 \\ \hline

\end{tabular}

\end{table*}

\subsection{Software Architecture}

\begin{figure}
    \centering
    \includegraphics[width=1\linewidth]{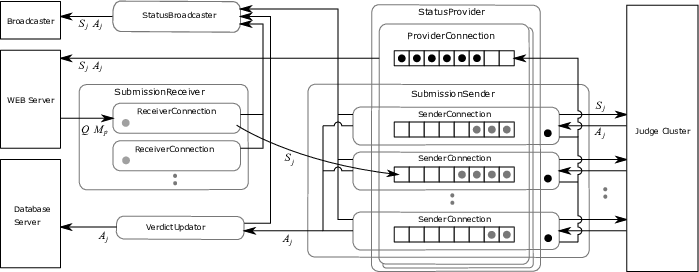}
    \caption{Internal software architecture of load balancer}
    \label{software_architecture_LB}
\end{figure}

In this section, we briefly describe the software architecture of the AOJ system, in particular the core components such as load balancer, broadcaster, and judge server. The load balancer is a core component of any OJ system. It receives submissions from the web server and distributes them to the judge server. Finally, it reports judging results using a variety of channels. Multithreaded processes are used to organize the components of the load balancer, which includes three instances such as \emph{SubmissionReceiver}, \emph{SubmissionSender}, and \emph{SubmissionProvider}. The software architecture, submission, and evaluation in the load balancer are shown in \cref{software_architecture_LB}. In addition, the details of the operation and flow of the load balancer are described in \cite{s_ref135}.

\begin{figure} [h]
    \centering
    \includegraphics[width=1\linewidth]{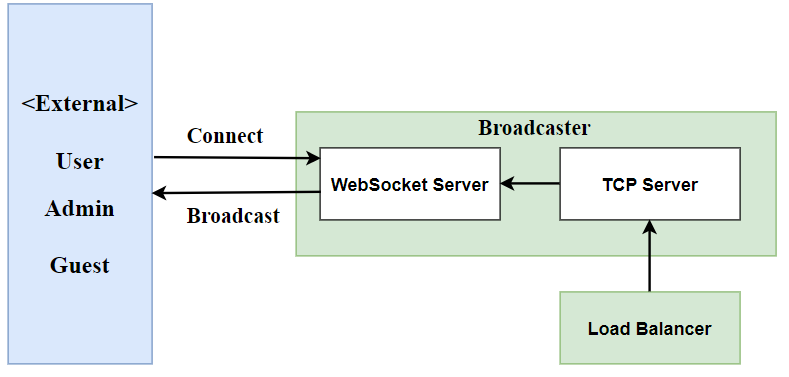}
    \caption{Software architecture of broadcaster}
    \label{software_architecture_broadcaster}
\end{figure}

\cref{software_architecture_broadcaster} shows the software architecture of the broadcaster, representing the communication between the broadcaster, load balancer, and external systems. Since the broadcaster is an asynchronous communication, WebSocket is used for push-type communication. First, the external system establishes a connection to send a request to the broadcaster via WebSocket in order to access the broadcaster. Next, the load balancer sends the change status to the Broadcaster through a TCP connection. The Broadcaster then receives the status using its internal TCP server. Finally, the Broadcaster broadcasts the status to all connected external systems through a pre-established WebSocket connection.

\begin{figure}
    \centering
    \includegraphics[width=1\linewidth]{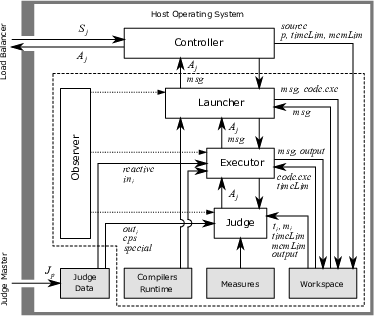}
    \caption{Software architecture of judge server}
    \label{software_architecture_judge_server}
\end{figure}

The internal software architecture of the Judge server is shown in \cref{software_architecture_judge_server}. An operating system is installed on the Judge server and there are five processes: Controller, Observer, Executor, Launcher, and Judge. The Observer is an administrator that monitors all submission processes. If there is something bad in the submissions, it reacts and terminates them.  A Controller is a communicator between the load balancer and Judge server. It receives submissions from the load balancer and provides the appropriate judging results. Launcher is used to prepare the submissions for the executor, then the executor is activated by the launcher to execute the program code. Finally, the judge evaluates the program code based on the judge data. However, problems can occur in the Launcher, Executor, and Judge phases when executing and evaluating program code. If the processes such as \texttt{code.exe} in the Executor phase and \emph{checker} and \emph{reactor} in the Judge phase take more time than usual, these processes are terminated by the Observer.

In addition, the entire operation of the AOJ system relies on a variety of software technologies~\cite{s_ref135}. The web server is implemented with the Spring framework (2018 and onwards) and Apache Tomcat (before 2018). Scala is used for the broadcaster and the load balancer runs entirely on a Java application. In contrast, judge servers are run on scripting languages.

\subsection{Operational Availability}
Since the AOJ system is active in a variety of programming activities, here are some brief operation-oriented statistics. AOJ has successfully held 100 on-site programming contests and over 3,000 virtual contests. AOJ is used as an academic tool in programming courses at the University of Aizu and has evaluated over 8 million submitted solution codes through its long-term service. The AOJ system has been in operation around the clock for more than a decade to learn and practice programming. We have calculated the availability of the AOJ system in hours from 2011 to 2022. The total possible hours (TPH) can be calculated using \cref{tph}.

\begin{equation} \label{tph}
    TPH=years_{operation} \times (365 \times 24)
\end{equation}

In each year, the services of the AOJ system are suspended for planned maintenance. Therefore, the total operating hours (TOH) can be calculated using \cref{toh}, where $y$ is the year and $MH_y$ stands for the total number of maintenance hours in a year $y$.

\begin{equation} \label{toh}
    TOH_y = (365 \times 24) - MH_y
\end{equation}

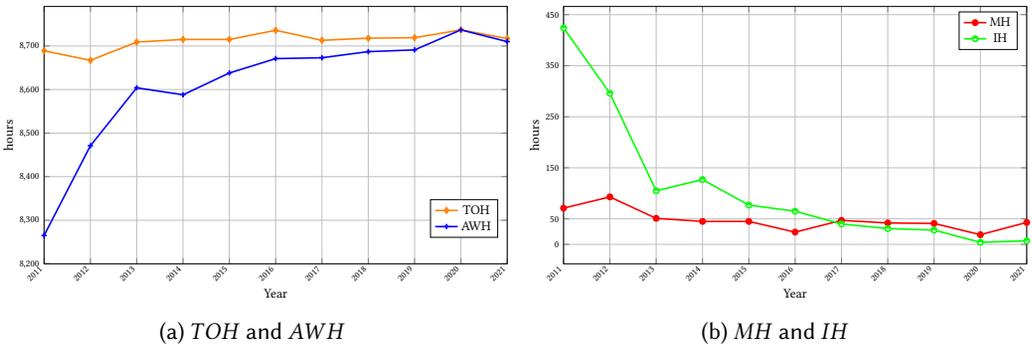
\begin{figure} [h]
\centering
 \subfloat[$TOH$ and $AWH$]{
   \begin{tikzpicture}[scale=0.5]
     \input{toh_awh}
      \label{toh_awh}
   \end{tikzpicture}
 }
 \subfloat[$MH$ and $IH$]{
   \begin{tikzpicture}[scale=.5]
     \input{mh_ih} 
     \label{mh_ih}
   \end{tikzpicture}
 }
   \hfill
    \caption{The availability of the AOJ system during 2011-2022} 
   \label{availability_AOJ}
\end{figure}

Assuming $MH_{2011}$ is 71, $TOH_{2011}$ would be 8,689 hours according to the above formula. We consider an hour as an actual working hour (AWH) if the AOJ system evaluates at least one submission in that hour, otherwise, it is considered an idle hour (IH). \cref{availability_AOJ} shows the $TOH_y$, $MH_y$, $AWH_y$, and $IH_y$ of the AOJ system during 2011-2022.

From \cref{availability_AOJ}, some observations can be made: ($i$) $AWH$ is relatively low in 2011 and 2012; ($ii$) in last few years, $AWH$ is close to $TOH$; ($iii$) $IH$ was also found to be very low in 2020, 2021, and 2022; ($iv$) $MH$ also decreases year by year. Furthermore, the $TPH_{(2011-2022)}$ for the past 12 years was 105,120 hours, in which the AOJ system was in operation (available) for about 
104,656 hours, which is 99.56\% of the $TPH_{(2011-2022)}$ excluding scheduled $MH$. In addition, the average $AWH$ is 98.74\% of $TOH$ over the past 12 years. This statistic demonstrates the high operational availability of the AOJ system throughout the years.

\subsection{Computational Performance}

Over the years, the AOJ system has been used in a variety of areas, including competitions, academics, and practice. As a result, the AOJ system has received millions of submissions for evaluation each year. \cref{{submission_received_1}} provides the statistics of submissions received by the AOJ system over the years, showing that the AOJ received approximately 1,059,756 and 1,096,846 submissions in 2021 and 2022, respectively. Moreover, the number of submissions appears to be increasing each year.
In addition, \cref{{day_wise_submission_1}} shows a further breakdown of submissions for October 2022, where we randomly selected the month. \cref{{day_wise_submission_1}} demonstrates that the AOJ received a large number of submissions on each day.

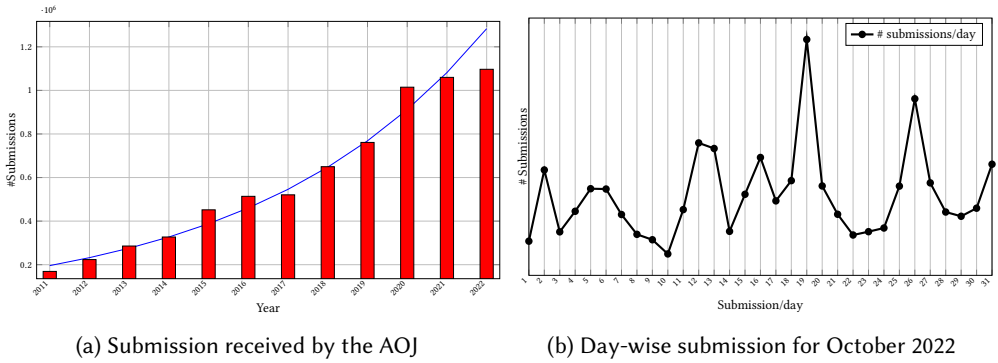
\begin{figure} [h]
\centering
 \subfloat[Submission received by the AOJ]{
   \begin{tikzpicture}[scale=0.5]
          
    \begin{axis}[
	x tick label style={
		/pgf/number format/1000 sep=},
	ylabel=\#Submissions,
	xlabel=Year,
	ylabel near ticks, ylabel shift={-3pt},
	width=\linewidth,
    height=.6\textwidth,
	enlargelimits=0.03,
	xmajorgrids=true,
    grid style=solid,
    grid=both,
    symbolic x coords={2011, 2012, 2013, 2014, 2015,2016, 2017, 2018, 2019, 2020, 2021, 2022},
    xtick=data,
    x tick label style={rotate=45,anchor=east},
    nodes near coords align={vertical},
    tick label style={font=\scriptsize},
]
\addplot[color=blue] 
    coordinates {(2011,195771.2352) (2012,232264.5534) (2013,275560.5168) (2014,326927.1929) (2015,387869.0266) (2016,460170.9037) (2017,545950.4269) (2018,647719.9366) (2019,768460.0939) (2020,911707.1785) (2021,1081656.661) (2022, 1283286.082)};
\addplot[ybar, fill=red] 
    coordinates {(2011,169123) (2012,223468) (2013,285646) (2014,327074) (2015,451803)(2016,513747) (2017,520642) (2018,649768) (2019,761161) (2020,1014683) (2021,1059756) (2022, 1096846)};
\end{axis}
    
      \label{submission_received_1}
   \end{tikzpicture}
 }
 \subfloat[Day-wise submission for October 2022]{
   \begin{tikzpicture}[scale=.5]
     \input{day_wise_submission} 
     \label{day_wise_submission_1}
   \end{tikzpicture}
 }
   \hfill
    \caption{The statistics of submissions received by the AOJ system over the years} 
   \label{fig:judge-performance}
\end{figure}

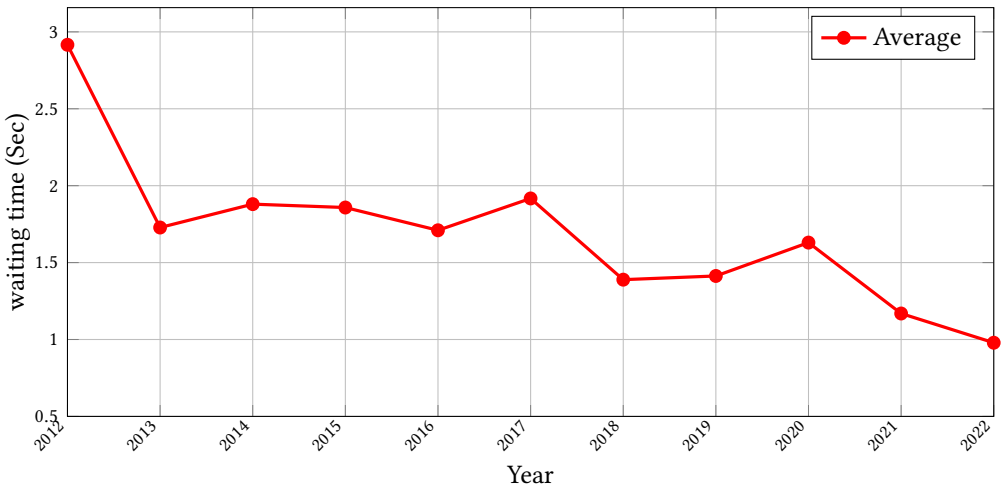
\begin{figure}[h] \centering
     \begin{tikzpicture}
\begin{axis}[
    xlabel={Year},
    ylabel={waiting time (Sec)},
    ylabel near ticks, ylabel shift={-5pt},
   	width=\linewidth,
    height=.5\textwidth,
    grid,
    grid style={gray!50},
    grid=both,
    xmin=2012,
    xmax=2022,
    symbolic x coords={2012, 2013, 2014, 2015,2016, 2017, 2018, 2019, 2020, 2021, 2022},
    xtick=data,
    ymin=.5,
    x tick label style={rotate=45,anchor=east},
    nodes near coords align={vertical},
    tick label style={font=\scriptsize},
    ytick={0,0.5,1.0,1.5,2.0,2.5,3.0},
    ymajorgrids=true,
    grid style=solid,
    tick label style={font=\scriptsize}, 
    ]


\addplot[mark=*, red, solid, very thick]
    coordinates {(2012,2.916) (2013,1.728) (2014,1.880) (2015,1.858) (2016,1.710) (2017,1.917) (2018,1.389) (2019,1.413) (2020,1.630) (2021,1.169) (2022,0.979)};\addlegendentry{Average}

    \end{axis}
    \end{tikzpicture}
     \caption{Year-wise waiting time to evaluate submissions by AOJ} 
   \label{average_waiting_time}
\end{figure}

The average waiting time for the AOJ system to evaluate a submission was calculated. \cref{average_waiting_time} shows the waiting time for the AOJ to evaluate submissions by year. The results show that most submissions were processed within one second (approximately 79.11\%). The waiting time has tended to decrease in recent years (see \cref{average_waiting_time}), which means that the effectiveness of the AOJ in evaluating the submitted code is increasing. However, the number of submissions has increased significantly in recent years. These statistics demonstrate that the AOJ system can be a good candidate for a state-of-the-art AES system.

\subsection{Research}

The real-world resources (e.g., source codes, logs, and educational data) of the AOJ system have become important for research in learning analytics, educational data mining, coding analysis, and programming support. The data resources of the AOJ system are freely available for research\footnote{\url{http://developers.u-aizu.ac.jp/index}}. Meanwhile, the solution code of the AOJ system has been used by IBM for their CodeNET project \cite{s_ref41} and by Google for their AlphaCode project \cite{li2022alphacode}. \cref{published_research} shows the list of published research papers using AOJ data resources. It can be seen that the published research papers using AOJ data resources have attracted the attention of researchers and received numerous citations. It is noted that the citations are counted based on Google scholar. In addition, other research papers can be found  on the AOJ website\footnote{\url{https://onlinejudge.u-aizu.ac.jp/papers}}.


\begin{table} []
\centering
\caption{A list of published research using AOJ data\label{published_research}}
\begin{tabular}{|p{2cm}|p{8cm}|c|c|}
\hline
\textbf{Category of Study} & \textbf{Title} & \textbf{\#Citations} & \textbf{Year of Pub.} \\
\hline

\multirow{6}{*}{ \adjustbox{stack=ll}{Learning Analytics and Data Mining}} &	Impact of Practical Skills on Academic Performance: A Data-Driven Analysis \cite{s_ref8} &	12 & 2021  \\ \cline{2-4}

& Learning Path Recommendation System for Programming Education Based on Neural Networks \cite{s_ref136} & 73 & 2020 \\ \cline{2-4}

& A Novel Rule-Based Online Judge Recommender System to Promote Computer Programming Education \cite{s_ref59} & 8 & 2021  \\ \cline{2-4}

& Data Analysis and Code Assessment Using Machine Learning Techniques for Programming Activities \cite{rahman2022data} & 1 & 2022 \\ \cline{2-4}

& Educational Data Mining to Support Programming Learning Using Problem-Solving Data \cite{s_ref9} & 12 & 2022 \\ \cline{2-4}
& Categorization of frequent errors in solution codes created by novice programmers \cite{nref52} & 3 & 2021 \\ \hline

\multirow{2}{*}{Dataset} & Project CodeNet: A Large-Scale AI for Code Dataset for Learning a Diversity of Coding Tasks \cite{s_ref41} &	44 & 2021  \\ \cline{2-4}

& Competition-Level Code Generation with AlphaCode \cite{li2022alphacode} & 155 & 2022 \\ \hline

\multirow{9}{*}{\adjustbox{stack=ll}{Programming Support}} & A Model with Iterative Trials for Correcting Logic Errors in Source Code \cite{nref53} &	5 & 2021  \\ \cline{2-4}

& A Bidirectional LSTM Language Model for Code Evaluation and Repair \cite{s_ref12} & 50 & 2021 \\\cline{2-4}

& Prompt Sensitivity of Language Model for Solving Programming Problems \cite{nref54} & - & 2022 \\ \cline{2-4}

& Source Code Assessment and Classification Based on Estimated Error Probability Using Attentive LSTM Language Model and Its Application in Programming Education \cite{s_ref17} & 43 & 2020 \\ \cline{2-4}

& Algorithm to Determine Extended Edit Distance between Program Codes \cite{nref58} & 5 & 2019 \\ \cline{2-4}

& Code Completion for Programming Education based on Deep Learning \cite{nref55} & 6 & 2021 \\ \cline{2-4}

& Identifying Algorithm in Program Code based on Structural Features Using CNN Classification Model \cite{s_ref131} & 6 & 2022 \\ \cline{2-4}

& Automatic Generation of Fill-in-the-Blank Programming Problems \cite{nref56} & 14 & 2019 \\ \cline{2-4}

& Logic Error Detection System Based On Structure Pattern And Error Degree \cite{nref57} & 7 & 2019 \\ \hline

OJ System & Online Judge System: Requirements, Architecture, and Experiences \cite{s_ref135} &	9 & 2022  \\ \hline

\end{tabular}

\end{table}

\section{Conclusion} \label{sec:conclusion}

The extensive utilization of automated code evaluation systems (AESs) across various domains is undeniably significant, and the reach of these systems continues to expand. Consequently, the classification of AESs according to their specific applications holds immense value for users. Additionally, the data resources generated by AESs offer considerable utility for diverse research and development tasks. In this survey paper, we have presented a thorough and concise examination of AESs and their associated data resources. Our initial emphasis was on categorizing AESs according to their respective application domains. Subsequently, we provided an overview of the available resources and datasets within AESs, facilitating users in their research and development endeavors. Furthermore, we summarized the application of machine learning in various coding analysis tasks aimed at problem-solving in programming. Lastly, we briefly discussed the Aizu Online Judge System as an illustrative real-world example of an AES, considering factors such as hardware and software development, operation, maintenance, and research aspects.

\begin{acks}
This research was supported by the Japan Society for the Promotion of Science (JSPS) KAKENHI 
 (\url{https://kaken.nii.ac.jp/en/grant/KAKENHI-PROJECT-23H03508/} accessed on 15 June 2023). Grant Number:~23H03508.
\end{acks}

\bibliographystyle{ACM-Reference-Format}


\end{document}

%% file: toh_awh.tex
\begin{axis}[
    xlabel={Year},
    ylabel={hours},
    ylabel near ticks, ylabel shift={-2pt},
   	width=\linewidth,
    height=.6\textwidth,
    grid,
    legend style={at={(0.98,0.25)},anchor=north east},
    grid style={gray!50},
    grid=both,
    xmin=2011,
    xmax=2021,
    ymin=8200,
    symbolic x coords={2011, 2012, 2013, 2014, 2015,2016, 2017, 2018, 2019, 2020, 2021},
    xtick=data,
    x tick label style={rotate=45,anchor=east},
    nodes near coords align={vertical},
    tick label style={font=\scriptsize},
    ytick={8200, 8300,8400,8500,8600,8700,8800},
    ymajorgrids=true,
    grid style=solid,
    tick label style={font=\scriptsize}, 
    ]


\addplot[mark=diamond, orange, solid, very thick]
    coordinates {(2011,8689)(2012,8667) (2013,8709) (2014,8715) (2015,8715) (2016,8736) (2017,8713) (2018,8718) (2019,8719) (2020,8737) (2021,8717)};\addlegendentry{TOH}

\addplot[mark=+, blue, solid, very thick]
    coordinates { (2011,8265) (2012,8471) (2013,8604) (2014,8588) (2015,8638) (2016,8671) (2017,8673) (2018,8687) (2019,8691) (2020,8737) (2021,8710)};\addlegendentry{AWH}

    \end{axis}

%% file: mh_ih.tex
\begin{axis}[
    xlabel={Year},
    ylabel={hours},
    ylabel near ticks, ylabel shift={-2pt},
   	width=\linewidth,
    height=.6\textwidth,
    grid,
    grid style={gray!50},
    grid=both,
    xmin=2011,
    xmax=2021,
    symbolic x coords={2011, 2012, 2013, 2014, 2015,2016, 2017, 2018, 2019, 2020, 2021},
    xtick=data,
    x tick label style={rotate=45,anchor=east},
    nodes near coords align={vertical},
    tick label style={font=\scriptsize},
    ytick={0,50,150,250,350,450},
    ymajorgrids=true,
    grid style=solid,
    tick label style={font=\scriptsize}, 
    ]


\addplot[mark=*, red, solid, very thick]
    coordinates {(2011,71) (2012,93) (2013,51) (2014,45) (2015,45) (2016,24) (2017,47) (2018,42) (2019,41) (2020,19) (2021,43)};\addlegendentry{MH}

\addplot[mark=halfcircle*, green, solid, very thick]
coordinates {(2011,424) (2012,296) (2013,105) (2014,127) (2015,77) (2016,65) (2017,40) (2018,31) (2019,28) (2020,4) (2021,7)};\addlegendentry{IH}
    \end{axis}

%% file: day_wise_submission.tex
\begin{axis}[
    xlabel={ Submission/day},
    ylabel={\# Submissions},
    ylabel near ticks, ylabel shift={-2pt},
   	width=\linewidth,
    height=.6\textwidth,
    grid,
    grid style={gray!50},
    grid=both,
    xmin=1,
    xmax=31,
    symbolic x coords={1, 2, 3, 4, 5, 6, 7, 8, 9, 10, 11, 12,13,14,15,16,17,18,19,20,21,22,23,24,25,26,27,28,29,30,31},
    xtick=data,
    x tick label style={rotate=45,anchor=east},
    nodes near coords align={vertical},
    tick label style={font=\scriptsize},
    ytick={0,50,150,250,350,450},
    ymajorgrids=true,
    grid style=solid,
    tick label style={font=\scriptsize}, 
    ]


\addplot[mark=*, black, solid, ultra  thick]
    coordinates {(1,1550) (2,2405) (3,1662) (4,1909) (5,2180) (6,2176) (7,1870) (8,1632) (9,1567) (10,1397) (11,1929) (12,2730) (13,2663) (14,1668) (15,2113) (16,2555) (17,2034) (18,2277) (19,3972) (20,2213) (21,1873) (22,1624) (23,1664) (24,1708) (25,2211) (26,3260) (27,2250) (28,1900) (29,1849) (30,1946) (31,2474)};\addlegendentry{\# submissions/day}

    \end{axis}

%% file: sample-acmsmall.bbl
\begin{thebibliography}{217}


\ifx \showCODEN    \undefined \def \showCODEN     #1{\unskip}     \fi
\ifx \showDOI      \undefined \def \showDOI       #1{#1}\fi
\ifx \showISBNx    \undefined \def \showISBNx     #1{\unskip}     \fi
\ifx \showISBNxiii \undefined \def \showISBNxiii  #1{\unskip}     \fi
\ifx \showISSN     \undefined \def \showISSN      #1{\unskip}     \fi
\ifx \showLCCN     \undefined \def \showLCCN      #1{\unskip}     \fi
\ifx \shownote     \undefined \def \shownote      #1{#1}          \fi
\ifx \showarticletitle \undefined \def \showarticletitle #1{#1}   \fi
\ifx \showURL      \undefined \def \showURL       {\relax}        \fi
\providecommand\bibfield[2]{#2}
\providecommand\bibinfo[2]{#2}
\providecommand\natexlab[1]{#1}
\providecommand\showeprint[2][]{arXiv:#2}

\bibitem[Abdeljaber et~al\mbox{.}(2017)]%
        {s_ref18}
\bibfield{author}{\bibinfo{person}{Osama Abdeljaber}, \bibinfo{person}{Onur
  Avci}, \bibinfo{person}{Serkan Kiranyaz}, \bibinfo{person}{Moncef Gabbouj},
  {and} \bibinfo{person}{Daniel~J. Inman}.} \bibinfo{year}{2017}\natexlab{}.
\newblock \showarticletitle{Real-time vibration-based structural damage
  detection using one-dimensional convolutional neural networks}.
\newblock \bibinfo{journal}{\emph{Journal of Sound and Vibration}}
  \bibinfo{volume}{388} (\bibinfo{year}{2017}), \bibinfo{pages}{154--170}.
\newblock
\showISSN{0022-460X}
\urldef\tempurl%
\url{https://doi.org/10.1016/j.jsv.2016.10.043}
\showDOI{\tempurl}


\bibitem[Abunadi and Alenezi(2015)]%
        {s_ref11}
\bibfield{author}{\bibinfo{person}{Ibrahim Abunadi} {and}
  \bibinfo{person}{Mamdouh Alenezi}.} \bibinfo{year}{2015}\natexlab{}.
\newblock \showarticletitle{Towards Cross Project Vulnerability Prediction in
  Open Source Web Applications}. In \bibinfo{booktitle}{\emph{Proceedings of
  The International Conference on Engineering' MIS 2015}} (Istanbul, Turkey)
  \emph{(\bibinfo{series}{ICEMIS '15})}. \bibinfo{publisher}{Association for
  Computing Machinery}, \bibinfo{address}{New York, NY, USA}, Article
  \bibinfo{articleno}{42}, \bibinfo{numpages}{5}~pages.
\newblock
\showISBNx{9781450334181}
\urldef\tempurl%
\url{https://doi.org/10.1145/2832987.2833051}
\showDOI{\tempurl}


\bibitem[Aggarwal(2020)]%
        {s_ref10}
\bibfield{author}{\bibinfo{person}{Simran Aggarwal}.}
  \bibinfo{year}{2020}\natexlab{}.
\newblock \showarticletitle{Software Code Analysis Using Ensemble Learning
  Techniques} \emph{(\bibinfo{series}{AISS '19})}.
  \bibinfo{publisher}{Association for Computing Machinery},
  \bibinfo{address}{New York, NY, USA}, Article \bibinfo{articleno}{9},
  \bibinfo{numpages}{7}~pages.
\newblock
\showISBNx{9781450372916}
\urldef\tempurl%
\url{https://doi.org/10.1145/3373477.3373486}
\showDOI{\tempurl}


\bibitem[Ahmad et~al\mbox{.}(2020)]%
        {nref47}
\bibfield{author}{\bibinfo{person}{Wasi~Uddin Ahmad}, \bibinfo{person}{Saikat
  Chakraborty}, \bibinfo{person}{Baishakhi Ray}, {and} \bibinfo{person}{Kai-Wei
  Chang}.} \bibinfo{year}{2020}\natexlab{}.
\newblock \showarticletitle{A transformer-based approach for source code
  summarization}.
\newblock \bibinfo{journal}{\emph{arXiv preprint arXiv:2005.00653}}
  (\bibinfo{year}{2020}).
\newblock


\bibitem[Ain et~al\mbox{.}(2019)]%
        {nref66}
\bibfield{author}{\bibinfo{person}{Qurat~Ul Ain}, \bibinfo{person}{Wasi~Haider
  Butt}, \bibinfo{person}{Muhammad~Waseem Anwar}, \bibinfo{person}{Farooque
  Azam}, {and} \bibinfo{person}{Bilal Maqbool}.}
  \bibinfo{year}{2019}\natexlab{}.
\newblock \showarticletitle{A Systematic Review on Code Clone Detection}.
\newblock \bibinfo{journal}{\emph{IEEE Access}}  \bibinfo{volume}{7}
  (\bibinfo{year}{2019}), \bibinfo{pages}{86121--86144}.
\newblock
\urldef\tempurl%
\url{https://doi.org/10.1109/ACCESS.2019.2918202}
\showDOI{\tempurl}


\bibitem[Al-Ashwal et~al\mbox{.}(2018)]%
        {nref65}
\bibfield{author}{\bibinfo{person}{Deena Al-Ashwal}, \bibinfo{person}{Eman~Zaid
  Al-Sewari}, {and} \bibinfo{person}{Asma~Abdulghani Al-Shargabi}.}
  \bibinfo{year}{2018}\natexlab{}.
\newblock \showarticletitle{A CASE tool for JAVA programs logical errors
  detection: Static and dynamic testing}. In \bibinfo{booktitle}{\emph{2018
  International Arab Conference on Information Technology (ACIT)}}. IEEE,
  \bibinfo{pages}{1--6}.
\newblock


\bibitem[Ala-Mutka(2005)]%
        {s_ref54}
\bibfield{author}{\bibinfo{person}{Kirsti~M Ala-Mutka}.}
  \bibinfo{year}{2005}\natexlab{}.
\newblock \showarticletitle{A survey of automated assessment approaches for
  programming assignments}.
\newblock \bibinfo{journal}{\emph{Computer science education}}
  \bibinfo{volume}{15}, \bibinfo{number}{2} (\bibinfo{year}{2005}),
  \bibinfo{pages}{83--102}.
\newblock


\bibitem[Allamanis et~al\mbox{.}(2015)]%
        {s_ref16}
\bibfield{author}{\bibinfo{person}{Miltiadis Allamanis},
  \bibinfo{person}{Earl~T. Barr}, \bibinfo{person}{Christian Bird}, {and}
  \bibinfo{person}{Charles Sutton}.} \bibinfo{year}{2015}\natexlab{}.
\newblock \showarticletitle{Suggesting Accurate Method and Class Names}. In
  \bibinfo{booktitle}{\emph{Proceedings of the 2015 10th Joint Meeting on
  Foundations of Software Engineering}} (Bergamo, Italy)
  \emph{(\bibinfo{series}{ESEC/FSE 2015})}. \bibinfo{publisher}{Association for
  Computing Machinery}, \bibinfo{address}{New York, NY, USA},
  \bibinfo{pages}{38–49}.
\newblock
\showISBNx{9781450336758}
\urldef\tempurl%
\url{https://doi.org/10.1145/2786805.2786849}
\showDOI{\tempurl}


\bibitem[Allamanis et~al\mbox{.}(2018)]%
        {s_ref25}
\bibfield{author}{\bibinfo{person}{Miltiadis Allamanis},
  \bibinfo{person}{Earl~T. Barr}, \bibinfo{person}{Premkumar Devanbu}, {and}
  \bibinfo{person}{Charles Sutton}.} \bibinfo{year}{2018}\natexlab{}.
\newblock \showarticletitle{A Survey of Machine Learning for Big Code and
  Naturalness}.
\newblock \bibinfo{journal}{\emph{ACM Comput. Surv.}} \bibinfo{volume}{51},
  \bibinfo{number}{4}, Article \bibinfo{articleno}{81} (\bibinfo{date}{jul}
  \bibinfo{year}{2018}), \bibinfo{numpages}{37}~pages.
\newblock
\showISSN{0360-0300}
\urldef\tempurl%
\url{https://doi.org/10.1145/3212695}
\showDOI{\tempurl}


\bibitem[Allamanis et~al\mbox{.}(2017)]%
        {nref36}
\bibfield{author}{\bibinfo{person}{Miltiadis Allamanis}, \bibinfo{person}{Marc
  Brockschmidt}, {and} \bibinfo{person}{Mahmoud Khademi}.}
  \bibinfo{year}{2017}\natexlab{}.
\newblock \showarticletitle{Learning to represent programs with graphs}.
\newblock \bibinfo{journal}{\emph{arXiv preprint arXiv:1711.00740}}
  (\bibinfo{year}{2017}).
\newblock


\bibitem[Allamanis et~al\mbox{.}(2016)]%
        {s_ref118}
\bibfield{author}{\bibinfo{person}{Miltiadis Allamanis}, \bibinfo{person}{Hao
  Peng}, {and} \bibinfo{person}{Charles Sutton}.}
  \bibinfo{year}{2016}\natexlab{}.
\newblock \showarticletitle{A convolutional attention network for extreme
  summarization of source code}. In \bibinfo{booktitle}{\emph{International
  conference on machine learning}}. PMLR, \bibinfo{pages}{2091--2100}.
\newblock


\bibitem[Allamanis and Sutton(2013)]%
        {s_ref13}
\bibfield{author}{\bibinfo{person}{Miltiadis Allamanis} {and}
  \bibinfo{person}{Charles Sutton}.} \bibinfo{year}{2013}\natexlab{}.
\newblock \showarticletitle{Mining Source Code Repositories at Massive Scale
  Using Language Modeling}. In \bibinfo{booktitle}{\emph{Proceedings of the
  10th Working Conference on Mining Software Repositories}} (San Francisco, CA,
  USA) \emph{(\bibinfo{series}{MSR '13})}. \bibinfo{publisher}{IEEE Press},
  \bibinfo{pages}{207–216}.
\newblock
\showISBNx{9781467329361}


\bibitem[Alon et~al\mbox{.}(2019)]%
        {nref19}
\bibfield{author}{\bibinfo{person}{Uri Alon}, \bibinfo{person}{Meital
  Zilberstein}, \bibinfo{person}{Omer Levy}, {and} \bibinfo{person}{Eran
  Yahav}.} \bibinfo{year}{2019}\natexlab{}.
\newblock \showarticletitle{code2vec: Learning distributed representations of
  code}.
\newblock \bibinfo{journal}{\emph{Proceedings of the ACM on Programming
  Languages}} \bibinfo{volume}{3}, \bibinfo{number}{POPL}
  (\bibinfo{year}{2019}), \bibinfo{pages}{1--29}.
\newblock


\bibitem[Alsolai and Roper(2020)]%
        {s_ref26}
\bibfield{author}{\bibinfo{person}{Hadeel Alsolai} {and} \bibinfo{person}{Marc
  Roper}.} \bibinfo{year}{2020}\natexlab{}.
\newblock \showarticletitle{A systematic literature review of machine learning
  techniques for software maintainability prediction}.
\newblock \bibinfo{journal}{\emph{Information and Software Technology}}
  \bibinfo{volume}{119} (\bibinfo{year}{2020}), \bibinfo{pages}{106214}.
\newblock
\showISSN{0950-5849}
\urldef\tempurl%
\url{https://doi.org/10.1016/j.infsof.2019.106214}
\showDOI{\tempurl}


\bibitem[Amorim et~al\mbox{.}(2018)]%
        {s_ref106}
\bibfield{author}{\bibinfo{person}{Leonardo~Afonso Amorim},
  \bibinfo{person}{Mateus~F. Freitas}, \bibinfo{person}{Altino Dantas},
  \bibinfo{person}{Eduardo~F. de Souza}, \bibinfo{person}{Celso~G.
  Camilo-Junior}, {and} \bibinfo{person}{Wellington~S. Martins}.}
  \bibinfo{year}{2018}\natexlab{}.
\newblock \showarticletitle{A New Word Embedding Approach to Evaluate Potential
  Fixes for Automated Program Repair}. In \bibinfo{booktitle}{\emph{2018
  International Joint Conference on Neural Networks (IJCNN)}}.
  \bibinfo{pages}{1--8}.
\newblock
\urldef\tempurl%
\url{https://doi.org/10.1109/IJCNN.2018.8489079}
\showDOI{\tempurl}


\bibitem[Aniche et~al\mbox{.}(2022)]%
        {s_ref15}
\bibfield{author}{\bibinfo{person}{Maurício Aniche}, \bibinfo{person}{Erick
  Maziero}, \bibinfo{person}{Rafael Durelli}, {and} \bibinfo{person}{Vinicius
  H.~S. Durelli}.} \bibinfo{year}{2022}\natexlab{}.
\newblock \showarticletitle{The Effectiveness of Supervised Machine Learning
  Algorithms in Predicting Software Refactoring}.
\newblock \bibinfo{journal}{\emph{IEEE Transactions on Software Engineering}}
  \bibinfo{volume}{48}, \bibinfo{number}{4} (\bibinfo{year}{2022}),
  \bibinfo{pages}{1432--1450}.
\newblock
\urldef\tempurl%
\url{https://doi.org/10.1109/TSE.2020.3021736}
\showDOI{\tempurl}


\bibitem[Anzai and Watanobe(2019)]%
        {nref58}
\bibfield{author}{\bibinfo{person}{Kazuki Anzai} {and} \bibinfo{person}{Yutaka
  Watanobe}.} \bibinfo{year}{2019}\natexlab{}.
\newblock \showarticletitle{Algorithm to determine extended edit distance
  between program codes}. In \bibinfo{booktitle}{\emph{2019 IEEE 13th
  International symposium on embedded multicore/many-core systems-on-chip
  (MCSoC)}}. IEEE, \bibinfo{pages}{180--186}.
\newblock


\bibitem[Athiwaratkun et~al\mbox{.}(2022)]%
        {athiwaratkun2022multilingual}
\bibfield{author}{\bibinfo{person}{Ben Athiwaratkun},
  \bibinfo{person}{Sanjay~Krishna Gouda}, \bibinfo{person}{Zijian Wang},
  \bibinfo{person}{Xiaopeng Li}, \bibinfo{person}{Yuchen Tian},
  \bibinfo{person}{Ming Tan}, \bibinfo{person}{Wasi~Uddin Ahmad},
  \bibinfo{person}{Shiqi Wang}, \bibinfo{person}{Qing Sun},
  \bibinfo{person}{Mingyue Shang}, \bibinfo{person}{Sujan~Kumar Gonugondla},
  \bibinfo{person}{Hantian Ding}, \bibinfo{person}{Varun Kumar},
  \bibinfo{person}{Nathan Fulton}, \bibinfo{person}{Arash Farahani},
  \bibinfo{person}{Siddhartha Jain}, \bibinfo{person}{Robert Giaquinto},
  \bibinfo{person}{Haifeng Qian}, \bibinfo{person}{Murali~Krishna Ramanathan},
  \bibinfo{person}{Ramesh Nallapati}, \bibinfo{person}{Baishakhi Ray},
  \bibinfo{person}{Parminder Bhatia}, \bibinfo{person}{Sudipta Sengupta},
  \bibinfo{person}{Dan Roth}, {and} \bibinfo{person}{Bing Xiang}.}
  \bibinfo{year}{2022}\natexlab{}.
\newblock \bibinfo{title}{Multi-lingual Evaluation of Code Generation Models}.
\newblock
\newblock
\showeprint[arxiv]{2210.14868}~[cs.LG]


\bibitem[Austin et~al\mbox{.}(2021)]%
        {austin2021program}
\bibfield{author}{\bibinfo{person}{Jacob Austin}, \bibinfo{person}{Augustus
  Odena}, \bibinfo{person}{Maxwell Nye}, \bibinfo{person}{Maarten Bosma},
  \bibinfo{person}{Henryk Michalewski}, \bibinfo{person}{David Dohan},
  \bibinfo{person}{Ellen Jiang}, \bibinfo{person}{Carrie Cai},
  \bibinfo{person}{Michael Terry}, \bibinfo{person}{Quoc Le}, {and}
  \bibinfo{person}{Charles Sutton}.} \bibinfo{year}{2021}\natexlab{}.
\newblock \bibinfo{title}{Program Synthesis with Large Language Models}.
\newblock
\newblock
\showeprint[arxiv]{2108.07732}~[cs.PL]


\bibitem[Azeem et~al\mbox{.}(2019)]%
        {s_ref37}
\bibfield{author}{\bibinfo{person}{Muhammad~Ilyas Azeem},
  \bibinfo{person}{Fabio Palomba}, \bibinfo{person}{Lin Shi}, {and}
  \bibinfo{person}{Qing Wang}.} \bibinfo{year}{2019}\natexlab{}.
\newblock \showarticletitle{Machine learning techniques for code smell
  detection: A systematic literature review and meta-analysis}.
\newblock \bibinfo{journal}{\emph{Information and Software Technology}}
  \bibinfo{volume}{108} (\bibinfo{year}{2019}), \bibinfo{pages}{115--138}.
\newblock


\bibitem[Bandara and Wijayarathna(2011)]%
        {s_ref90}
\bibfield{author}{\bibinfo{person}{Upul Bandara} {and} \bibinfo{person}{Gamini
  Wijayarathna}.} \bibinfo{year}{2011}\natexlab{}.
\newblock \showarticletitle{A machine learning based tool for source code
  plagiarism detection}.
\newblock \bibinfo{journal}{\emph{International Journal of Machine Learning and
  Computing}} \bibinfo{volume}{1}, \bibinfo{number}{4} (\bibinfo{year}{2011}),
  \bibinfo{pages}{337}.
\newblock


\bibitem[Barbez et~al\mbox{.}(2020)]%
        {nref41}
\bibfield{author}{\bibinfo{person}{Antoine Barbez}, \bibinfo{person}{Foutse
  Khomh}, {and} \bibinfo{person}{Yann-Ga{\"e}l Gu{\'e}h{\'e}neuc}.}
  \bibinfo{year}{2020}\natexlab{}.
\newblock \showarticletitle{A machine-learning based ensemble method for
  anti-patterns detection}.
\newblock \bibinfo{journal}{\emph{Journal of Systems and Software}}
  \bibinfo{volume}{161} (\bibinfo{year}{2020}), \bibinfo{pages}{110486}.
\newblock


\bibitem[Bavarian et~al\mbox{.}(2022)]%
        {bavarian2022efficient}
\bibfield{author}{\bibinfo{person}{Mohammad Bavarian}, \bibinfo{person}{Heewoo
  Jun}, \bibinfo{person}{Nikolas Tezak}, \bibinfo{person}{John Schulman},
  \bibinfo{person}{Christine McLeavey}, \bibinfo{person}{Jerry Tworek}, {and}
  \bibinfo{person}{Mark Chen}.} \bibinfo{year}{2022}\natexlab{}.
\newblock \bibinfo{title}{Efficient Training of Language Models to Fill in the
  Middle}.
\newblock
\newblock
\showeprint[arxiv]{2207.14255}~[cs.CL]


\bibitem[Ben-Nun et~al\mbox{.}(2018)]%
        {s_ref44}
\bibfield{author}{\bibinfo{person}{Tal Ben-Nun},
  \bibinfo{person}{Alice~Shoshana Jakobovits}, {and} \bibinfo{person}{Torsten
  Hoefler}.} \bibinfo{year}{2018}\natexlab{}.
\newblock \showarticletitle{Neural Code Comprehension: A Learnable
  Representation of Code Semantics}. In \bibinfo{booktitle}{\emph{Proceedings
  of the 32nd International Conference on Neural Information Processing
  Systems}} \emph{(\bibinfo{series}{NIPS'18})}. \bibinfo{publisher}{Curran
  Associates Inc.}, \bibinfo{address}{Red Hook, NY, USA},
  \bibinfo{pages}{3589–3601}.
\newblock


\bibitem[Bielik et~al\mbox{.}(2016)]%
        {s_ref64}
\bibfield{author}{\bibinfo{person}{Pavol Bielik}, \bibinfo{person}{Veselin
  Raychev}, {and} \bibinfo{person}{Martin Vechev}.}
  \bibinfo{year}{2016}\natexlab{}.
\newblock \showarticletitle{PHOG: probabilistic model for code}. In
  \bibinfo{booktitle}{\emph{International conference on machine learning}}.
  PMLR, \bibinfo{pages}{2933--2942}.
\newblock


\bibitem[Brown et~al\mbox{.}(2020)]%
        {brown2020gpt3}
\bibfield{author}{\bibinfo{person}{Tom Brown}, \bibinfo{person}{Benjamin Mann},
  \bibinfo{person}{Nick Ryder}, \bibinfo{person}{Melanie Subbiah},
  \bibinfo{person}{Jared~D Kaplan}, \bibinfo{person}{Prafulla Dhariwal},
  \bibinfo{person}{Arvind Neelakantan}, \bibinfo{person}{Pranav Shyam},
  \bibinfo{person}{Girish Sastry}, \bibinfo{person}{Amanda Askell},
  \bibinfo{person}{Sandhini Agarwal}, \bibinfo{person}{Ariel Herbert-Voss},
  \bibinfo{person}{Gretchen Krueger}, \bibinfo{person}{Tom Henighan},
  \bibinfo{person}{Rewon Child}, \bibinfo{person}{Aditya Ramesh},
  \bibinfo{person}{Daniel Ziegler}, \bibinfo{person}{Jeffrey Wu},
  \bibinfo{person}{Clemens Winter}, \bibinfo{person}{Chris Hesse},
  \bibinfo{person}{Mark Chen}, \bibinfo{person}{Eric Sigler},
  \bibinfo{person}{Mateusz Litwin}, \bibinfo{person}{Scott Gray},
  \bibinfo{person}{Benjamin Chess}, \bibinfo{person}{Jack Clark},
  \bibinfo{person}{Christopher Berner}, \bibinfo{person}{Sam McCandlish},
  \bibinfo{person}{Alec Radford}, \bibinfo{person}{Ilya Sutskever}, {and}
  \bibinfo{person}{Dario Amodei}.} \bibinfo{year}{2020}\natexlab{}.
\newblock \showarticletitle{Language Models are Few-Shot Learners}. In
  \bibinfo{booktitle}{\emph{Proceedings of Advances in Neural Information
  Processing Systems (NeurIPS)}}, Vol.~\bibinfo{volume}{33}.
  \bibinfo{pages}{1877--1901}.
\newblock


\bibitem[Butgereit(2019)]%
        {s_ref81}
\bibfield{author}{\bibinfo{person}{Laurie Butgereit}.}
  \bibinfo{year}{2019}\natexlab{}.
\newblock \showarticletitle{Using Machine Learning to Prioritize Automated
  Testing in an Agile Environment}. In \bibinfo{booktitle}{\emph{2019
  Conference on Information Communications Technology and Society (ICTAS)}}.
  \bibinfo{pages}{1--6}.
\newblock
\urldef\tempurl%
\url{https://doi.org/10.1109/ICTAS.2019.8703639}
\showDOI{\tempurl}


\bibitem[Büch and Andrzejak(2019)]%
        {s_ref88}
\bibfield{author}{\bibinfo{person}{Lutz Büch} {and} \bibinfo{person}{Artur
  Andrzejak}.} \bibinfo{year}{2019}\natexlab{}.
\newblock \showarticletitle{Learning-Based Recursive Aggregation of Abstract
  Syntax Trees for Code Clone Detection}. In \bibinfo{booktitle}{\emph{2019
  IEEE 26th International Conference on Software Analysis, Evolution and
  Reengineering (SANER)}}. \bibinfo{pages}{95--104}.
\newblock
\urldef\tempurl%
\url{https://doi.org/10.1109/SANER.2019.8668039}
\showDOI{\tempurl}


\bibitem[Cheang et~al\mbox{.}(2003)]%
        {s_ref53}
\bibfield{author}{\bibinfo{person}{Brenda Cheang}, \bibinfo{person}{Andy
  Kurnia}, \bibinfo{person}{Andrew Lim}, {and} \bibinfo{person}{Wee-Chong
  Oon}.} \bibinfo{year}{2003}\natexlab{}.
\newblock \showarticletitle{On Automated Grading of Programming Assignments in
  an Academic Institution}.
\newblock \bibinfo{journal}{\emph{Comput. Educ.}} \bibinfo{volume}{41},
  \bibinfo{number}{2} (\bibinfo{date}{sep} \bibinfo{year}{2003}),
  \bibinfo{pages}{121–131}.
\newblock
\showISSN{0360-1315}
\urldef\tempurl%
\url{https://doi.org/10.1016/S0360-1315(03)00030-7}
\showDOI{\tempurl}


\bibitem[Chen et~al\mbox{.}(2020a)]%
        {nref30}
\bibfield{author}{\bibinfo{person}{Jinyin Chen}, \bibinfo{person}{Keke Hu},
  \bibinfo{person}{Yue Yu}, \bibinfo{person}{Zhuangzhi Chen},
  \bibinfo{person}{Qi Xuan}, \bibinfo{person}{Yi Liu}, {and}
  \bibinfo{person}{Vladimir Filkov}.} \bibinfo{year}{2020}\natexlab{a}.
\newblock \showarticletitle{Software visualization and deep transfer learning
  for effective software defect prediction}. In
  \bibinfo{booktitle}{\emph{Proceedings of the ACM/IEEE 42nd international
  conference on software engineering}}. \bibinfo{pages}{578--589}.
\newblock


\bibitem[Chen et~al\mbox{.}(2020b)]%
        {s_ref82}
\bibfield{author}{\bibinfo{person}{Jinyin Chen}, \bibinfo{person}{Keke Hu},
  \bibinfo{person}{Yue Yu}, \bibinfo{person}{Zhuangzhi Chen},
  \bibinfo{person}{Qi Xuan}, \bibinfo{person}{Yi Liu}, {and}
  \bibinfo{person}{Vladimir Filkov}.} \bibinfo{year}{2020}\natexlab{b}.
\newblock \showarticletitle{Software Visualization and Deep Transfer Learning
  for Effective Software Defect Prediction}. In \bibinfo{booktitle}{\emph{2020
  IEEE/ACM 42nd International Conference on Software Engineering (ICSE)}}.
  \bibinfo{pages}{578--589}.
\newblock


\bibitem[Chen et~al\mbox{.}(2021)]%
        {chen2021codex}
\bibfield{author}{\bibinfo{person}{Mark Chen}, \bibinfo{person}{Jerry Tworek},
  \bibinfo{person}{Heewoo Jun}, \bibinfo{person}{Qiming Yuan},
  \bibinfo{person}{Henrique~Ponde de Oliveira~Pinto}, \bibinfo{person}{Jared
  Kaplan}, \bibinfo{person}{Harri Edwards}, \bibinfo{person}{Yuri Burda},
  \bibinfo{person}{Nicholas Joseph}, \bibinfo{person}{Greg Brockman},
  \bibinfo{person}{Alex Ray}, \bibinfo{person}{Raul Puri},
  \bibinfo{person}{Gretchen Krueger}, \bibinfo{person}{Michael Petrov},
  \bibinfo{person}{Heidy Khlaaf}, \bibinfo{person}{Girish Sastry},
  \bibinfo{person}{Pamela Mishkin}, \bibinfo{person}{Brooke Chan},
  \bibinfo{person}{Scott Gray}, \bibinfo{person}{Nick Ryder},
  \bibinfo{person}{Mikhail Pavlov}, \bibinfo{person}{Alethea Power},
  \bibinfo{person}{Lukasz Kaiser}, \bibinfo{person}{Mohammad Bavarian},
  \bibinfo{person}{Clemens Winter}, \bibinfo{person}{Philippe Tillet},
  \bibinfo{person}{Felipe~Petroski Such}, \bibinfo{person}{Dave Cummings},
  \bibinfo{person}{Matthias Plappert}, \bibinfo{person}{Fotios Chantzis},
  \bibinfo{person}{Elizabeth Barnes}, \bibinfo{person}{Ariel Herbert-Voss},
  \bibinfo{person}{William~Hebgen Guss}, \bibinfo{person}{Alex Nichol},
  \bibinfo{person}{Alex Paino}, \bibinfo{person}{Nikolas Tezak},
  \bibinfo{person}{Jie Tang}, \bibinfo{person}{Igor Babuschkin},
  \bibinfo{person}{Suchir Balaji}, \bibinfo{person}{Shantanu Jain},
  \bibinfo{person}{William Saunders}, \bibinfo{person}{Christopher Hesse},
  \bibinfo{person}{Andrew~N. Carr}, \bibinfo{person}{Jan Leike},
  \bibinfo{person}{Josh Achiam}, \bibinfo{person}{Vedant Misra},
  \bibinfo{person}{Evan Morikawa}, \bibinfo{person}{Alec Radford},
  \bibinfo{person}{Matthew Knight}, \bibinfo{person}{Miles Brundage},
  \bibinfo{person}{Mira Murati}, \bibinfo{person}{Katie Mayer},
  \bibinfo{person}{Peter Welinder}, \bibinfo{person}{Bob McGrew},
  \bibinfo{person}{Dario Amodei}, \bibinfo{person}{Sam McCandlish},
  \bibinfo{person}{Ilya Sutskever}, {and} \bibinfo{person}{Wojciech Zaremba}.}
  \bibinfo{year}{2021}\natexlab{}.
\newblock \showarticletitle{Evaluating Large Language Models Trained on Code}.
\newblock \bibinfo{journal}{\emph{arXiv preprint}}
  \bibinfo{volume}{arXiv:2107.03374} (\bibinfo{year}{2021}).
\newblock
\urldef\tempurl%
\url{https://doi.org/10.48550/arXiv.2107.03374}
\showDOI{\tempurl}


\bibitem[Clement et~al\mbox{.}(2020)]%
        {s_ref119}
\bibfield{author}{\bibinfo{person}{Colin~B Clement}, \bibinfo{person}{Dawn
  Drain}, \bibinfo{person}{Jonathan Timcheck}, \bibinfo{person}{Alexey
  Svyatkovskiy}, {and} \bibinfo{person}{Neel Sundaresan}.}
  \bibinfo{year}{2020}\natexlab{}.
\newblock \showarticletitle{PyMT5: multi-mode translation of natural language
  and Python code with transformers}.
\newblock \bibinfo{journal}{\emph{arXiv preprint arXiv:2010.03150}}
  (\bibinfo{year}{2020}).
\newblock


\bibitem[Comb{\'e}fis and Wautelet(2014)]%
        {s_ref49}
\bibfield{author}{\bibinfo{person}{S{\'e}bastien Comb{\'e}fis} {and}
  \bibinfo{person}{J{\'e}r{\'e}my Wautelet}.} \bibinfo{year}{2014}\natexlab{}.
\newblock \showarticletitle{Programming Trainings and Informatics Teaching
  Through Online Contests.}
\newblock \bibinfo{journal}{\emph{Olympiads in Informatics}}
  \bibinfo{volume}{8} (\bibinfo{year}{2014}).
\newblock


\bibitem[Cormack et~al\mbox{.}(2006)]%
        {s_ref2}
\bibfield{author}{\bibinfo{person}{Gordon Cormack}, \bibinfo{person}{Ian
  Munro}, \bibinfo{person}{Troy Vasiga}, {and} \bibinfo{person}{Graeme
  Kemkes}.} \bibinfo{year}{2006}\natexlab{}.
\newblock \showarticletitle{Structure, Scoring and Purpose of Computing
  Competitions}.
\newblock \bibinfo{journal}{\emph{Informatics in Education}}
  \bibinfo{volume}{5}, \bibinfo{number}{1} (\bibinfo{date}{jan}
  \bibinfo{year}{2006}), \bibinfo{pages}{15–36}.
\newblock
\showISSN{1648-5831}


\bibitem[Cort{\'e}s-Coy et~al\mbox{.}(2014)]%
        {nref45}
\bibfield{author}{\bibinfo{person}{Luis~Fernando Cort{\'e}s-Coy},
  \bibinfo{person}{Mario Linares-V{\'a}squez}, \bibinfo{person}{Jairo Aponte},
  {and} \bibinfo{person}{Denys Poshyvanyk}.} \bibinfo{year}{2014}\natexlab{}.
\newblock \showarticletitle{On automatically generating commit messages via
  summarization of source code changes}. In \bibinfo{booktitle}{\emph{2014 IEEE
  14th International Working Conference on Source Code Analysis and
  Manipulation}}. IEEE, \bibinfo{pages}{275--284}.
\newblock


\bibitem[Cosma and Joy(2008)]%
        {nref2}
\bibfield{author}{\bibinfo{person}{Georgina Cosma} {and} \bibinfo{person}{Mike
  Joy}.} \bibinfo{year}{2008}\natexlab{}.
\newblock \showarticletitle{Towards a definition of source-code plagiarism}.
\newblock \bibinfo{journal}{\emph{IEEE Transactions on Education}}
  \bibinfo{volume}{51}, \bibinfo{number}{2} (\bibinfo{year}{2008}),
  \bibinfo{pages}{195--200}.
\newblock


\bibitem[Costello and Stolovitzky(2013)]%
        {nref17}
\bibfield{author}{\bibinfo{person}{JC Costello} {and} \bibinfo{person}{G
  Stolovitzky}.} \bibinfo{year}{2013}\natexlab{}.
\newblock \showarticletitle{Seeking the wisdom of crowds through
  challenge-based competitions in biomedical research}.
\newblock \bibinfo{journal}{\emph{Clinical Pharmacology \& Therapeutics}}
  \bibinfo{volume}{93}, \bibinfo{number}{5} (\bibinfo{year}{2013}),
  \bibinfo{pages}{396--398}.
\newblock


\bibitem[Cvitkovic et~al\mbox{.}(2019)]%
        {nref37}
\bibfield{author}{\bibinfo{person}{Milan Cvitkovic}, \bibinfo{person}{Badal
  Singh}, {and} \bibinfo{person}{Animashree Anandkumar}.}
  \bibinfo{year}{2019}\natexlab{}.
\newblock \showarticletitle{Open vocabulary learning on source code with a
  graph-structured cache}. In \bibinfo{booktitle}{\emph{International
  Conference on Machine Learning}}. PMLR, \bibinfo{pages}{1475--1485}.
\newblock


\bibitem[Devlin et~al\mbox{.}(2019)]%
        {devlin2019bert}
\bibfield{author}{\bibinfo{person}{Jacob Devlin}, \bibinfo{person}{Ming-Wei
  Chang}, \bibinfo{person}{Kenton Lee}, {and} \bibinfo{person}{Kristina
  Toutanova}.} \bibinfo{year}{2019}\natexlab{}.
\newblock \showarticletitle{{BERT}: Pre-training of Deep Bidirectional
  Transformers for Language Understanding}. In
  \bibinfo{booktitle}{\emph{Proceedings of the 2019 Conference of the North
  {A}merican Chapter of the Association for Computational Linguistics: Human
  Language Technologies, Volume 1 (Long and Short Papers)}}.
  \bibinfo{publisher}{Association for Computational Linguistics},
  \bibinfo{address}{Minneapolis, Minnesota}, \bibinfo{pages}{4171--4186}.
\newblock
\urldef\tempurl%
\url{https://doi.org/10.18653/v1/N19-1423}
\showDOI{\tempurl}


\bibitem[Došilović and Mekterović(2020)]%
        {nref59}
\bibfield{author}{\bibinfo{person}{Herman~Zvonimir Došilović} {and}
  \bibinfo{person}{Igor Mekterović}.} \bibinfo{year}{2020}\natexlab{}.
\newblock \showarticletitle{Robust and Scalable Online Code Execution System}.
  In \bibinfo{booktitle}{\emph{2020 43rd International Convention on
  Information, Communication and Electronic Technology (MIPRO)}}.
  \bibinfo{pages}{1627--1632}.
\newblock
\urldef\tempurl%
\url{https://doi.org/10.23919/MIPRO48935.2020.9245310}
\showDOI{\tempurl}


\bibitem[Fang et~al\mbox{.}(2020)]%
        {nref70}
\bibfield{author}{\bibinfo{person}{Chunrong Fang}, \bibinfo{person}{Zixi Liu},
  \bibinfo{person}{Yangyang Shi}, \bibinfo{person}{Jeff Huang}, {and}
  \bibinfo{person}{Qingkai Shi}.} \bibinfo{year}{2020}\natexlab{}.
\newblock \showarticletitle{Functional Code Clone Detection with Syntax and
  Semantics Fusion Learning}. In \bibinfo{booktitle}{\emph{Proceedings of the
  29th ACM SIGSOFT International Symposium on Software Testing and Analysis}}
  (Virtual Event, USA) \emph{(\bibinfo{series}{ISSTA 2020})}.
  \bibinfo{publisher}{Association for Computing Machinery},
  \bibinfo{address}{New York, NY, USA}, \bibinfo{pages}{516–527}.
\newblock
\showISBNx{9781450380089}
\urldef\tempurl%
\url{https://doi.org/10.1145/3395363.3397362}
\showDOI{\tempurl}


\bibitem[Felter et~al\mbox{.}(2015)]%
        {s_ref6}
\bibfield{author}{\bibinfo{person}{Wes Felter}, \bibinfo{person}{Alexandre
  Ferreira}, \bibinfo{person}{Ram Rajamony}, {and} \bibinfo{person}{Juan
  Rubio}.} \bibinfo{year}{2015}\natexlab{}.
\newblock \showarticletitle{An updated performance comparison of virtual
  machines and Linux containers}. In \bibinfo{booktitle}{\emph{2015 IEEE
  International Symposium on Performance Analysis of Systems and Software
  (ISPASS)}}. \bibinfo{pages}{171--172}.
\newblock
\urldef\tempurl%
\url{https://doi.org/10.1109/ISPASS.2015.7095802}
\showDOI{\tempurl}


\bibitem[Feng et~al\mbox{.}(2020)]%
        {feng2020codebert}
\bibfield{author}{\bibinfo{person}{Zhangyin Feng}, \bibinfo{person}{Daya Guo},
  \bibinfo{person}{Duyu Tang}, \bibinfo{person}{Nan Duan},
  \bibinfo{person}{Xiaocheng Feng}, \bibinfo{person}{Ming Gong},
  \bibinfo{person}{Linjun Shou}, \bibinfo{person}{Bing Qin},
  \bibinfo{person}{Ting Liu}, \bibinfo{person}{Daxin Jiang}, {and}
  \bibinfo{person}{Ming Zhou}.} \bibinfo{year}{2020}\natexlab{}.
\newblock \showarticletitle{{C}ode{BERT}: A Pre-Trained Model for Programming
  and Natural Languages}. In \bibinfo{booktitle}{\emph{Findings of the
  Association for Computational Linguistics: EMNLP 2020}}.
  \bibinfo{publisher}{Association for Computational Linguistics},
  \bibinfo{address}{Online}, \bibinfo{pages}{1536--1547}.
\newblock
\urldef\tempurl%
\url{https://doi.org/10.18653/v1/2020.findings-emnlp.139}
\showDOI{\tempurl}


\bibitem[Fernandes et~al\mbox{.}(2018)]%
        {s_ref120}
\bibfield{author}{\bibinfo{person}{Patrick Fernandes},
  \bibinfo{person}{Miltiadis Allamanis}, {and} \bibinfo{person}{Marc
  Brockschmidt}.} \bibinfo{year}{2018}\natexlab{}.
\newblock \showarticletitle{Structured neural summarization}.
\newblock \bibinfo{journal}{\emph{arXiv preprint arXiv:1811.01824}}
  (\bibinfo{year}{2018}).
\newblock


\bibitem[Fonte et~al\mbox{.}(2013)]%
        {s_ref56}
\bibfield{author}{\bibinfo{person}{Daniela Fonte}, \bibinfo{person}{Daniela~da
  Cruz}, \bibinfo{person}{Alda~Lopes Gan{\c{c}}arski}, {and}
  \bibinfo{person}{Pedro~Rangel Henriques}.} \bibinfo{year}{2013}\natexlab{}.
\newblock \showarticletitle{A flexible dynamic system for automatic grading of
  programming exercises}.
\newblock  (\bibinfo{year}{2013}).
\newblock


\bibitem[Forisek(2007)]%
        {s_ref4}
\bibfield{author}{\bibinfo{person}{Michal Forisek}.}
  \bibinfo{year}{2007}\natexlab{}.
\newblock \showarticletitle{Security of Programming Contest Systems}.
\newblock


\bibitem[Goldbloom(2010)]%
        {nref14}
\bibfield{author}{\bibinfo{person}{Anthony Goldbloom}.}
  \bibinfo{year}{2010}\natexlab{}.
\newblock \showarticletitle{Data Prediction Competitions -- Far More than Just
  a Bit of Fun}. In \bibinfo{booktitle}{\emph{2010 IEEE International
  Conference on Data Mining Workshops}}. \bibinfo{pages}{1385--1386}.
\newblock
\urldef\tempurl%
\url{https://doi.org/10.1109/ICDMW.2010.56}
\showDOI{\tempurl}


\bibitem[Google(2021)]%
        {s_ref46}
\bibfield{author}{\bibinfo{person}{Google}.} \bibinfo{year}{2021}\natexlab{}.
\newblock \bibinfo{title}{gcj-dataset}.
\newblock
\newblock
\newblock
\shownote{Available:
  \url{https://openreview.net/attachment?id=AZ4vmLoJft\&name=supplementary\_material}}.


\bibitem[Gopinath et~al\mbox{.}(2014)]%
        {s_ref109}
\bibfield{author}{\bibinfo{person}{Divya Gopinath}, \bibinfo{person}{Sarfraz
  Khurshid}, \bibinfo{person}{Diptikalyan Saha}, {and} \bibinfo{person}{Satish
  Chandra}.} \bibinfo{year}{2014}\natexlab{}.
\newblock \showarticletitle{Data-guided repair of selection statements}. In
  \bibinfo{booktitle}{\emph{Proceedings of the 36th International Conference on
  Software Engineering}}. \bibinfo{pages}{243--253}.
\newblock


\bibitem[Goues et~al\mbox{.}(2019)]%
        {s_ref103}
\bibfield{author}{\bibinfo{person}{Claire~Le Goues}, \bibinfo{person}{Michael
  Pradel}, {and} \bibinfo{person}{Abhik Roychoudhury}.}
  \bibinfo{year}{2019}\natexlab{}.
\newblock \showarticletitle{Automated program repair}.
\newblock \bibinfo{journal}{\emph{Commun. ACM}} \bibinfo{volume}{62},
  \bibinfo{number}{12} (\bibinfo{year}{2019}), \bibinfo{pages}{56--65}.
\newblock


\bibitem[Gousios(2013)]%
        {nref20}
\bibfield{author}{\bibinfo{person}{Georgios Gousios}.}
  \bibinfo{year}{2013}\natexlab{}.
\newblock \showarticletitle{The GHTorent dataset and tool suite}. In
  \bibinfo{booktitle}{\emph{2013 10th Working Conference on Mining Software
  Repositories (MSR)}}. IEEE, \bibinfo{pages}{233--236}.
\newblock


\bibitem[Graves et~al\mbox{.}(2013)]%
        {s_ref24}
\bibfield{author}{\bibinfo{person}{Alex Graves}, \bibinfo{person}{Navdeep
  Jaitly}, {and} \bibinfo{person}{Abdel-rahman Mohamed}.}
  \bibinfo{year}{2013}\natexlab{}.
\newblock \showarticletitle{Hybrid speech recognition with Deep Bidirectional
  LSTM}. In \bibinfo{booktitle}{\emph{2013 IEEE Workshop on Automatic Speech
  Recognition and Understanding}}. \bibinfo{pages}{273--278}.
\newblock
\urldef\tempurl%
\url{https://doi.org/10.1109/ASRU.2013.6707742}
\showDOI{\tempurl}


\bibitem[Greff et~al\mbox{.}(2017)]%
        {s_ref23}
\bibfield{author}{\bibinfo{person}{Klaus Greff}, \bibinfo{person}{Rupesh~K.
  Srivastava}, \bibinfo{person}{Jan Koutník}, \bibinfo{person}{Bas~R.
  Steunebrink}, {and} \bibinfo{person}{Jürgen Schmidhuber}.}
  \bibinfo{year}{2017}\natexlab{}.
\newblock \showarticletitle{LSTM: A Search Space Odyssey}.
\newblock \bibinfo{journal}{\emph{IEEE Transactions on Neural Networks and
  Learning Systems}} \bibinfo{volume}{28}, \bibinfo{number}{10}
  (\bibinfo{year}{2017}), \bibinfo{pages}{2222--2232}.
\newblock
\urldef\tempurl%
\url{https://doi.org/10.1109/TNNLS.2016.2582924}
\showDOI{\tempurl}


\bibitem[Gu et~al\mbox{.}(2018)]%
        {s_ref61}
\bibfield{author}{\bibinfo{person}{Xiaodong Gu}, \bibinfo{person}{Hongyu
  Zhang}, {and} \bibinfo{person}{Sunghun Kim}.}
  \bibinfo{year}{2018}\natexlab{}.
\newblock \showarticletitle{Deep Code Search}. In
  \bibinfo{booktitle}{\emph{2018 IEEE/ACM 40th International Conference on
  Software Engineering (ICSE)}}. \bibinfo{pages}{933--944}.
\newblock
\urldef\tempurl%
\url{https://doi.org/10.1145/3180155.3180167}
\showDOI{\tempurl}


\bibitem[Guggulothu and Moiz(2020)]%
        {nref22}
\bibfield{author}{\bibinfo{person}{Thirupathi Guggulothu} {and}
  \bibinfo{person}{Salman~Abdul Moiz}.} \bibinfo{year}{2020}\natexlab{}.
\newblock \showarticletitle{Code smell detection using multi-label
  classification approach}.
\newblock \bibinfo{journal}{\emph{Software Quality Journal}}
  \bibinfo{volume}{28} (\bibinfo{year}{2020}), \bibinfo{pages}{1063--1086}.
\newblock


\bibitem[Guo et~al\mbox{.}(2021)]%
        {guo2021graphcodebert}
\bibfield{author}{\bibinfo{person}{Daya Guo}, \bibinfo{person}{Shuo Ren},
  \bibinfo{person}{Shuai Lu}, \bibinfo{person}{Zhangyin Feng},
  \bibinfo{person}{Duyu Tang}, \bibinfo{person}{Shujie LIU},
  \bibinfo{person}{Long Zhou}, \bibinfo{person}{Nan Duan},
  \bibinfo{person}{Alexey Svyatkovskiy}, \bibinfo{person}{Shengyu Fu},
  \bibinfo{person}{Michele Tufano}, \bibinfo{person}{Shao~Kun Deng},
  \bibinfo{person}{Colin Clement}, \bibinfo{person}{Dawn Drain},
  \bibinfo{person}{Neel Sundaresan}, \bibinfo{person}{Jian Yin},
  \bibinfo{person}{Daxin Jiang}, {and} \bibinfo{person}{Ming Zhou}.}
  \bibinfo{year}{2021}\natexlab{}.
\newblock \showarticletitle{{GraphCodeBERT}: Pre-training Code Representations
  with Data Flow}. In \bibinfo{booktitle}{\emph{International Conference on
  Learning Representations}}.
\newblock
\urldef\tempurl%
\url{https://openreview.net/forum?id=jLoC4ez43PZ}
\showURL{%
\tempurl}


\bibitem[Guo et~al\mbox{.}(2019)]%
        {s_ref126}
\bibfield{author}{\bibinfo{person}{Daya Guo}, \bibinfo{person}{Duyu Tang},
  \bibinfo{person}{Nan Duan}, \bibinfo{person}{Ming Zhou}, {and}
  \bibinfo{person}{Jian Yin}.} \bibinfo{year}{2019}\natexlab{}.
\newblock \showarticletitle{Coupling retrieval and meta-learning for
  context-dependent semantic parsing}.
\newblock \bibinfo{journal}{\emph{arXiv preprint arXiv:1906.07108}}
  (\bibinfo{year}{2019}).
\newblock


\bibitem[Gupta et~al\mbox{.}(2017)]%
        {s_ref104}
\bibfield{author}{\bibinfo{person}{Rahul Gupta}, \bibinfo{person}{Soham Pal},
  \bibinfo{person}{Aditya Kanade}, {and} \bibinfo{person}{Shirish Shevade}.}
  \bibinfo{year}{2017}\natexlab{}.
\newblock \showarticletitle{DeepFix: Fixing Common C Language Errors by Deep
  Learning}. In \bibinfo{booktitle}{\emph{Proceedings of the Thirty-First AAAI
  Conference on Artificial Intelligence}} (San Francisco, California, USA)
  \emph{(\bibinfo{series}{AAAI'17})}. \bibinfo{publisher}{AAAI Press},
  \bibinfo{pages}{1345–1351}.
\newblock


\bibitem[Hellendoorn and Devanbu(2017)]%
        {s_ref30}
\bibfield{author}{\bibinfo{person}{Vincent~J. Hellendoorn} {and}
  \bibinfo{person}{Premkumar Devanbu}.} \bibinfo{year}{2017}\natexlab{}.
\newblock \showarticletitle{Are Deep Neural Networks the Best Choice for
  Modeling Source Code?}. In \bibinfo{booktitle}{\emph{Proceedings of the 2017
  11th Joint Meeting on Foundations of Software Engineering}} (Paderborn,
  Germany) \emph{(\bibinfo{series}{ESEC/FSE 2017})}.
  \bibinfo{publisher}{Association for Computing Machinery},
  \bibinfo{address}{New York, NY, USA}, \bibinfo{pages}{763–773}.
\newblock
\showISBNx{9781450351058}
\urldef\tempurl%
\url{https://doi.org/10.1145/3106237.3106290}
\showDOI{\tempurl}


\bibitem[Hendrycks et~al\mbox{.}(2021)]%
        {hendrycks2021apps}
\bibfield{author}{\bibinfo{person}{Dan Hendrycks}, \bibinfo{person}{Steven
  Basart}, \bibinfo{person}{Saurav Kadavath}, \bibinfo{person}{Mantas Mazeika},
  \bibinfo{person}{Akul Arora}, \bibinfo{person}{Ethan Guo},
  \bibinfo{person}{Collin Burns}, \bibinfo{person}{Samir Puranik},
  \bibinfo{person}{Horace He}, \bibinfo{person}{Dawn Song}, {and}
  \bibinfo{person}{Jacob Steinhardt}.} \bibinfo{year}{2021}\natexlab{}.
\newblock \showarticletitle{Measuring Coding Challenge Competence With {APPS}}.
\newblock \bibinfo{journal}{\emph{In: Proceedings of Conference on Neural
  Information Processing Systems (NeurIPS) Track on Datasets and Benchmarks
  (Round 2)}} (\bibinfo{year}{2021}).
\newblock


\bibitem[Hoang et~al\mbox{.}(2020)]%
        {nref38}
\bibfield{author}{\bibinfo{person}{Thong Hoang}, \bibinfo{person}{Hong~Jin
  Kang}, \bibinfo{person}{David Lo}, {and} \bibinfo{person}{Julia Lawall}.}
  \bibinfo{year}{2020}\natexlab{}.
\newblock \showarticletitle{CC2Vec: Distributed representations of code
  changes}. In \bibinfo{booktitle}{\emph{Proceedings of the ACM/IEEE 42nd
  International Conference on Software Engineering}}.
  \bibinfo{pages}{518--529}.
\newblock


\bibitem[Hsiao et~al\mbox{.}(2020)]%
        {s_ref60}
\bibfield{author}{\bibinfo{person}{I-Han Hsiao}, \bibinfo{person}{Po-Kai
  Huang}, {and} \bibinfo{person}{Hannah Murphy}.}
  \bibinfo{year}{2020}\natexlab{}.
\newblock \showarticletitle{Integrating Programming Learning Analytics Across
  Physical and Digital Space}.
\newblock \bibinfo{journal}{\emph{IEEE Transactions on Emerging Topics in
  Computing}} \bibinfo{volume}{8}, \bibinfo{number}{1} (\bibinfo{year}{2020}),
  \bibinfo{pages}{206--217}.
\newblock
\urldef\tempurl%
\url{https://doi.org/10.1109/TETC.2017.2701201}
\showDOI{\tempurl}


\bibitem[Hu et~al\mbox{.}(2018b)]%
        {nref50}
\bibfield{author}{\bibinfo{person}{Gang Hu}, \bibinfo{person}{Linjie Zhu},
  {and} \bibinfo{person}{Junfeng Yang}.} \bibinfo{year}{2018}\natexlab{b}.
\newblock \showarticletitle{AppFlow: using machine learning to synthesize
  robust, reusable UI tests}. In \bibinfo{booktitle}{\emph{Proceedings of the
  2018 26th ACM Joint Meeting on European Software Engineering Conference and
  Symposium on the Foundations of Software Engineering}}.
  \bibinfo{pages}{269--282}.
\newblock


\bibitem[Hu et~al\mbox{.}(2018a)]%
        {s_ref121}
\bibfield{author}{\bibinfo{person}{Xing Hu}, \bibinfo{person}{Ge Li},
  \bibinfo{person}{Xin Xia}, \bibinfo{person}{David Lo}, \bibinfo{person}{Shuai
  Lu}, {and} \bibinfo{person}{Zhi Jin}.} \bibinfo{year}{2018}\natexlab{a}.
\newblock \showarticletitle{Summarizing Source Code with Transferred API
  Knowledge}. In \bibinfo{booktitle}{\emph{Proceedings of the 27th
  International Joint Conference on Artificial Intelligence}} (Stockholm,
  Sweden) \emph{(\bibinfo{series}{IJCAI'18})}. \bibinfo{publisher}{AAAI Press},
  \bibinfo{pages}{2269–2275}.
\newblock
\showISBNx{9780999241127}


\bibitem[Husain et~al\mbox{.}(2019)]%
        {s_ref62}
\bibfield{author}{\bibinfo{person}{Hamel Husain}, \bibinfo{person}{Ho-Hsiang
  Wu}, \bibinfo{person}{Tiferet Gazit}, \bibinfo{person}{Miltiadis Allamanis},
  {and} \bibinfo{person}{Marc Brockschmidt}.} \bibinfo{year}{2019}\natexlab{}.
\newblock \showarticletitle{Codesearchnet challenge: Evaluating the state of
  semantic code search}.
\newblock \bibinfo{journal}{\emph{arXiv preprint arXiv:1909.09436}}
  (\bibinfo{year}{2019}).
\newblock


\bibitem[Ihantola et~al\mbox{.}(2010)]%
        {s_ref55}
\bibfield{author}{\bibinfo{person}{Petri Ihantola}, \bibinfo{person}{Tuukka
  Ahoniemi}, \bibinfo{person}{Ville Karavirta}, {and} \bibinfo{person}{Otto
  Sepp\"{a}l\"{a}}.} \bibinfo{year}{2010}\natexlab{}.
\newblock \showarticletitle{Review of Recent Systems for Automatic Assessment
  of Programming Assignments}. In \bibinfo{booktitle}{\emph{Proceedings of the
  10th Koli Calling International Conference on Computing Education Research}}
  (Koli, Finland) \emph{(\bibinfo{series}{Koli Calling '10})}.
  \bibinfo{publisher}{Association for Computing Machinery},
  \bibinfo{address}{New York, NY, USA}, \bibinfo{pages}{86–93}.
\newblock
\showISBNx{9781450305204}
\urldef\tempurl%
\url{https://doi.org/10.1145/1930464.1930480}
\showDOI{\tempurl}


\bibitem[Iyer et~al\mbox{.}(2019)]%
        {s_ref127}
\bibfield{author}{\bibinfo{person}{Srinivasan Iyer}, \bibinfo{person}{Alvin
  Cheung}, {and} \bibinfo{person}{Luke Zettlemoyer}.}
  \bibinfo{year}{2019}\natexlab{}.
\newblock \showarticletitle{Learning programmatic idioms for scalable semantic
  parsing}.
\newblock \bibinfo{journal}{\emph{arXiv preprint arXiv:1904.09086}}
  (\bibinfo{year}{2019}).
\newblock


\bibitem[Iyer et~al\mbox{.}(2016)]%
        {s_ref122}
\bibfield{author}{\bibinfo{person}{Srinivasan Iyer}, \bibinfo{person}{Ioannis
  Konstas}, \bibinfo{person}{Alvin Cheung}, {and} \bibinfo{person}{Luke
  Zettlemoyer}.} \bibinfo{year}{2016}\natexlab{}.
\newblock \showarticletitle{Summarizing source code using a neural attention
  model}. In \bibinfo{booktitle}{\emph{Proceedings of the 54th Annual Meeting
  of the Association for Computational Linguistics (Volume 1: Long Papers)}}.
  \bibinfo{pages}{2073--2083}.
\newblock


\bibitem[Iyer et~al\mbox{.}(2018)]%
        {iyer2018concode}
\bibfield{author}{\bibinfo{person}{Srinivasan Iyer}, \bibinfo{person}{Ioannis
  Konstas}, \bibinfo{person}{Alvin Cheung}, {and} \bibinfo{person}{Luke
  Zettlemoyer}.} \bibinfo{year}{2018}\natexlab{}.
\newblock \showarticletitle{Mapping Language to Code in Programmatic Context}.
  In \bibinfo{booktitle}{\emph{Proceedings of the 2018 Conference on Empirical
  Methods in Natural Language Processing}}. \bibinfo{publisher}{Association for
  Computational Linguistics}, \bibinfo{address}{Brussels, Belgium},
  \bibinfo{pages}{1643--1652}.
\newblock
\urldef\tempurl%
\url{https://doi.org/10.18653/v1/D18-1192}
\showDOI{\tempurl}


\bibitem[Ji et~al\mbox{.}(2018)]%
        {s_ref134}
\bibfield{author}{\bibinfo{person}{Tao Ji}, \bibinfo{person}{Jinkun Pan},
  \bibinfo{person}{Liqian Chen}, {and} \bibinfo{person}{Xiaoguang Mao}.}
  \bibinfo{year}{2018}\natexlab{}.
\newblock \showarticletitle{Identifying Supplementary Bug-fix Commits}. In
  \bibinfo{booktitle}{\emph{2018 IEEE 42nd Annual Computer Software and
  Applications Conference (COMPSAC)}}, Vol.~\bibinfo{volume}{01}.
  \bibinfo{pages}{184--193}.
\newblock
\urldef\tempurl%
\url{https://doi.org/10.1109/COMPSAC.2018.00031}
\showDOI{\tempurl}


\bibitem[Jiang et~al\mbox{.}(2021)]%
        {jiang2021treebert}
\bibfield{author}{\bibinfo{person}{Xue Jiang}, \bibinfo{person}{Zhuoran Zheng},
  \bibinfo{person}{Chen Lyu}, \bibinfo{person}{Liang Li}, {and}
  \bibinfo{person}{Lei Lyu}.} \bibinfo{year}{2021}\natexlab{}.
\newblock \showarticletitle{{TreeBERT}: A tree-based pre-trained model for
  programming language}. In \bibinfo{booktitle}{\emph{Uncertainty in Artificial
  Intelligence}}. PMLR, \bibinfo{pages}{54--63}.
\newblock


\bibitem[Just et~al\mbox{.}(2014)]%
        {nref31}
\bibfield{author}{\bibinfo{person}{Ren{\'e} Just}, \bibinfo{person}{Darioush
  Jalali}, {and} \bibinfo{person}{Michael~D Ernst}.}
  \bibinfo{year}{2014}\natexlab{}.
\newblock \showarticletitle{Defects4J: A database of existing faults to enable
  controlled testing studies for Java programs}. In
  \bibinfo{booktitle}{\emph{Proceedings of the 2014 international symposium on
  software testing and analysis}}. \bibinfo{pages}{437--440}.
\newblock


\bibitem[Kanade et~al\mbox{.}(2020)]%
        {kanade2020cubert}
\bibfield{author}{\bibinfo{person}{Aditya Kanade}, \bibinfo{person}{Petros
  Maniatis}, \bibinfo{person}{Gogul Balakrishnan}, {and}
  \bibinfo{person}{Kensen Shi}.} \bibinfo{year}{2020}\natexlab{}.
\newblock \showarticletitle{Learning and Evaluating Contextual Embedding of
  Source Code}. In \bibinfo{booktitle}{\emph{Proceedings of the 37th
  International Conference on Machine Learning}}
  \emph{(\bibinfo{series}{ICML'20})}. \bibinfo{publisher}{JMLR.org}, Article
  \bibinfo{articleno}{474}, \bibinfo{numpages}{12}~pages.
\newblock


\bibitem[Karaivanov et~al\mbox{.}(2014)]%
        {s_ref68}
\bibfield{author}{\bibinfo{person}{Svetoslav Karaivanov},
  \bibinfo{person}{Veselin Raychev}, {and} \bibinfo{person}{Martin Vechev}.}
  \bibinfo{year}{2014}\natexlab{}.
\newblock \showarticletitle{Phrase-Based Statistical Translation of Programming
  Languages}. In \bibinfo{booktitle}{\emph{Proceedings of the 2014 ACM
  International Symposium on New Ideas, New Paradigms, and Reflections on
  Programming \& Software}} (Portland, Oregon, USA)
  \emph{(\bibinfo{series}{Onward! 2014})}. \bibinfo{publisher}{Association for
  Computing Machinery}, \bibinfo{address}{New York, NY, USA},
  \bibinfo{pages}{173–184}.
\newblock
\showISBNx{9781450332101}
\urldef\tempurl%
\url{https://doi.org/10.1145/2661136.2661148}
\showDOI{\tempurl}


\bibitem[Krizhevsky et~al\mbox{.}(2017)]%
        {s_ref20}
\bibfield{author}{\bibinfo{person}{Alex Krizhevsky}, \bibinfo{person}{Ilya
  Sutskever}, {and} \bibinfo{person}{Geoffrey~E. Hinton}.}
  \bibinfo{year}{2017}\natexlab{}.
\newblock \showarticletitle{ImageNet Classification with Deep Convolutional
  Neural Networks}.
\newblock \bibinfo{journal}{\emph{Commun. ACM}} \bibinfo{volume}{60},
  \bibinfo{number}{6} (\bibinfo{date}{may} \bibinfo{year}{2017}),
  \bibinfo{pages}{84–90}.
\newblock
\showISSN{0001-0782}
\urldef\tempurl%
\url{https://doi.org/10.1145/3065386}
\showDOI{\tempurl}


\bibitem[Kurnia et~al\mbox{.}(2001)]%
        {s_ref3}
\bibfield{author}{\bibinfo{person}{Andy Kurnia}, \bibinfo{person}{Andrew Lim},
  {and} \bibinfo{person}{Brenda Cheang}.} \bibinfo{year}{2001}\natexlab{}.
\newblock \showarticletitle{Online Judge}.
\newblock \bibinfo{journal}{\emph{Computers \& Education}}
  \bibinfo{volume}{36}, \bibinfo{number}{4} (\bibinfo{year}{2001}),
  \bibinfo{pages}{299--315}.
\newblock
\showISSN{0360-1315}
\urldef\tempurl%
\url{https://doi.org/10.1016/S0360-1315(01)00018-5}
\showDOI{\tempurl}


\bibitem[Le et~al\mbox{.}(2020)]%
        {s_ref27}
\bibfield{author}{\bibinfo{person}{Triet H.~M. Le}, \bibinfo{person}{Hao Chen},
  {and} \bibinfo{person}{Muhammad~Ali Babar}.} \bibinfo{year}{2020}\natexlab{}.
\newblock \showarticletitle{Deep Learning for Source Code Modeling and
  Generation: Models, Applications, and Challenges}.
\newblock \bibinfo{journal}{\emph{ACM Comput. Surv.}} \bibinfo{volume}{53},
  \bibinfo{number}{3}, Article \bibinfo{articleno}{62} (\bibinfo{date}{jun}
  \bibinfo{year}{2020}), \bibinfo{numpages}{38}~pages.
\newblock
\showISSN{0360-0300}
\urldef\tempurl%
\url{https://doi.org/10.1145/3383458}
\showDOI{\tempurl}


\bibitem[Le et~al\mbox{.}(2015)]%
        {s_ref111}
\bibfield{author}{\bibinfo{person}{Xuan-Bach~D. Le},
  \bibinfo{person}{Tien-Duy~B. Le}, {and} \bibinfo{person}{David Lo}.}
  \bibinfo{year}{2015}\natexlab{}.
\newblock \showarticletitle{Should fixing these failures be delegated to
  automated program repair?}. In \bibinfo{booktitle}{\emph{2015 IEEE 26th
  International Symposium on Software Reliability Engineering (ISSRE)}}.
  \bibinfo{pages}{427--437}.
\newblock
\urldef\tempurl%
\url{https://doi.org/10.1109/ISSRE.2015.7381836}
\showDOI{\tempurl}


\bibitem[Le~Goues et~al\mbox{.}(2015)]%
        {nref35}
\bibfield{author}{\bibinfo{person}{Claire Le~Goues}, \bibinfo{person}{Neal
  Holtschulte}, \bibinfo{person}{Edward~K Smith}, \bibinfo{person}{Yuriy Brun},
  \bibinfo{person}{Premkumar Devanbu}, \bibinfo{person}{Stephanie Forrest},
  {and} \bibinfo{person}{Westley Weimer}.} \bibinfo{year}{2015}\natexlab{}.
\newblock \showarticletitle{The ManyBugs and IntroClass benchmarks for
  automated repair of C programs}.
\newblock \bibinfo{journal}{\emph{IEEE Transactions on Software Engineering}}
  \bibinfo{volume}{41}, \bibinfo{number}{12} (\bibinfo{year}{2015}),
  \bibinfo{pages}{1236--1256}.
\newblock


\bibitem[Leal and Silva(2003)]%
        {s_ref57}
\bibfield{author}{\bibinfo{person}{Jos{\'e}~Paulo Leal} {and}
  \bibinfo{person}{Fernando Silva}.} \bibinfo{year}{2003}\natexlab{}.
\newblock \showarticletitle{Mooshak: A Web-based multi-site programming contest
  system}.
\newblock \bibinfo{journal}{\emph{Software: Practice and Experience}}
  \bibinfo{volume}{33}, \bibinfo{number}{6} (\bibinfo{year}{2003}),
  \bibinfo{pages}{567--581}.
\newblock


\bibitem[LeClair et~al\mbox{.}(2019)]%
        {s_ref36}
\bibfield{author}{\bibinfo{person}{Alexander LeClair}, \bibinfo{person}{Siyuan
  Jiang}, {and} \bibinfo{person}{Collin McMillan}.}
  \bibinfo{year}{2019}\natexlab{}.
\newblock \showarticletitle{A Neural Model for Generating Natural Language
  Summaries of Program Subroutines}. In \bibinfo{booktitle}{\emph{Proceedings
  of the 41st International Conference on Software Engineering}} (Montreal,
  Quebec, Canada) \emph{(\bibinfo{series}{ICSE '19})}. \bibinfo{publisher}{IEEE
  Press}, \bibinfo{pages}{795–806}.
\newblock
\urldef\tempurl%
\url{https://doi.org/10.1109/ICSE.2019.00087}
\showDOI{\tempurl}


\bibitem[Lee et~al\mbox{.}(2017)]%
        {s_ref19}
\bibfield{author}{\bibinfo{person}{Song-Mi Lee}, \bibinfo{person}{Sang~Min
  Yoon}, {and} \bibinfo{person}{Heeryon Cho}.} \bibinfo{year}{2017}\natexlab{}.
\newblock \showarticletitle{Human activity recognition from accelerometer data
  using Convolutional Neural Network}. In \bibinfo{booktitle}{\emph{2017 IEEE
  International Conference on Big Data and Smart Computing (BigComp)}}.
  \bibinfo{pages}{131--134}.
\newblock
\urldef\tempurl%
\url{https://doi.org/10.1109/BIGCOMP.2017.7881728}
\showDOI{\tempurl}


\bibitem[Lei et~al\mbox{.}(2022)]%
        {nref69}
\bibfield{author}{\bibinfo{person}{Maggie Lei}, \bibinfo{person}{Hao Li},
  \bibinfo{person}{Ji Li}, \bibinfo{person}{Namrata Aundhkar}, {and}
  \bibinfo{person}{Dae-Kyoo Kim}.} \bibinfo{year}{2022}\natexlab{}.
\newblock \showarticletitle{Deep learning application on code clone detection:
  A review of current knowledge}.
\newblock \bibinfo{journal}{\emph{Journal of Systems and Software}}
  \bibinfo{volume}{184} (\bibinfo{year}{2022}), \bibinfo{pages}{111141}.
\newblock
\showISSN{0164-1212}
\urldef\tempurl%
\url{https://doi.org/10.1016/j.jss.2021.111141}
\showDOI{\tempurl}


\bibitem[Li et~al\mbox{.}(2018)]%
        {s_ref102}
\bibfield{author}{\bibinfo{person}{Jian Li}, \bibinfo{person}{Yue Wang},
  \bibinfo{person}{Michael~R. Lyu}, {and} \bibinfo{person}{Irwin King}.}
  \bibinfo{year}{2018}\natexlab{}.
\newblock \showarticletitle{Code Completion with Neural Attention and Pointer
  Networks}. In \bibinfo{booktitle}{\emph{Proceedings of the 27th International
  Joint Conference on Artificial Intelligence}} (Stockholm, Sweden)
  \emph{(\bibinfo{series}{IJCAI'18})}. \bibinfo{publisher}{AAAI Press},
  \bibinfo{pages}{4159–25}.
\newblock
\showISBNx{9780999241127}


\bibitem[Li et~al\mbox{.}(2017)]%
        {nref67}
\bibfield{author}{\bibinfo{person}{Liuqing Li}, \bibinfo{person}{He Feng},
  \bibinfo{person}{Wenjie Zhuang}, \bibinfo{person}{Na Meng}, {and}
  \bibinfo{person}{Barbara Ryder}.} \bibinfo{year}{2017}\natexlab{}.
\newblock \showarticletitle{CCLearner: A Deep Learning-Based Clone Detection
  Approach}. In \bibinfo{booktitle}{\emph{2017 IEEE International Conference on
  Software Maintenance and Evolution (ICSME)}}. \bibinfo{pages}{249--260}.
\newblock
\urldef\tempurl%
\url{https://doi.org/10.1109/ICSME.2017.46}
\showDOI{\tempurl}


\bibitem[Li et~al\mbox{.}(2022a)]%
        {li2022alphacode}
\bibfield{author}{\bibinfo{person}{Yujia Li}, \bibinfo{person}{David Choi},
  \bibinfo{person}{Junyoung Chung}, \bibinfo{person}{Nate Kushman},
  \bibinfo{person}{Julian Schrittwieser}, \bibinfo{person}{Rémi Leblond},
  \bibinfo{person}{Tom Eccles}, \bibinfo{person}{James Keeling},
  \bibinfo{person}{Felix Gimeno}, \bibinfo{person}{Agustin~Dal Lago},
  \bibinfo{person}{Thomas Hubert}, \bibinfo{person}{Peter Choy},
  \bibinfo{person}{Cyprien de Masson~d’Autume}, \bibinfo{person}{Igor
  Babuschkin}, \bibinfo{person}{Xinyun Chen}, \bibinfo{person}{Po-Sen Huang},
  \bibinfo{person}{Johannes Welbl}, \bibinfo{person}{Sven Gowal},
  \bibinfo{person}{Alexey Cherepanov}, \bibinfo{person}{James Molloy},
  \bibinfo{person}{Daniel~J. Mankowitz}, \bibinfo{person}{Esme~Sutherland
  Robson}, \bibinfo{person}{Pushmeet Kohli}, \bibinfo{person}{Nando de
  Freitas}, \bibinfo{person}{Koray Kavukcuoglu}, {and} \bibinfo{person}{Oriol
  Vinyals}.} \bibinfo{year}{2022}\natexlab{a}.
\newblock \showarticletitle{Competition-level code generation with
  {AlphaCode}}.
\newblock \bibinfo{journal}{\emph{Science}} \bibinfo{volume}{378},
  \bibinfo{number}{6624} (\bibinfo{year}{2022}), \bibinfo{pages}{1092--1097}.
\newblock
\urldef\tempurl%
\url{https://doi.org/10.1126/science.abq1158}
\showDOI{\tempurl}


\bibitem[Li et~al\mbox{.}(2022b)]%
        {nref71}
\bibfield{author}{\bibinfo{person}{Yi Li}, \bibinfo{person}{Shaohua Wang},
  {and} \bibinfo{person}{Tien~N. Nguyen}.} \bibinfo{year}{2022}\natexlab{b}.
\newblock \showarticletitle{DEAR: A Novel Deep Learning-Based Approach for
  Automated Program Repair}. In \bibinfo{booktitle}{\emph{Proceedings of the
  44th International Conference on Software Engineering}} (Pittsburgh,
  Pennsylvania) \emph{(\bibinfo{series}{ICSE '22})}.
  \bibinfo{publisher}{Association for Computing Machinery},
  \bibinfo{address}{New York, NY, USA}, \bibinfo{pages}{511–523}.
\newblock
\showISBNx{9781450392211}
\urldef\tempurl%
\url{https://doi.org/10.1145/3510003.3510177}
\showDOI{\tempurl}


\bibitem[Li et~al\mbox{.}(2019a)]%
        {nref32}
\bibfield{author}{\bibinfo{person}{Yi Li}, \bibinfo{person}{Shaohua Wang},
  \bibinfo{person}{Tien~N Nguyen}, {and} \bibinfo{person}{Son Van~Nguyen}.}
  \bibinfo{year}{2019}\natexlab{a}.
\newblock \showarticletitle{Improving bug detection via context-based code
  representation learning and attention-based neural networks}.
\newblock \bibinfo{journal}{\emph{Proceedings of the ACM on Programming
  Languages}} \bibinfo{volume}{3}, \bibinfo{number}{OOPSLA}
  (\bibinfo{year}{2019}), \bibinfo{pages}{1--30}.
\newblock


\bibitem[Li et~al\mbox{.}(2019b)]%
        {s_ref75}
\bibfield{author}{\bibinfo{person}{Yi Li}, \bibinfo{person}{Shaohua Wang},
  \bibinfo{person}{Tien~N. Nguyen}, {and} \bibinfo{person}{Son Van~Nguyen}.}
  \bibinfo{year}{2019}\natexlab{b}.
\newblock \showarticletitle{Improving Bug Detection via Context-Based Code
  Representation Learning and Attention-Based Neural Networks}.
\newblock \bibinfo{journal}{\emph{Proc. ACM Program. Lang.}}
  \bibinfo{volume}{3}, \bibinfo{number}{OOPSLA}, Article
  \bibinfo{articleno}{162} (\bibinfo{date}{oct} \bibinfo{year}{2019}),
  \bibinfo{numpages}{30}~pages.
\newblock
\urldef\tempurl%
\url{https://doi.org/10.1145/3360588}
\showDOI{\tempurl}


\bibitem[Lima et~al\mbox{.}(2020)]%
        {s_ref31}
\bibfield{author}{\bibinfo{person}{Rui Lima}, \bibinfo{person}{António
  Miguel~Rosado da Cruz}, {and} \bibinfo{person}{Jorge Ribeiro}.}
  \bibinfo{year}{2020}\natexlab{}.
\newblock \showarticletitle{Artificial Intelligence Applied to Software
  Testing: A Literature Review}. In \bibinfo{booktitle}{\emph{2020 15th Iberian
  Conference on Information Systems and Technologies (CISTI)}}.
  \bibinfo{pages}{1--6}.
\newblock
\urldef\tempurl%
\url{https://doi.org/10.23919/CISTI49556.2020.9141124}
\showDOI{\tempurl}


\bibitem[Lin et~al\mbox{.}(2017)]%
        {lin2017quixbugs}
\bibfield{author}{\bibinfo{person}{Derrick Lin}, \bibinfo{person}{James
  Koppel}, \bibinfo{person}{Angela Chen}, {and} \bibinfo{person}{Armando
  Solar-Lezama}.} \bibinfo{year}{2017}\natexlab{}.
\newblock \showarticletitle{QuixBugs: A Multi-Lingual Program Repair Benchmark
  Set Based on the Quixey Challenge}. In \bibinfo{booktitle}{\emph{Proceedings
  Companion of the 2017 ACM SIGPLAN International Conference on Systems,
  Programming, Languages, and Applications: Software for Humanity}} (Vancouver,
  BC, Canada) \emph{(\bibinfo{series}{SPLASH Companion 2017})}.
  \bibinfo{publisher}{Association for Computing Machinery},
  \bibinfo{address}{New York, NY, USA}, \bibinfo{pages}{55–56}.
\newblock
\showISBNx{9781450355148}
\urldef\tempurl%
\url{https://doi.org/10.1145/3135932.3135941}
\showDOI{\tempurl}


\bibitem[Lin et~al\mbox{.}(2018)]%
        {nref26}
\bibfield{author}{\bibinfo{person}{Guanjun Lin}, \bibinfo{person}{Jun Zhang},
  \bibinfo{person}{Wei Luo}, \bibinfo{person}{Lei Pan}, \bibinfo{person}{Yang
  Xiang}, \bibinfo{person}{Olivier De~Vel}, {and} \bibinfo{person}{Paul
  Montague}.} \bibinfo{year}{2018}\natexlab{}.
\newblock \showarticletitle{Cross-project transfer representation learning for
  vulnerable function discovery}.
\newblock \bibinfo{journal}{\emph{IEEE Transactions on Industrial Informatics}}
  \bibinfo{volume}{14}, \bibinfo{number}{7} (\bibinfo{year}{2018}),
  \bibinfo{pages}{3289--3297}.
\newblock


\bibitem[Ling et~al\mbox{.}(2020)]%
        {nref72}
\bibfield{author}{\bibinfo{person}{Chunyang Ling}, \bibinfo{person}{Zeqi Lin},
  \bibinfo{person}{Yanzhen Zou}, {and} \bibinfo{person}{Bing Xie}.}
  \bibinfo{year}{2020}\natexlab{}.
\newblock \showarticletitle{Adaptive Deep Code Search}. In
  \bibinfo{booktitle}{\emph{Proceedings of the 28th International Conference on
  Program Comprehension}} (Seoul, Republic of Korea)
  \emph{(\bibinfo{series}{ICPC '20})}. \bibinfo{publisher}{Association for
  Computing Machinery}, \bibinfo{address}{New York, NY, USA},
  \bibinfo{pages}{48–59}.
\newblock
\showISBNx{9781450379588}
\urldef\tempurl%
\url{https://doi.org/10.1145/3387904.3389278}
\showDOI{\tempurl}


\bibitem[Liu et~al\mbox{.}(2019c)]%
        {nref73}
\bibfield{author}{\bibinfo{person}{Bohong Liu}, \bibinfo{person}{Tao Wang},
  \bibinfo{person}{Xunhui Zhang}, \bibinfo{person}{Qiang Fan},
  \bibinfo{person}{Gang Yin}, {and} \bibinfo{person}{Jinsheng Deng}.}
  \bibinfo{year}{2019}\natexlab{c}.
\newblock \showarticletitle{A Neural-Network Based Code Summarization Approach
  by Using Source Code and Its Call Dependencies}. In
  \bibinfo{booktitle}{\emph{Proceedings of the 11th Asia-Pacific Symposium on
  Internetware}} (Fukuoka, Japan) \emph{(\bibinfo{series}{Internetware '19})}.
  \bibinfo{publisher}{Association for Computing Machinery},
  \bibinfo{address}{New York, NY, USA}, Article \bibinfo{articleno}{12},
  \bibinfo{numpages}{10}~pages.
\newblock
\showISBNx{9781450377010}
\urldef\tempurl%
\url{https://doi.org/10.1145/3361242.3362774}
\showDOI{\tempurl}


\bibitem[Liu et~al\mbox{.}(2016)]%
        {s_ref108}
\bibfield{author}{\bibinfo{person}{Chang Liu}, \bibinfo{person}{Xinyun Chen},
  \bibinfo{person}{Eui~Chul Shin}, \bibinfo{person}{Mingcheng Chen}, {and}
  \bibinfo{person}{Dawn Song}.} \bibinfo{year}{2016}\natexlab{}.
\newblock \showarticletitle{Latent attention for if-then program synthesis}.
\newblock \bibinfo{journal}{\emph{Advances in Neural Information Processing
  Systems}}  \bibinfo{volume}{29} (\bibinfo{year}{2016}).
\newblock


\bibitem[Liu et~al\mbox{.}(2020)]%
        {s_ref100}
\bibfield{author}{\bibinfo{person}{Fang Liu}, \bibinfo{person}{Ge Li},
  \bibinfo{person}{Bolin Wei}, \bibinfo{person}{Xin Xia},
  \bibinfo{person}{Zhiyi Fu}, {and} \bibinfo{person}{Zhi Jin}.}
  \bibinfo{year}{2020}\natexlab{}.
\newblock \showarticletitle{A self-attentional neural architecture for code
  completion with multi-task learning}. In
  \bibinfo{booktitle}{\emph{Proceedings of the 28th International Conference on
  Program Comprehension}}. \bibinfo{pages}{37--47}.
\newblock


\bibitem[Liu et~al\mbox{.}(2021)]%
        {s_ref35_ref101}
\bibfield{author}{\bibinfo{person}{Fang Liu}, \bibinfo{person}{Ge Li},
  \bibinfo{person}{Yunfei Zhao}, {and} \bibinfo{person}{Zhi Jin}.}
  \bibinfo{year}{2021}\natexlab{}.
\newblock \showarticletitle{Multi-Task Learning Based Pre-Trained Language
  Model for Code Completion}. In \bibinfo{booktitle}{\emph{Proceedings of the
  35th IEEE/ACM International Conference on Automated Software Engineering}}
  (Virtual Event, Australia) \emph{(\bibinfo{series}{ASE '20})}.
  \bibinfo{publisher}{Association for Computing Machinery},
  \bibinfo{address}{New York, NY, USA}, \bibinfo{pages}{473–485}.
\newblock
\showISBNx{9781450367684}
\urldef\tempurl%
\url{https://doi.org/10.1145/3324884.3416591}
\showDOI{\tempurl}


\bibitem[Liu et~al\mbox{.}(2019a)]%
        {nref51}
\bibfield{author}{\bibinfo{person}{Xiao Liu}, \bibinfo{person}{Xiaoting Li},
  \bibinfo{person}{Rupesh Prajapati}, {and} \bibinfo{person}{Dinghao Wu}.}
  \bibinfo{year}{2019}\natexlab{a}.
\newblock \showarticletitle{Deepfuzz: Automatic generation of syntax valid c
  programs for fuzz testing}. In \bibinfo{booktitle}{\emph{Proceedings of the
  AAAI Conference on Artificial Intelligence}}, Vol.~\bibinfo{volume}{33}.
  \bibinfo{pages}{1044--1051}.
\newblock


\bibitem[Liu et~al\mbox{.}(2019b)]%
        {liu2019roberta}
\bibfield{author}{\bibinfo{person}{Yinhan Liu}, \bibinfo{person}{Myle Ott},
  \bibinfo{person}{Naman Goyal}, \bibinfo{person}{Jingfei Du},
  \bibinfo{person}{Mandar Joshi}, \bibinfo{person}{Danqi Chen},
  \bibinfo{person}{Omer Levy}, \bibinfo{person}{Mike Lewis},
  \bibinfo{person}{Luke Zettlemoyer}, {and} \bibinfo{person}{Veselin
  Stoyanov}.} \bibinfo{year}{2019}\natexlab{b}.
\newblock \showarticletitle{{RoBERTa}: A Robustly Optimized BERT Pretraining
  Approach}.
\newblock \bibinfo{journal}{\emph{arXiv preprint}}
  \bibinfo{volume}{arXiv:1907.11692} (\bibinfo{year}{2019}).
\newblock
\urldef\tempurl%
\url{https://doi.org/10.48550/arXiv.1907.11692}
\showDOI{\tempurl}


\bibitem[Lu et~al\mbox{.}(2021)]%
        {s_ref47}
\bibfield{author}{\bibinfo{person}{Shuai Lu}, \bibinfo{person}{Daya Guo},
  \bibinfo{person}{Shuo Ren}, \bibinfo{person}{Junjie Huang},
  \bibinfo{person}{Alexey Svyatkovskiy}, \bibinfo{person}{Ambrosio Blanco},
  \bibinfo{person}{Colin Clement}, \bibinfo{person}{Dawn Drain},
  \bibinfo{person}{Daxin Jiang}, \bibinfo{person}{Duyu Tang}, {et~al\mbox{.}}}
  \bibinfo{year}{2021}\natexlab{}.
\newblock \showarticletitle{Codexglue: A machine learning benchmark dataset for
  code understanding and generation}.
\newblock \bibinfo{journal}{\emph{arXiv preprint arXiv:2102.04664}}
  (\bibinfo{year}{2021}).
\newblock


\bibitem[Lu et~al\mbox{.}(2017)]%
        {s_ref58}
\bibfield{author}{\bibinfo{person}{Xudong Lu}, \bibinfo{person}{Dongyu Zheng},
  {and} \bibinfo{person}{Lei Liu}.} \bibinfo{year}{2017}\natexlab{}.
\newblock \showarticletitle{Data Driven Analysis on the Effect of Online Judge
  System}. In \bibinfo{booktitle}{\emph{2017 IEEE International Conference on
  Internet of Things (iThings) and IEEE Green Computing and Communications
  (GreenCom) and IEEE Cyber, Physical and Social Computing (CPSCom) and IEEE
  Smart Data (SmartData)}}. \bibinfo{pages}{573--577}.
\newblock
\urldef\tempurl%
\url{https://doi.org/10.1109/iThings-GreenCom-CPSCom-SmartData.2017.90}
\showDOI{\tempurl}


\bibitem[M.~Mostafizer et~al\mbox{.}(2020)]%
        {s_ref14}
\bibfield{author}{\bibinfo{person}{Rahman M.~Mostafizer},
  \bibinfo{person}{Yutaka Watanobe}, {and} \bibinfo{person}{Keita Nakamura}.}
  \bibinfo{year}{2020}\natexlab{}.
\newblock \showarticletitle{A Neural Network Based Intelligent Support Model
  for Program Code Completion}.
\newblock \bibinfo{journal}{\emph{Scientific Programming}}
  \bibinfo{volume}{2020} (\bibinfo{year}{2020}).
\newblock
\urldef\tempurl%
\url{https://doi.org/10.1155/2020/7426461}
\showDOI{\tempurl}


\bibitem[M.~Mostafizer et~al\mbox{.}(2021)]%
        {s_ref12}
\bibfield{author}{\bibinfo{person}{Rahman M.~Mostafizer},
  \bibinfo{person}{Yutaka Watanobe}, {and} \bibinfo{person}{Keita Nakamura}.}
  \bibinfo{year}{2021}\natexlab{}.
\newblock \showarticletitle{A Bidirectional LSTM Language Model for Code
  Evaluation and Repair}.
\newblock \bibinfo{journal}{\emph{Symmetry}} \bibinfo{volume}{13},
  \bibinfo{number}{2} (\bibinfo{year}{2021}).
\newblock
\showISSN{2073-8994}
\urldef\tempurl%
\url{https://doi.org/10.3390/sym13020247}
\showDOI{\tempurl}


\bibitem[Ma et~al\mbox{.}(2018)]%
        {s_ref132}
\bibfield{author}{\bibinfo{person}{Yuzhan Ma}, \bibinfo{person}{Sarah
  Fakhoury}, \bibinfo{person}{Michael Christensen}, \bibinfo{person}{Venera
  Arnaoudova}, \bibinfo{person}{Waleed Zogaan}, {and} \bibinfo{person}{Mehdi
  Mirakhorli}.} \bibinfo{year}{2018}\natexlab{}.
\newblock \showarticletitle{Automatic Classification of Software Artifacts in
  Open-Source Applications}. In \bibinfo{booktitle}{\emph{Proceedings of the
  15th International Conference on Mining Software Repositories}} (Gothenburg,
  Sweden) \emph{(\bibinfo{series}{MSR '18})}. \bibinfo{publisher}{Association
  for Computing Machinery}, \bibinfo{address}{New York, NY, USA},
  \bibinfo{pages}{414–425}.
\newblock
\showISBNx{9781450357166}
\urldef\tempurl%
\url{https://doi.org/10.1145/3196398.3196446}
\showDOI{\tempurl}


\bibitem[Matsumoto et~al\mbox{.}(2021)]%
        {nref53}
\bibfield{author}{\bibinfo{person}{Taku Matsumoto}, \bibinfo{person}{Yutaka
  Watanobe}, {and} \bibinfo{person}{Keita Nakamura}.}
  \bibinfo{year}{2021}\natexlab{}.
\newblock \showarticletitle{A model with iterative trials for correcting logic
  errors in source code}.
\newblock \bibinfo{journal}{\emph{Applied Sciences}} \bibinfo{volume}{11},
  \bibinfo{number}{11} (\bibinfo{year}{2021}), \bibinfo{pages}{4755}.
\newblock


\bibitem[Medeiros et~al\mbox{.}(2013)]%
        {nref44}
\bibfield{author}{\bibinfo{person}{Ib{\'e}ria Medeiros},
  \bibinfo{person}{Nuno~F Neves}, {and} \bibinfo{person}{Miguel Correia}.}
  \bibinfo{year}{2013}\natexlab{}.
\newblock \showarticletitle{Securing energy metering software with automatic
  source code correction}. In \bibinfo{booktitle}{\emph{2013 11th IEEE
  International Conference on Industrial Informatics (INDIN)}}. IEEE,
  \bibinfo{pages}{701--706}.
\newblock


\bibitem[Merkel(2014)]%
        {s_ref7}
\bibfield{author}{\bibinfo{person}{Dirk Merkel}.}
  \bibinfo{year}{2014}\natexlab{}.
\newblock \showarticletitle{Docker: Lightweight Linux Containers for Consistent
  Development and Deployment}.
\newblock \bibinfo{journal}{\emph{Linux J.}} \bibinfo{volume}{2014},
  \bibinfo{number}{239}, Article \bibinfo{articleno}{2} (\bibinfo{date}{mar}
  \bibinfo{year}{2014}).
\newblock


\bibitem[Meyer and Saez-Rodriguez(2021)]%
        {nref16}
\bibfield{author}{\bibinfo{person}{Pablo Meyer} {and} \bibinfo{person}{Julio
  Saez-Rodriguez}.} \bibinfo{year}{2021}\natexlab{}.
\newblock \showarticletitle{Advances in systems biology modeling: 10 years of
  crowdsourcing DREAM challenges}.
\newblock \bibinfo{journal}{\emph{Cell Systems}} \bibinfo{volume}{12},
  \bibinfo{number}{6} (\bibinfo{year}{2021}), \bibinfo{pages}{636--653}.
\newblock


\bibitem[Miceli~Barone and Sennrich(2017)]%
        {miceli2017parallel}
\bibfield{author}{\bibinfo{person}{Antonio~Valerio Miceli~Barone} {and}
  \bibinfo{person}{Rico Sennrich}.} \bibinfo{year}{2017}\natexlab{}.
\newblock \showarticletitle{A Parallel Corpus of Python Functions and
  Documentation Strings for Automated Code Documentation and Code Generation}.
  In \bibinfo{booktitle}{\emph{Proceedings of the Eighth International Joint
  Conference on Natural Language Processing (Volume 2: Short Papers)}}.
  \bibinfo{publisher}{Asian Federation of Natural Language Processing},
  \bibinfo{address}{Taipei, Taiwan}, \bibinfo{pages}{314--319}.
\newblock
\urldef\tempurl%
\url{https://aclanthology.org/I17-2053}
\showURL{%
\tempurl}


\bibitem[Mirzayanov(2020)]%
        {nref62}
\bibfield{author}{\bibinfo{person}{Mike Mirzayanov}.}
  \bibinfo{year}{2020}\natexlab{}.
\newblock \bibinfo{title}{Codeforces: Results of 2020.}
\newblock
  \bibinfo{howpublished}{\url{https://codeforces.com/blog/entry/89502}}.
\newblock
\newblock
\shownote{Accessed: 2023-05-23}.


\bibitem[Mostaeen et~al\mbox{.}(2020)]%
        {s_ref94}
\bibfield{author}{\bibinfo{person}{Golam Mostaeen}, \bibinfo{person}{Banani
  Roy}, \bibinfo{person}{Chanchal~K Roy}, \bibinfo{person}{Kevin Schneider},
  {and} \bibinfo{person}{Jeffrey Svajlenko}.} \bibinfo{year}{2020}\natexlab{}.
\newblock \showarticletitle{A machine learning based framework for code clone
  validation}.
\newblock \bibinfo{journal}{\emph{Journal of Systems and Software}}
  \bibinfo{volume}{169} (\bibinfo{year}{2020}), \bibinfo{pages}{110686}.
\newblock


\bibitem[Mostaeen et~al\mbox{.}(2018)]%
        {s_ref96}
\bibfield{author}{\bibinfo{person}{Golam Mostaeen}, \bibinfo{person}{Jeffrey
  Svajlenko}, \bibinfo{person}{Banani Roy}, \bibinfo{person}{Chanchal~K. Roy},
  {and} \bibinfo{person}{Kevin~A. Schneider}.} \bibinfo{year}{2018}\natexlab{}.
\newblock \showarticletitle{On the Use of Machine Learning Techniques Towards
  the Design of Cloud Based Automatic Code Clone Validation Tools}. In
  \bibinfo{booktitle}{\emph{2018 IEEE 18th International Working Conference on
  Source Code Analysis and Manipulation (SCAM)}}. \bibinfo{pages}{155--164}.
\newblock
\urldef\tempurl%
\url{https://doi.org/10.1109/SCAM.2018.00025}
\showDOI{\tempurl}


\bibitem[Mou et~al\mbox{.}(2016)]%
        {s_ref89}
\bibfield{author}{\bibinfo{person}{Lili Mou}, \bibinfo{person}{Ge Li},
  \bibinfo{person}{Lu Zhang}, \bibinfo{person}{Tao Wang}, {and}
  \bibinfo{person}{Zhi Jin}.} \bibinfo{year}{2016}\natexlab{}.
\newblock \showarticletitle{Convolutional Neural Networks over Tree Structures
  for Programming Language Processing}. In
  \bibinfo{booktitle}{\emph{Proceedings of the Thirtieth AAAI Conference on
  Artificial Intelligence}} (Phoenix, Arizona)
  \emph{(\bibinfo{series}{AAAI'16})}. \bibinfo{publisher}{AAAI Press},
  \bibinfo{pages}{1287–1293}.
\newblock


\bibitem[Naohiro(2012)]%
        {s_ref43}
\bibfield{author}{\bibinfo{person}{Takahashi Naohiro}.}
  \bibinfo{year}{2012}\natexlab{}.
\newblock \bibinfo{title}{AtCoder Inc.}
\newblock
\newblock
\newblock
\shownote{Available: \url{https://atcoder.jp/}}.


\bibitem[N{\'E}METH and ZSAK{\'O}(2015)]%
        {s_ref50}
\bibfield{author}{\bibinfo{person}{{\'A}gnes~Erd{\H{o}}sn{\'e} N{\'E}METH}
  {and} \bibinfo{person}{L{\'a}szl{\'o} ZSAK{\'O}}.}
  \bibinfo{year}{2015}\natexlab{}.
\newblock \showarticletitle{Online training and contests for informatics
  contestants of secondary school age}.
\newblock \bibinfo{journal}{\emph{Edukacja-Technika-Informatyka}}
  \bibinfo{volume}{6}, \bibinfo{number}{1} (\bibinfo{year}{2015}),
  \bibinfo{pages}{273--280}.
\newblock


\bibitem[Nguyen et~al\mbox{.}(2015)]%
        {s_ref70}
\bibfield{author}{\bibinfo{person}{Anh~Tuan Nguyen},
  \bibinfo{person}{Tung~Thanh Nguyen}, {and} \bibinfo{person}{Tien~N. Nguyen}.}
  \bibinfo{year}{2015}\natexlab{}.
\newblock \showarticletitle{Divide-and-Conquer Approach for Multi-phase
  Statistical Migration for Source Code (T)}. In \bibinfo{booktitle}{\emph{2015
  30th IEEE/ACM International Conference on Automated Software Engineering
  (ASE)}}. \bibinfo{pages}{585--596}.
\newblock
\urldef\tempurl%
\url{https://doi.org/10.1109/ASE.2015.74}
\showDOI{\tempurl}


\bibitem[Nikolaou(2021)]%
        {nref8}
\bibfield{author}{\bibinfo{person}{Ioannis Nikolaou}.}
  \bibinfo{year}{2021}\natexlab{}.
\newblock \showarticletitle{What is the Role of Technology in Recruitment and
  Selection?}
\newblock \bibinfo{journal}{\emph{The Spanish journal of psychology}}
  \bibinfo{volume}{24} (\bibinfo{year}{2021}), \bibinfo{pages}{e2}.
\newblock


\bibitem[Oksanen(2018)]%
        {nref7}
\bibfield{author}{\bibinfo{person}{Reija Oksanen}.}
  \bibinfo{year}{2018}\natexlab{}.
\newblock \emph{\bibinfo{title}{New technology-based recruitment methods}}.
\newblock \bibinfo{thesistype}{Master's\ thesis}.
\newblock


\bibitem[Omri and Sinz(2020)]%
        {s_ref32}
\bibfield{author}{\bibinfo{person}{Safa Omri} {and} \bibinfo{person}{Carsten
  Sinz}.} \bibinfo{year}{2020}\natexlab{}.
\newblock \showarticletitle{Deep Learning for Software Defect Prediction: A
  Survey}. In \bibinfo{booktitle}{\emph{Proceedings of the IEEE/ACM 42nd
  International Conference on Software Engineering Workshops}} (Seoul, Republic
  of Korea) \emph{(\bibinfo{series}{ICSEW'20})}.
  \bibinfo{publisher}{Association for Computing Machinery},
  \bibinfo{address}{New York, NY, USA}, \bibinfo{pages}{209–214}.
\newblock
\showISBNx{9781450379632}
\urldef\tempurl%
\url{https://doi.org/10.1145/3387940.3391463}
\showDOI{\tempurl}


\bibitem[Ouyang et~al\mbox{.}(2022)]%
        {ouyang2022instructgpt}
\bibfield{author}{\bibinfo{person}{Long Ouyang}, \bibinfo{person}{Jeffrey Wu},
  \bibinfo{person}{Xu Jiang}, \bibinfo{person}{Diogo Almeida},
  \bibinfo{person}{Carroll Wainwright}, \bibinfo{person}{Pamela Mishkin},
  \bibinfo{person}{Chong Zhang}, \bibinfo{person}{Sandhini Agarwal},
  \bibinfo{person}{Katarina Slama}, \bibinfo{person}{Alex Gray},
  \bibinfo{person}{John Schulman}, \bibinfo{person}{Jacob Hilton},
  \bibinfo{person}{Fraser Kelton}, \bibinfo{person}{Luke Miller},
  \bibinfo{person}{Maddie Simens}, \bibinfo{person}{Amanda Askell},
  \bibinfo{person}{Peter Welinder}, \bibinfo{person}{Paul Christiano},
  \bibinfo{person}{Jan Leike}, {and} \bibinfo{person}{Ryan Lowe}.}
  \bibinfo{year}{2022}\natexlab{}.
\newblock \showarticletitle{Training language models to follow instructions
  with human feedback}. In \bibinfo{booktitle}{\emph{Advances in Neural
  Information Processing Systems}}, \bibfield{editor}{\bibinfo{person}{Alice~H.
  Oh}, \bibinfo{person}{Alekh Agarwal}, \bibinfo{person}{Danielle Belgrave},
  {and} \bibinfo{person}{Kyunghyun Cho}} (Eds.).
\newblock
\urldef\tempurl%
\url{https://openreview.net/forum?id=TG8KACxEON}
\showURL{%
\tempurl}


\bibitem[Panichella et~al\mbox{.}(2012)]%
        {nref48}
\bibfield{author}{\bibinfo{person}{Sebastiano Panichella},
  \bibinfo{person}{Jairo Aponte}, \bibinfo{person}{Massimiliano Di~Penta},
  \bibinfo{person}{Andrian Marcus}, {and} \bibinfo{person}{Gerardo Canfora}.}
  \bibinfo{year}{2012}\natexlab{}.
\newblock \showarticletitle{Mining source code descriptions from developer
  communications}. In \bibinfo{booktitle}{\emph{2012 20th IEEE International
  Conference on Program Comprehension (ICPC)}}. IEEE, \bibinfo{pages}{63--72}.
\newblock


\bibitem[Perl et~al\mbox{.}(2015)]%
        {nref42}
\bibfield{author}{\bibinfo{person}{Henning Perl}, \bibinfo{person}{Sergej
  Dechand}, \bibinfo{person}{Matthew Smith}, \bibinfo{person}{Daniel Arp},
  \bibinfo{person}{Fabian Yamaguchi}, \bibinfo{person}{Konrad Rieck},
  \bibinfo{person}{Sascha Fahl}, {and} \bibinfo{person}{Yasemin Acar}.}
  \bibinfo{year}{2015}\natexlab{}.
\newblock \showarticletitle{Vccfinder: Finding potential vulnerabilities in
  open-source projects to assist code audits}. In
  \bibinfo{booktitle}{\emph{Proceedings of the 22nd ACM SIGSAC Conference on
  Computer and Communications Security}}. \bibinfo{pages}{426--437}.
\newblock


\bibitem[Piskachev et~al\mbox{.}(2019)]%
        {nref43}
\bibfield{author}{\bibinfo{person}{Goran Piskachev}, \bibinfo{person}{Lisa
  Nguyen~Quang Do}, {and} \bibinfo{person}{Eric Bodden}.}
  \bibinfo{year}{2019}\natexlab{}.
\newblock \showarticletitle{Codebase-adaptive detection of security-relevant
  methods}. In \bibinfo{booktitle}{\emph{Proceedings of the 28th ACM SIGSOFT
  International Symposium on Software Testing and Analysis}}.
  \bibinfo{pages}{181--191}.
\newblock


\bibitem[Pohl(2006)]%
        {s_ref48}
\bibfield{author}{\bibinfo{person}{Wolfgang Pohl}.}
  \bibinfo{year}{2006}\natexlab{}.
\newblock \showarticletitle{Computer Science Contests for Secondary School
  Students: Approaches to Classification}.
\newblock \bibinfo{journal}{\emph{Informatics in Education}}
  \bibinfo{volume}{5}, \bibinfo{number}{1} (\bibinfo{date}{jan}
  \bibinfo{year}{2006}), \bibinfo{pages}{125–132}.
\newblock
\showISSN{1648-5831}


\bibitem[Ponta et~al\mbox{.}(2019)]%
        {nref27}
\bibfield{author}{\bibinfo{person}{Serena~Elisa Ponta}, \bibinfo{person}{Henrik
  Plate}, \bibinfo{person}{Antonino Sabetta}, \bibinfo{person}{Michele Bezzi},
  {and} \bibinfo{person}{C{\'e}dric Dangremont}.}
  \bibinfo{year}{2019}\natexlab{}.
\newblock \showarticletitle{A manually-curated dataset of fixes to
  vulnerabilities of open-source software}. In \bibinfo{booktitle}{\emph{2019
  IEEE/ACM 16th International Conference on Mining Software Repositories
  (MSR)}}. IEEE, \bibinfo{pages}{383--387}.
\newblock


\bibitem[Pradel and Sen(2018a)]%
        {nref33}
\bibfield{author}{\bibinfo{person}{Michael Pradel} {and}
  \bibinfo{person}{Koushik Sen}.} \bibinfo{year}{2018}\natexlab{a}.
\newblock \showarticletitle{Deepbugs: A learning approach to name-based bug
  detection}.
\newblock \bibinfo{journal}{\emph{Proceedings of the ACM on Programming
  Languages}} \bibinfo{volume}{2}, \bibinfo{number}{OOPSLA}
  (\bibinfo{year}{2018}), \bibinfo{pages}{1--25}.
\newblock


\bibitem[Pradel and Sen(2018b)]%
        {s_ref76}
\bibfield{author}{\bibinfo{person}{Michael Pradel} {and}
  \bibinfo{person}{Koushik Sen}.} \bibinfo{year}{2018}\natexlab{b}.
\newblock \showarticletitle{DeepBugs: A Learning Approach to Name-Based Bug
  Detection}.
\newblock \bibinfo{journal}{\emph{Proc. ACM Program. Lang.}}
  \bibinfo{volume}{2}, \bibinfo{number}{OOPSLA}, Article
  \bibinfo{articleno}{147} (\bibinfo{date}{oct} \bibinfo{year}{2018}),
  \bibinfo{numpages}{25}~pages.
\newblock
\urldef\tempurl%
\url{https://doi.org/10.1145/3276517}
\showDOI{\tempurl}


\bibitem[Premtoon et~al\mbox{.}(2020)]%
        {s_ref63}
\bibfield{author}{\bibinfo{person}{Varot Premtoon}, \bibinfo{person}{James
  Koppel}, {and} \bibinfo{person}{Armando Solar-Lezama}.}
  \bibinfo{year}{2020}\natexlab{}.
\newblock \showarticletitle{Semantic Code Search via Equational Reasoning}. In
  \bibinfo{booktitle}{\emph{Proceedings of the 41st ACM SIGPLAN Conference on
  Programming Language Design and Implementation}} (London, UK)
  \emph{(\bibinfo{series}{PLDI 2020})}. \bibinfo{publisher}{Association for
  Computing Machinery}, \bibinfo{address}{New York, NY, USA},
  \bibinfo{pages}{1066–1082}.
\newblock
\showISBNx{9781450376136}
\urldef\tempurl%
\url{https://doi.org/10.1145/3385412.3386001}
\showDOI{\tempurl}


\bibitem[Pu et~al\mbox{.}(2016)]%
        {nref3}
\bibfield{author}{\bibinfo{person}{Yewen Pu}, \bibinfo{person}{Karthik
  Narasimhan}, \bibinfo{person}{Armando Solar-Lezama}, {and}
  \bibinfo{person}{Regina Barzilay}.} \bibinfo{year}{2016}\natexlab{}.
\newblock \showarticletitle{sk\_p: a neural program corrector for MOOCs}. In
  \bibinfo{booktitle}{\emph{Companion Proceedings of the 2016 ACM SIGPLAN
  International Conference on Systems, Programming, Languages and Applications:
  Software for Humanity}}. \bibinfo{pages}{39--40}.
\newblock


\bibitem[Puri et~al\mbox{.}(2021)]%
        {s_ref41}
\bibfield{author}{\bibinfo{person}{Ruchir Puri}, \bibinfo{person}{David~S
  Kung}, \bibinfo{person}{Geert Janssen}, \bibinfo{person}{Wei Zhang},
  \bibinfo{person}{Giacomo Domeniconi}, \bibinfo{person}{Vladmir Zolotov},
  \bibinfo{person}{Julian Dolby}, \bibinfo{person}{Jie Chen},
  \bibinfo{person}{Mihir Choudhury}, \bibinfo{person}{Lindsey Decker},
  {et~al\mbox{.}}} \bibinfo{year}{2021}\natexlab{}.
\newblock \showarticletitle{Project codenet: A large-scale ai for code dataset
  for learning a diversity of coding tasks}.
\newblock \bibinfo{journal}{\emph{arXiv preprint arXiv:2105.12655}}
  \bibinfo{volume}{1035} (\bibinfo{year}{2021}).
\newblock


\bibitem[Qasem et~al\mbox{.}(2020)]%
        {s_ref80}
\bibfield{author}{\bibinfo{person}{Osama~Al Qasem}, \bibinfo{person}{Mohammed
  Akour}, {and} \bibinfo{person}{Mamdouh Alenezi}.}
  \bibinfo{year}{2020}\natexlab{}.
\newblock \showarticletitle{The Influence of Deep Learning Algorithms Factors
  in Software Fault Prediction}.
\newblock \bibinfo{journal}{\emph{IEEE Access}}  \bibinfo{volume}{8}
  (\bibinfo{year}{2020}), \bibinfo{pages}{63945--63960}.
\newblock
\urldef\tempurl%
\url{https://doi.org/10.1109/ACCESS.2020.2985290}
\showDOI{\tempurl}


\bibitem[Radford et~al\mbox{.}(2018)]%
        {radford2018gpt}
\bibfield{author}{\bibinfo{person}{Alec Radford}, \bibinfo{person}{Karthik
  Narasimhan}, \bibinfo{person}{Tim Salimans}, {and} \bibinfo{person}{Ilya
  Sutskever}.} \bibinfo{year}{2018}\natexlab{}.
\newblock \showarticletitle{Improving language understanding by generative
  pre-training}.
\newblock  (\bibinfo{year}{2018}).
\newblock


\bibitem[Radford et~al\mbox{.}(2019)]%
        {radford2019gpt2}
\bibfield{author}{\bibinfo{person}{Alec Radford}, \bibinfo{person}{Jeff Wu},
  \bibinfo{person}{Rewon Child}, \bibinfo{person}{David Luan},
  \bibinfo{person}{Dario Amodei}, {and} \bibinfo{person}{Ilya Sutskever}.}
  \bibinfo{year}{2019}\natexlab{}.
\newblock \showarticletitle{Language Models are Unsupervised Multitask
  Learners}.
\newblock  (\bibinfo{year}{2019}).
\newblock


\bibitem[Rahman(2022)]%
        {rahman2022data}
\bibfield{author}{\bibinfo{person}{Md~Mostafizer Rahman}.}
  \bibinfo{year}{2022}\natexlab{}.
\newblock \bibinfo{title}{Data Analysis and Code Assessment Using Machine
  Learning Techniques for Programming Activities}.
\newblock
\newblock
\urldef\tempurl%
\url{https://doi.org/10.15016/00000215}
\showDOI{\tempurl}


\bibitem[Rahman et~al\mbox{.}(2021a)]%
        {nref52}
\bibfield{author}{\bibinfo{person}{Md~Mostafizer Rahman},
  \bibinfo{person}{Shunsuke Kawabayashi}, {and} \bibinfo{person}{Yutaka
  Watanobe}.} \bibinfo{year}{2021}\natexlab{a}.
\newblock \showarticletitle{Categorization of frequent errors in solution codes
  created by novice programmers}. In \bibinfo{booktitle}{\emph{SHS Web of
  Conferences}}, Vol.~\bibinfo{volume}{102}. EDP Sciences,
  \bibinfo{pages}{04014}.
\newblock


\bibitem[Rahman et~al\mbox{.}(2021b)]%
        {s_ref8}
\bibfield{author}{\bibinfo{person}{Md.~Mostafizer Rahman},
  \bibinfo{person}{Yutaka Watanobe}, \bibinfo{person}{Rage~Uday Kiran},
  \bibinfo{person}{Truong~Cong Thang}, {and} \bibinfo{person}{Incheon Paik}.}
  \bibinfo{year}{2021}\natexlab{b}.
\newblock \showarticletitle{Impact of Practical Skills on Academic Performance:
  A Data-Driven Analysis}.
\newblock \bibinfo{journal}{\emph{IEEE Access}}  \bibinfo{volume}{9}
  (\bibinfo{year}{2021}), \bibinfo{pages}{139975--139993}.
\newblock
\urldef\tempurl%
\url{https://doi.org/10.1109/ACCESS.2021.3119145}
\showDOI{\tempurl}


\bibitem[Rahman et~al\mbox{.}(2022)]%
        {s_ref9}
\bibfield{author}{\bibinfo{person}{Md.~Mostafizer Rahman},
  \bibinfo{person}{Yutaka Watanobe}, \bibinfo{person}{Taku Matsumoto},
  \bibinfo{person}{Rage~Uday Kiran}, {and} \bibinfo{person}{Keita Nakamura}.}
  \bibinfo{year}{2022}\natexlab{}.
\newblock \showarticletitle{Educational Data Mining to Support Programming
  Learning Using Problem-Solving Data}.
\newblock \bibinfo{journal}{\emph{IEEE Access}}  \bibinfo{volume}{10}
  (\bibinfo{year}{2022}), \bibinfo{pages}{26186--26202}.
\newblock
\urldef\tempurl%
\url{https://doi.org/10.1109/ACCESS.2022.3157288}
\showDOI{\tempurl}


\bibitem[Rahman et~al\mbox{.}(2020)]%
        {s_ref17}
\bibfield{author}{\bibinfo{person}{Md.~Mostafizer Rahman},
  \bibinfo{person}{Yutaka Watanobe}, {and} \bibinfo{person}{Keita Nakamura}.}
  \bibinfo{year}{2020}\natexlab{}.
\newblock \showarticletitle{Source Code Assessment and Classification Based on
  Estimated Error Probability Using Attentive LSTM Language Model and Its
  Application in Programming Education}.
\newblock \bibinfo{journal}{\emph{Applied Sciences}} \bibinfo{volume}{10},
  \bibinfo{number}{8} (\bibinfo{year}{2020}).
\newblock
\showISSN{2076-3417}
\urldef\tempurl%
\url{https://doi.org/10.3390/app10082973}
\showDOI{\tempurl}


\bibitem[Rahman et~al\mbox{.}(2021c)]%
        {s_ref59}
\bibfield{author}{\bibinfo{person}{Md~Mostafizer Rahman},
  \bibinfo{person}{Yutaka Watanobe}, \bibinfo{person}{Uday~Kiran Rage}, {and}
  \bibinfo{person}{Keita Nakamura}.} \bibinfo{year}{2021}\natexlab{c}.
\newblock \showarticletitle{A novel rule-based online judge recommender system
  to promote computer programming education}. In
  \bibinfo{booktitle}{\emph{Advances and Trends in Artificial Intelligence.
  From Theory to Practice: 34th International Conference on Industrial,
  Engineering and Other Applications of Applied Intelligent Systems, IEA/AIE
  2021, Kuala Lumpur, Malaysia, July 26--29, 2021, Proceedings, Part II 34}}.
  Springer, \bibinfo{pages}{15--27}.
\newblock


\bibitem[Ray et~al\mbox{.}(2016)]%
        {s_ref77}
\bibfield{author}{\bibinfo{person}{Baishakhi Ray}, \bibinfo{person}{Vincent
  Hellendoorn}, \bibinfo{person}{Saheel Godhane}, \bibinfo{person}{Zhaopeng
  Tu}, \bibinfo{person}{Alberto Bacchelli}, {and} \bibinfo{person}{Premkumar
  Devanbu}.} \bibinfo{year}{2016}\natexlab{}.
\newblock \showarticletitle{On the "Naturalness" of Buggy Code}. In
  \bibinfo{booktitle}{\emph{Proceedings of the 38th International Conference on
  Software Engineering}} (Austin, Texas) \emph{(\bibinfo{series}{ICSE '16})}.
  \bibinfo{publisher}{Association for Computing Machinery},
  \bibinfo{address}{New York, NY, USA}, \bibinfo{pages}{428–439}.
\newblock
\showISBNx{9781450339001}
\urldef\tempurl%
\url{https://doi.org/10.1145/2884781.2884848}
\showDOI{\tempurl}


\bibitem[Raychev et~al\mbox{.}(2016)]%
        {s_ref72}
\bibfield{author}{\bibinfo{person}{Veselin Raychev}, \bibinfo{person}{Pavol
  Bielik}, {and} \bibinfo{person}{Martin Vechev}.}
  \bibinfo{year}{2016}\natexlab{}.
\newblock \showarticletitle{Probabilistic Model for Code with Decision Trees}.
\newblock \bibinfo{journal}{\emph{SIGPLAN Not.}} \bibinfo{volume}{51},
  \bibinfo{number}{10} (\bibinfo{date}{oct} \bibinfo{year}{2016}),
  \bibinfo{pages}{731–747}.
\newblock
\showISSN{0362-1340}
\urldef\tempurl%
\url{https://doi.org/10.1145/3022671.2984041}
\showDOI{\tempurl}


\bibitem[Raychev et~al\mbox{.}(2014)]%
        {s_ref65}
\bibfield{author}{\bibinfo{person}{Veselin Raychev}, \bibinfo{person}{Martin
  Vechev}, {and} \bibinfo{person}{Eran Yahav}.}
  \bibinfo{year}{2014}\natexlab{}.
\newblock \showarticletitle{Code Completion with Statistical Language Models}.
  In \bibinfo{booktitle}{\emph{Proceedings of the 35th ACM SIGPLAN Conference
  on Programming Language Design and Implementation}} (Edinburgh, United
  Kingdom) \emph{(\bibinfo{series}{PLDI '14})}. \bibinfo{publisher}{Association
  for Computing Machinery}, \bibinfo{address}{New York, NY, USA},
  \bibinfo{pages}{419–428}.
\newblock
\showISBNx{9781450327848}
\urldef\tempurl%
\url{https://doi.org/10.1145/2594291.2594321}
\showDOI{\tempurl}


\bibitem[Revilla et~al\mbox{.}(2008)]%
        {s_ref51}
\bibfield{author}{\bibinfo{person}{Miguel~A Revilla}, \bibinfo{person}{Shahriar
  Manzoor}, {and} \bibinfo{person}{Rujia Liu}.}
  \bibinfo{year}{2008}\natexlab{}.
\newblock \showarticletitle{Competitive learning in informatics: The UVa online
  judge experience}.
\newblock \bibinfo{journal}{\emph{Olympiads in Informatics}}
  \bibinfo{volume}{2}, \bibinfo{number}{10} (\bibinfo{year}{2008}),
  \bibinfo{pages}{131--148}.
\newblock


\bibitem[Rivers and Koedinger(2013)]%
        {nref4}
\bibfield{author}{\bibinfo{person}{Kelly Rivers} {and}
  \bibinfo{person}{Kenneth~R Koedinger}.} \bibinfo{year}{2013}\natexlab{}.
\newblock \showarticletitle{Automatic generation of programming feedback: A
  data-driven approach}. In \bibinfo{booktitle}{\emph{The First Workshop on
  AI-supported Education for Computer Science (AIEDCS 2013)}},
  Vol.~\bibinfo{volume}{50}. \bibinfo{pages}{50--59}.
\newblock


\bibitem[Rodr{\'\i}guez-del Pino et~al\mbox{.}(2012)]%
        {nref12}
\bibfield{author}{\bibinfo{person}{Juan~Carlos Rodr{\'\i}guez-del Pino},
  \bibinfo{person}{Enrique Rubio~Royo}, {and} \bibinfo{person}{Zen{\'o}n
  Hern{\'a}ndez~Figueroa}.} \bibinfo{year}{2012}\natexlab{}.
\newblock \showarticletitle{A Virtual Programming Lab for Moodle with automatic
  assessment and anti-plagiarism features}.
\newblock  (\bibinfo{year}{2012}).
\newblock


\bibitem[Roziere et~al\mbox{.}(2020)]%
        {s_ref69}
\bibfield{author}{\bibinfo{person}{Baptiste Roziere},
  \bibinfo{person}{Marie-Anne Lachaux}, \bibinfo{person}{Lowik Chanussot},
  {and} \bibinfo{person}{Guillaume Lample}.} \bibinfo{year}{2020}\natexlab{}.
\newblock \showarticletitle{Unsupervised Translation of Programming Languages}.
  In \bibinfo{booktitle}{\emph{Proceedings of the 34th International Conference
  on Neural Information Processing Systems}} (Vancouver, BC, Canada)
  \emph{(\bibinfo{series}{NIPS'20})}. \bibinfo{publisher}{Curran Associates
  Inc.}, \bibinfo{address}{Red Hook, NY, USA}, Article
  \bibinfo{articleno}{1730}, \bibinfo{numpages}{11}~pages.
\newblock
\showISBNx{9781713829546}


\bibitem[Russell et~al\mbox{.}(2018)]%
        {nref28}
\bibfield{author}{\bibinfo{person}{Rebecca Russell}, \bibinfo{person}{Louis
  Kim}, \bibinfo{person}{Lei Hamilton}, \bibinfo{person}{Tomo Lazovich},
  \bibinfo{person}{Jacob Harer}, \bibinfo{person}{Onur Ozdemir},
  \bibinfo{person}{Paul Ellingwood}, {and} \bibinfo{person}{Marc McConley}.}
  \bibinfo{year}{2018}\natexlab{}.
\newblock \showarticletitle{Automated Vulnerability Detection in Source Code
  Using Deep Representation Learning}. In \bibinfo{booktitle}{\emph{2018 17th
  IEEE International Conference on Machine Learning and Applications (ICMLA)}}.
  \bibinfo{pages}{757--762}.
\newblock
\urldef\tempurl%
\url{https://doi.org/10.1109/ICMLA.2018.00120}
\showDOI{\tempurl}


\bibitem[Sachdev et~al\mbox{.}(2018)]%
        {s_ref113}
\bibfield{author}{\bibinfo{person}{Saksham Sachdev}, \bibinfo{person}{Hongyu
  Li}, \bibinfo{person}{Sifei Luan}, \bibinfo{person}{Seohyun Kim},
  \bibinfo{person}{Koushik Sen}, {and} \bibinfo{person}{Satish Chandra}.}
  \bibinfo{year}{2018}\natexlab{}.
\newblock \showarticletitle{Retrieval on Source Code: A Neural Code Search}. In
  \bibinfo{booktitle}{\emph{Proceedings of the 2nd ACM SIGPLAN International
  Workshop on Machine Learning and Programming Languages}} (Philadelphia, PA,
  USA) \emph{(\bibinfo{series}{MAPL 2018})}. \bibinfo{publisher}{Association
  for Computing Machinery}, \bibinfo{address}{New York, NY, USA},
  \bibinfo{pages}{31–41}.
\newblock
\showISBNx{9781450358347}
\urldef\tempurl%
\url{https://doi.org/10.1145/3211346.3211353}
\showDOI{\tempurl}


\bibitem[Saez-Rodriguez et~al\mbox{.}(2016)]%
        {nref18}
\bibfield{author}{\bibinfo{person}{Julio Saez-Rodriguez},
  \bibinfo{person}{James~C Costello}, \bibinfo{person}{Stephen~H Friend},
  \bibinfo{person}{Michael~R Kellen}, \bibinfo{person}{Lara Mangravite},
  \bibinfo{person}{Pablo Meyer}, \bibinfo{person}{Thea Norman}, {and}
  \bibinfo{person}{Gustavo Stolovitzky}.} \bibinfo{year}{2016}\natexlab{}.
\newblock \showarticletitle{Crowdsourcing biomedical research: leveraging
  communities as innovation engines}.
\newblock \bibinfo{journal}{\emph{Nature Reviews Genetics}}
  \bibinfo{volume}{17}, \bibinfo{number}{8} (\bibinfo{year}{2016}),
  \bibinfo{pages}{470--486}.
\newblock


\bibitem[Sainath et~al\mbox{.}(2015)]%
        {s_ref22}
\bibfield{author}{\bibinfo{person}{Tara~N. Sainath}, \bibinfo{person}{Brian
  Kingsbury}, \bibinfo{person}{George Saon}, \bibinfo{person}{Hagen Soltau},
  \bibinfo{person}{Abdel rahman Mohamed}, \bibinfo{person}{George Dahl}, {and}
  \bibinfo{person}{Bhuvana Ramabhadran}.} \bibinfo{year}{2015}\natexlab{}.
\newblock \showarticletitle{Deep Convolutional Neural Networks for Large-scale
  Speech Tasks}.
\newblock \bibinfo{journal}{\emph{Neural Networks}}  \bibinfo{volume}{64}
  (\bibinfo{year}{2015}), \bibinfo{pages}{39--48}.
\newblock
\showISSN{0893-6080}
\urldef\tempurl%
\url{https://doi.org/10.1016/j.neunet.2014.08.005}
\showDOI{\tempurl}
\newblock
\shownote{Special Issue on “Deep Learning of Representations”}.


\bibitem[Saito and Watanobe(2020)]%
        {s_ref136}
\bibfield{author}{\bibinfo{person}{Tomohiro Saito} {and}
  \bibinfo{person}{Yutaka Watanobe}.} \bibinfo{year}{2020}\natexlab{}.
\newblock \showarticletitle{Learning Path Recommendation System for Programming
  Education Based on Neural Networks}.
\newblock \bibinfo{journal}{\emph{Int. J. Distance Educ. Technol.}}
  \bibinfo{volume}{18}, \bibinfo{number}{1} (\bibinfo{date}{jan}
  \bibinfo{year}{2020}), \bibinfo{pages}{36–64}.
\newblock
\showISSN{1539-3100}
\urldef\tempurl%
\url{https://doi.org/10.4018/IJDET.2020010103}
\showDOI{\tempurl}


\bibitem[Samara(2017)]%
        {nref63}
\bibfield{author}{\bibinfo{person}{Ghassan Samara}.}
  \bibinfo{year}{2017}\natexlab{}.
\newblock \showarticletitle{A practical approach for detecting logical error in
  object oriented environment}.
\newblock \bibinfo{journal}{\emph{arXiv preprint arXiv:1712.04189}}
  (\bibinfo{year}{2017}).
\newblock


\bibitem[Shabtai et~al\mbox{.}(2009)]%
        {s_ref38}
\bibfield{author}{\bibinfo{person}{Asaf Shabtai}, \bibinfo{person}{Robert
  Moskovitch}, \bibinfo{person}{Yuval Elovici}, {and} \bibinfo{person}{Chanan
  Glezer}.} \bibinfo{year}{2009}\natexlab{}.
\newblock \showarticletitle{Detection of malicious code by applying machine
  learning classifiers on static features: A state-of-the-art survey}.
\newblock \bibinfo{journal}{\emph{information security technical report}}
  \bibinfo{volume}{14}, \bibinfo{number}{1} (\bibinfo{year}{2009}),
  \bibinfo{pages}{16--29}.
\newblock


\bibitem[Sharma et~al\mbox{.}(2021a)]%
        {nref23}
\bibfield{author}{\bibinfo{person}{Tushar Sharma}, \bibinfo{person}{Vasiliki
  Efstathiou}, \bibinfo{person}{Panos Louridas}, {and}
  \bibinfo{person}{Diomidis Spinellis}.} \bibinfo{year}{2021}\natexlab{a}.
\newblock \showarticletitle{Code smell detection by deep direct-learning and
  transfer-learning}.
\newblock \bibinfo{journal}{\emph{Journal of Systems and Software}}
  \bibinfo{volume}{176} (\bibinfo{year}{2021}), \bibinfo{pages}{110936}.
\newblock


\bibitem[Sharma et~al\mbox{.}(2021b)]%
        {nref60}
\bibfield{author}{\bibinfo{person}{Tushar Sharma}, \bibinfo{person}{Maria
  Kechagia}, \bibinfo{person}{Stefanos Georgiou}, \bibinfo{person}{Rohit
  Tiwari}, \bibinfo{person}{Indira Vats}, \bibinfo{person}{Hadi Moazen}, {and}
  \bibinfo{person}{Federica Sarro}.} \bibinfo{year}{2021}\natexlab{b}.
\newblock \showarticletitle{A survey on machine learning techniques for source
  code analysis}.
\newblock \bibinfo{journal}{\emph{arXiv preprint arXiv:2110.09610}}
  (\bibinfo{year}{2021}).
\newblock


\bibitem[Sharma and Kessentini(2021)]%
        {nref24}
\bibfield{author}{\bibinfo{person}{Tushar Sharma} {and}
  \bibinfo{person}{Marouane Kessentini}.} \bibinfo{year}{2021}\natexlab{}.
\newblock \showarticletitle{Qscored: A large dataset of code smells and quality
  metrics}. In \bibinfo{booktitle}{\emph{2021 IEEE/ACM 18th International
  Conference on Mining Software Repositories (MSR)}}. IEEE,
  \bibinfo{pages}{590--594}.
\newblock


\bibitem[Shen and Chen(2020)]%
        {s_ref39}
\bibfield{author}{\bibinfo{person}{Zhidong Shen} {and} \bibinfo{person}{Si
  Chen}.} \bibinfo{year}{2020}\natexlab{}.
\newblock \showarticletitle{A survey of automatic software vulnerability
  detection, program repair, and defect prediction techniques}.
\newblock \bibinfo{journal}{\emph{Security and Communication Networks}}
  \bibinfo{volume}{2020} (\bibinfo{year}{2020}), \bibinfo{pages}{1--16}.
\newblock


\bibitem[Shirafuji et~al\mbox{.}(2022)]%
        {nref54}
\bibfield{author}{\bibinfo{person}{Atsushi Shirafuji}, \bibinfo{person}{Takumi
  Ito}, \bibinfo{person}{Makoto Morishita}, \bibinfo{person}{Yuki Nakamura},
  \bibinfo{person}{Yusuke Oda}, \bibinfo{person}{Jun Suzuki}, {and}
  \bibinfo{person}{Yutaka Watanobe}.} \bibinfo{year}{2022}\natexlab{}.
\newblock \showarticletitle{Prompt Sensitivity of Language Model for Solving
  Programming Problems}.
\newblock  (\bibinfo{year}{2022}), \bibinfo{pages}{346--359}.
\newblock
\urldef\tempurl%
\url{https://doi.org/10.3233/FAIA220264}
\showDOI{\tempurl}


\bibitem[Shuai et~al\mbox{.}(2020)]%
        {s_ref117}
\bibfield{author}{\bibinfo{person}{Jianhang Shuai}, \bibinfo{person}{Ling Xu},
  \bibinfo{person}{Chao Liu}, \bibinfo{person}{Meng Yan}, \bibinfo{person}{Xin
  Xia}, {and} \bibinfo{person}{Yan Lei}.} \bibinfo{year}{2020}\natexlab{}.
\newblock \showarticletitle{Improving Code Search with Co-Attentive
  Representation Learning}. In \bibinfo{booktitle}{\emph{Proceedings of the
  28th International Conference on Program Comprehension}} (Seoul, Republic of
  Korea) \emph{(\bibinfo{series}{ICPC '20})}. \bibinfo{publisher}{Association
  for Computing Machinery}, \bibinfo{address}{New York, NY, USA},
  \bibinfo{pages}{196–207}.
\newblock
\showISBNx{9781450379588}
\urldef\tempurl%
\url{https://doi.org/10.1145/3387904.3389269}
\showDOI{\tempurl}


\bibitem[Singh et~al\mbox{.}(2013)]%
        {nref5}
\bibfield{author}{\bibinfo{person}{Rishabh Singh}, \bibinfo{person}{Sumit
  Gulwani}, {and} \bibinfo{person}{Armando Solar-Lezama}.}
  \bibinfo{year}{2013}\natexlab{}.
\newblock \showarticletitle{Automated feedback generation for introductory
  programming assignments}. In \bibinfo{booktitle}{\emph{Proceedings of the
  34th ACM SIGPLAN conference on Programming language design and
  implementation}}. \bibinfo{pages}{15--26}.
\newblock


\bibitem[Skiena and Revilla(2003)]%
        {s_ref52}
\bibfield{author}{\bibinfo{person}{Steven~S Skiena} {and}
  \bibinfo{person}{Miguel~A Revilla}.} \bibinfo{year}{2003}\natexlab{}.
\newblock \showarticletitle{Programming challenges: The programming contest
  training manual}.
\newblock \bibinfo{journal}{\emph{Acm SIGACT News}} \bibinfo{volume}{34},
  \bibinfo{number}{3} (\bibinfo{year}{2003}), \bibinfo{pages}{68--74}.
\newblock


\bibitem[Song et~al\mbox{.}(2019)]%
        {nref64}
\bibfield{author}{\bibinfo{person}{Dowon Song}, \bibinfo{person}{Myungho Lee},
  {and} \bibinfo{person}{Hakjoo Oh}.} \bibinfo{year}{2019}\natexlab{}.
\newblock \showarticletitle{Automatic and scalable detection of logical errors
  in functional programming assignments}.
\newblock \bibinfo{journal}{\emph{Proceedings of the ACM on Programming
  Languages}} \bibinfo{volume}{3}, \bibinfo{number}{OOPSLA}
  (\bibinfo{year}{2019}), \bibinfo{pages}{1--30}.
\newblock


\bibitem[Spacco et~al\mbox{.}(2015)]%
        {nref9}
\bibfield{author}{\bibinfo{person}{Jaime Spacco}, \bibinfo{person}{Paul Denny},
  \bibinfo{person}{Brad Richards}, \bibinfo{person}{David Babcock},
  \bibinfo{person}{David Hovemeyer}, \bibinfo{person}{James Moscola}, {and}
  \bibinfo{person}{Robert Duvall}.} \bibinfo{year}{2015}\natexlab{}.
\newblock \showarticletitle{Analyzing student work patterns using programming
  exercise data}. In \bibinfo{booktitle}{\emph{Proceedings of the 46th ACM
  Technical Symposium on Computer Science Education}}. \bibinfo{pages}{18--23}.
\newblock


\bibitem[Sui and Xue(2016)]%
        {nref39}
\bibfield{author}{\bibinfo{person}{Yulei Sui} {and} \bibinfo{person}{Jingling
  Xue}.} \bibinfo{year}{2016}\natexlab{}.
\newblock \showarticletitle{SVF: interprocedural static value-flow analysis in
  LLVM}. In \bibinfo{booktitle}{\emph{Proceedings of the 25th international
  conference on compiler construction}}. \bibinfo{pages}{265--266}.
\newblock


\bibitem[Svajlenko et~al\mbox{.}(2014a)]%
        {nref25}
\bibfield{author}{\bibinfo{person}{Jeffrey Svajlenko},
  \bibinfo{person}{Judith~F Islam}, \bibinfo{person}{Iman Keivanloo},
  \bibinfo{person}{Chanchal~K Roy}, {and} \bibinfo{person}{Mohammad~Mamun
  Mia}.} \bibinfo{year}{2014}\natexlab{a}.
\newblock \showarticletitle{Towards a big data curated benchmark of
  inter-project code clones}. In \bibinfo{booktitle}{\emph{2014 IEEE
  International Conference on Software Maintenance and Evolution}}. IEEE,
  \bibinfo{pages}{476--480}.
\newblock


\bibitem[Svajlenko et~al\mbox{.}(2014b)]%
        {s_ref71}
\bibfield{author}{\bibinfo{person}{Jeffrey Svajlenko},
  \bibinfo{person}{Judith~F. Islam}, \bibinfo{person}{Iman Keivanloo},
  \bibinfo{person}{Chanchal~K. Roy}, {and} \bibinfo{person}{Mohammad~Mamun
  Mia}.} \bibinfo{year}{2014}\natexlab{b}.
\newblock \showarticletitle{Towards a Big Data Curated Benchmark of
  Inter-project Code Clones}. In \bibinfo{booktitle}{\emph{2014 IEEE
  International Conference on Software Maintenance and Evolution}}.
  \bibinfo{pages}{476--480}.
\newblock
\urldef\tempurl%
\url{https://doi.org/10.1109/ICSME.2014.77}
\showDOI{\tempurl}


\bibitem[Svyatkovskiy et~al\mbox{.}(2020)]%
        {s_ref66}
\bibfield{author}{\bibinfo{person}{Alexey Svyatkovskiy},
  \bibinfo{person}{Shao~Kun Deng}, \bibinfo{person}{Shengyu Fu}, {and}
  \bibinfo{person}{Neel Sundaresan}.} \bibinfo{year}{2020}\natexlab{}.
\newblock \showarticletitle{IntelliCode Compose: Code Generation Using
  Transformer}. In \bibinfo{booktitle}{\emph{Proceedings of the 28th ACM Joint
  Meeting on European Software Engineering Conference and Symposium on the
  Foundations of Software Engineering}} (Virtual Event, USA)
  \emph{(\bibinfo{series}{ESEC/FSE 2020})}. \bibinfo{publisher}{Association for
  Computing Machinery}, \bibinfo{address}{New York, NY, USA},
  \bibinfo{pages}{1433–1443}.
\newblock
\showISBNx{9781450370431}
\urldef\tempurl%
\url{https://doi.org/10.1145/3368089.3417058}
\showDOI{\tempurl}


\bibitem[Svyatkovskiy et~al\mbox{.}(2019)]%
        {s_ref67}
\bibfield{author}{\bibinfo{person}{Alexey Svyatkovskiy}, \bibinfo{person}{Ying
  Zhao}, \bibinfo{person}{Shengyu Fu}, {and} \bibinfo{person}{Neel
  Sundaresan}.} \bibinfo{year}{2019}\natexlab{}.
\newblock \showarticletitle{Pythia: AI-Assisted Code Completion System}. In
  \bibinfo{booktitle}{\emph{Proceedings of the 25th ACM SIGKDD International
  Conference on Knowledge Discovery \& Data Mining}} (Anchorage, AK, USA)
  \emph{(\bibinfo{series}{KDD '19})}. \bibinfo{publisher}{Association for
  Computing Machinery}, \bibinfo{address}{New York, NY, USA},
  \bibinfo{pages}{2727–2735}.
\newblock
\showISBNx{9781450362016}
\urldef\tempurl%
\url{https://doi.org/10.1145/3292500.3330699}
\showDOI{\tempurl}


\bibitem[Szegedy et~al\mbox{.}(2015)]%
        {s_ref21}
\bibfield{author}{\bibinfo{person}{Christian Szegedy}, \bibinfo{person}{Wei
  Liu}, \bibinfo{person}{Yangqing Jia}, \bibinfo{person}{Pierre Sermanet},
  \bibinfo{person}{Scott Reed}, \bibinfo{person}{Dragomir Anguelov},
  \bibinfo{person}{Dumitru Erhan}, \bibinfo{person}{Vincent Vanhoucke}, {and}
  \bibinfo{person}{Andrew Rabinovich}.} \bibinfo{year}{2015}\natexlab{}.
\newblock \showarticletitle{Going deeper with convolutions}. In
  \bibinfo{booktitle}{\emph{2015 IEEE Conference on Computer Vision and Pattern
  Recognition (CVPR)}}. \bibinfo{pages}{1--9}.
\newblock
\urldef\tempurl%
\url{https://doi.org/10.1109/CVPR.2015.7298594}
\showDOI{\tempurl}


\bibitem[Terada and Watanobe(2019)]%
        {nref56}
\bibfield{author}{\bibinfo{person}{Kenta Terada} {and} \bibinfo{person}{Yutaka
  Watanobe}.} \bibinfo{year}{2019}\natexlab{}.
\newblock \showarticletitle{Automatic generation of fill-in-the-blank
  programming problems}. In \bibinfo{booktitle}{\emph{2019 IEEE 13th
  International Symposium on Embedded Multicore/Many-core Systems-on-Chip
  (MCSoC)}}. IEEE, \bibinfo{pages}{187--193}.
\newblock


\bibitem[Terada and Watanobe(2021)]%
        {nref55}
\bibfield{author}{\bibinfo{person}{Kenta Terada} {and} \bibinfo{person}{Yutaka
  Watanobe}.} \bibinfo{year}{2021}\natexlab{}.
\newblock \showarticletitle{Code completion for programming education based on
  deep learning}.
\newblock \bibinfo{journal}{\emph{International Journal of Computational
  Intelligence Studies}} \bibinfo{volume}{10}, \bibinfo{number}{2-3}
  (\bibinfo{year}{2021}), \bibinfo{pages}{78--98}.
\newblock


\bibitem[Tufano et~al\mbox{.}(2019)]%
        {s_ref74}
\bibfield{author}{\bibinfo{person}{Michele Tufano}, \bibinfo{person}{Cody
  Watson}, \bibinfo{person}{Gabriele Bavota}, \bibinfo{person}{Massimiliano~Di
  Penta}, \bibinfo{person}{Martin White}, {and} \bibinfo{person}{Denys
  Poshyvanyk}.} \bibinfo{year}{2019}\natexlab{}.
\newblock \showarticletitle{An Empirical Study on Learning Bug-Fixing Patches
  in the Wild via Neural Machine Translation}.
\newblock \bibinfo{journal}{\emph{ACM Trans. Softw. Eng. Methodol.}}
  \bibinfo{volume}{28}, \bibinfo{number}{4}, Article \bibinfo{articleno}{19}
  (\bibinfo{date}{sep} \bibinfo{year}{2019}), \bibinfo{numpages}{29}~pages.
\newblock
\showISSN{1049-331X}
\urldef\tempurl%
\url{https://doi.org/10.1145/3340544}
\showDOI{\tempurl}


\bibitem[Ucci et~al\mbox{.}(2019)]%
        {s_ref40}
\bibfield{author}{\bibinfo{person}{Daniele Ucci}, \bibinfo{person}{Leonardo
  Aniello}, {and} \bibinfo{person}{Roberto Baldoni}.}
  \bibinfo{year}{2019}\natexlab{}.
\newblock \showarticletitle{Survey of machine learning techniques for malware
  analysis}.
\newblock \bibinfo{journal}{\emph{Computers \& Security}}  \bibinfo{volume}{81}
  (\bibinfo{year}{2019}), \bibinfo{pages}{123--147}.
\newblock


\bibitem[Ugurel et~al\mbox{.}(2002)]%
        {s_ref133}
\bibfield{author}{\bibinfo{person}{Secil Ugurel}, \bibinfo{person}{Robert
  Krovetz}, {and} \bibinfo{person}{C.~Lee Giles}.}
  \bibinfo{year}{2002}\natexlab{}.
\newblock \showarticletitle{What's the Code? Automatic Classification of Source
  Code Archives}. In \bibinfo{booktitle}{\emph{Proceedings of the Eighth ACM
  SIGKDD International Conference on Knowledge Discovery and Data Mining}}
  (Edmonton, Alberta, Canada) \emph{(\bibinfo{series}{KDD '02})}.
  \bibinfo{publisher}{Association for Computing Machinery},
  \bibinfo{address}{New York, NY, USA}, \bibinfo{pages}{632–638}.
\newblock
\showISBNx{158113567X}
\urldef\tempurl%
\url{https://doi.org/10.1145/775047.775141}
\showDOI{\tempurl}


\bibitem[Ullah et~al\mbox{.}(2019a)]%
        {s_ref45}
\bibfield{author}{\bibinfo{person}{Farhan Ullah}, \bibinfo{person}{Hamad
  Naeem}, \bibinfo{person}{Sohail Jabbar}, \bibinfo{person}{Shehzad Khalid},
  \bibinfo{person}{Muhammad~Ahsan Latif}, \bibinfo{person}{Fadi Al-turjman},
  {and} \bibinfo{person}{Leonardo Mostarda}.} \bibinfo{year}{2019}\natexlab{a}.
\newblock \showarticletitle{Cyber Security Threats Detection in Internet of
  Things Using Deep Learning Approach}.
\newblock \bibinfo{journal}{\emph{IEEE Access}}  \bibinfo{volume}{7}
  (\bibinfo{year}{2019}), \bibinfo{pages}{124379--124389}.
\newblock
\urldef\tempurl%
\url{https://doi.org/10.1109/ACCESS.2019.2937347}
\showDOI{\tempurl}


\bibitem[Ullah et~al\mbox{.}(2019b)]%
        {ullah2019gcj}
\bibfield{author}{\bibinfo{person}{Farhan Ullah}, \bibinfo{person}{Hamad
  Naeem}, \bibinfo{person}{Sohail Jabbar}, \bibinfo{person}{Shehzad Khalid},
  \bibinfo{person}{Muhammad~Ahsan Latif}, \bibinfo{person}{Fadi Al-turjman},
  {and} \bibinfo{person}{Leonardo Mostarda}.} \bibinfo{year}{2019}\natexlab{b}.
\newblock \showarticletitle{Cyber Security Threats Detection in Internet of
  Things Using Deep Learning Approach}.
\newblock \bibinfo{journal}{\emph{IEEE Access}}  \bibinfo{volume}{7}
  (\bibinfo{year}{2019}), \bibinfo{pages}{124379--124389}.
\newblock
\urldef\tempurl%
\url{https://doi.org/10.1109/ACCESS.2019.2937347}
\showDOI{\tempurl}


\bibitem[Van~Rijn et~al\mbox{.}(2013)]%
        {nref15}
\bibfield{author}{\bibinfo{person}{Jan~N Van~Rijn}, \bibinfo{person}{Bernd
  Bischl}, \bibinfo{person}{Luis Torgo}, \bibinfo{person}{Bo Gao},
  \bibinfo{person}{Venkatesh Umaashankar}, \bibinfo{person}{Simon Fischer},
  \bibinfo{person}{Patrick Winter}, \bibinfo{person}{Bernd Wiswedel},
  \bibinfo{person}{Michael~R Berthold}, {and} \bibinfo{person}{Joaquin
  Vanschoren}.} \bibinfo{year}{2013}\natexlab{}.
\newblock \showarticletitle{OpenML: A collaborative science platform}. In
  \bibinfo{booktitle}{\emph{Machine Learning and Knowledge Discovery in
  Databases: European Conference, ECML PKDD 2013, Prague, Czech Republic,
  September 23-27, 2013, Proceedings, Part III 13}}. Springer,
  \bibinfo{pages}{645--649}.
\newblock


\bibitem[Vasic et~al\mbox{.}(2019)]%
        {s_ref110}
\bibfield{author}{\bibinfo{person}{Marko Vasic}, \bibinfo{person}{Aditya
  Kanade}, \bibinfo{person}{Petros Maniatis}, \bibinfo{person}{David Bieber},
  {and} \bibinfo{person}{Rishabh Singh}.} \bibinfo{year}{2019}\natexlab{}.
\newblock \showarticletitle{Neural program repair by jointly learning to
  localize and repair}.
\newblock \bibinfo{journal}{\emph{arXiv preprint arXiv:1904.01720}}
  (\bibinfo{year}{2019}).
\newblock


\bibitem[Wan et~al\mbox{.}(2020)]%
        {s_ref115}
\bibfield{author}{\bibinfo{person}{Yao Wan}, \bibinfo{person}{Jingdong Shu},
  \bibinfo{person}{Yulei Sui}, \bibinfo{person}{Guandong Xu},
  \bibinfo{person}{Zhou Zhao}, \bibinfo{person}{Jian Wu}, {and}
  \bibinfo{person}{Philip~S. Yu}.} \bibinfo{year}{2020}\natexlab{}.
\newblock \showarticletitle{Multi-Modal Attention Network Learning for Semantic
  Source Code Retrieval}. In \bibinfo{booktitle}{\emph{Proceedings of the 34th
  IEEE/ACM International Conference on Automated Software Engineering}} (San
  Diego, California) \emph{(\bibinfo{series}{ASE '19})}.
  \bibinfo{publisher}{IEEE Press}, \bibinfo{pages}{13–25}.
\newblock
\showISBNx{9781728125084}
\urldef\tempurl%
\url{https://doi.org/10.1109/ASE.2019.00012}
\showDOI{\tempurl}


\bibitem[Wan et~al\mbox{.}(2018)]%
        {s_ref123}
\bibfield{author}{\bibinfo{person}{Yao Wan}, \bibinfo{person}{Zhou Zhao},
  \bibinfo{person}{Min Yang}, \bibinfo{person}{Guandong Xu},
  \bibinfo{person}{Haochao Ying}, \bibinfo{person}{Jian Wu}, {and}
  \bibinfo{person}{Philip~S. Yu}.} \bibinfo{year}{2018}\natexlab{}.
\newblock \showarticletitle{Improving Automatic Source Code Summarization via
  Deep Reinforcement Learning}. In \bibinfo{booktitle}{\emph{Proceedings of the
  33rd ACM/IEEE International Conference on Automated Software Engineering}}
  (Montpellier, France) \emph{(\bibinfo{series}{ASE '18})}.
  \bibinfo{publisher}{Association for Computing Machinery},
  \bibinfo{address}{New York, NY, USA}, \bibinfo{pages}{397–407}.
\newblock
\showISBNx{9781450359375}
\urldef\tempurl%
\url{https://doi.org/10.1145/3238147.3238206}
\showDOI{\tempurl}


\bibitem[Wan et~al\mbox{.}(2021)]%
        {s_ref28}
\bibfield{author}{\bibinfo{person}{Zhiyuan Wan}, \bibinfo{person}{Xin Xia},
  \bibinfo{person}{David Lo}, {and} \bibinfo{person}{Gail~C. Murphy}.}
  \bibinfo{year}{2021}\natexlab{}.
\newblock \showarticletitle{How does Machine Learning Change Software
  Development Practices?}
\newblock \bibinfo{journal}{\emph{IEEE Transactions on Software Engineering}}
  \bibinfo{volume}{47}, \bibinfo{number}{9} (\bibinfo{year}{2021}),
  \bibinfo{pages}{1857--1871}.
\newblock
\urldef\tempurl%
\url{https://doi.org/10.1109/TSE.2019.2937083}
\showDOI{\tempurl}


\bibitem[Wang et~al\mbox{.}(2016a)]%
        {s_ref78}
\bibfield{author}{\bibinfo{person}{Song Wang}, \bibinfo{person}{Devin Chollak},
  \bibinfo{person}{Dana Movshovitz-Attias}, {and} \bibinfo{person}{Lin Tan}.}
  \bibinfo{year}{2016}\natexlab{a}.
\newblock \showarticletitle{Bugram: Bug detection with n-gram language models}.
  In \bibinfo{booktitle}{\emph{2016 31st IEEE/ACM International Conference on
  Automated Software Engineering (ASE)}}. \bibinfo{pages}{708--719}.
\newblock


\bibitem[Wang et~al\mbox{.}(2016b)]%
        {s_ref79}
\bibfield{author}{\bibinfo{person}{Song Wang}, \bibinfo{person}{Taiyue Liu},
  {and} \bibinfo{person}{Lin Tan}.} \bibinfo{year}{2016}\natexlab{b}.
\newblock \showarticletitle{Automatically Learning Semantic Features for Defect
  Prediction}. In \bibinfo{booktitle}{\emph{Proceedings of the 38th
  International Conference on Software Engineering}} (Austin, Texas)
  \emph{(\bibinfo{series}{ICSE '16})}. \bibinfo{publisher}{Association for
  Computing Machinery}, \bibinfo{address}{New York, NY, USA},
  \bibinfo{pages}{297–308}.
\newblock
\showISBNx{9781450339001}
\urldef\tempurl%
\url{https://doi.org/10.1145/2884781.2884804}
\showDOI{\tempurl}


\bibitem[Wang et~al\mbox{.}(2020a)]%
        {s_ref93}
\bibfield{author}{\bibinfo{person}{Wenhan Wang}, \bibinfo{person}{Ge Li},
  \bibinfo{person}{Bo Ma}, \bibinfo{person}{Xin Xia}, {and}
  \bibinfo{person}{Zhi Jin}.} \bibinfo{year}{2020}\natexlab{a}.
\newblock \showarticletitle{Detecting code clones with graph neural network and
  flow-augmented abstract syntax tree}. In \bibinfo{booktitle}{\emph{2020 IEEE
  27th International Conference on Software Analysis, Evolution and
  Reengineering (SANER)}}. IEEE, \bibinfo{pages}{261--271}.
\newblock


\bibitem[Wang et~al\mbox{.}(2020b)]%
        {s_ref112}
\bibfield{author}{\bibinfo{person}{Wenhua Wang}, \bibinfo{person}{Yuqun Zhang},
  \bibinfo{person}{Zhengran Zeng}, {and} \bibinfo{person}{Guandong Xu}.}
  \bibinfo{year}{2020}\natexlab{b}.
\newblock \showarticletitle{TranSˆ3: A Transformer-based Framework for
  Unifying Code Summarization and Code Search. CoRR abs/2003.03238 (2020)}.
\newblock \bibinfo{journal}{\emph{arXiv preprint arXiv:2003.03238}}
  (\bibinfo{year}{2020}).
\newblock


\bibitem[Wang et~al\mbox{.}(2021)]%
        {wang2021syncobert}
\bibfield{author}{\bibinfo{person}{Xin Wang}, \bibinfo{person}{Yasheng Wang},
  \bibinfo{person}{Fei Mi}, \bibinfo{person}{Pingyi Zhou}, \bibinfo{person}{Yao
  Wan}, \bibinfo{person}{Xiao Liu}, \bibinfo{person}{Li Li},
  \bibinfo{person}{Hao Wu}, \bibinfo{person}{Jin Liu}, {and}
  \bibinfo{person}{Xin Jiang}.} \bibinfo{year}{2021}\natexlab{}.
\newblock \showarticletitle{{SynCoBERT}: Syntax-Guided Multi-Modal Contrastive
  Pre-Training for Code Representation}.
\newblock \bibinfo{journal}{\emph{arXiv preprint}}
  \bibinfo{volume}{arXiv:2108.04556} (\bibinfo{year}{2021}).
\newblock
\urldef\tempurl%
\url{https://doi.org/10.48550/arXiv.2108.04556}
\showDOI{\tempurl}


\bibitem[Wang et~al\mbox{.}(2023)]%
        {wang2023mconala}
\bibfield{author}{\bibinfo{person}{Zhiruo Wang}, \bibinfo{person}{Grace
  Cuenca}, \bibinfo{person}{Shuyan Zhou}, \bibinfo{person}{Frank~F. Xu}, {and}
  \bibinfo{person}{Graham Neubig}.} \bibinfo{year}{2023}\natexlab{}.
\newblock \bibinfo{title}{MCoNaLa: A Benchmark for Code Generation from
  Multiple Natural Languages}.
\newblock
\newblock
\showeprint[arxiv]{2203.08388}~[cs.CL]


\bibitem[Wasik et~al\mbox{.}(2018)]%
        {s_ref1}
\bibfield{author}{\bibinfo{person}{Szymon Wasik}, \bibinfo{person}{Maciej
  Antczak}, \bibinfo{person}{Jan Badura}, \bibinfo{person}{Artur Laskowski},
  {and} \bibinfo{person}{Tomasz Sternal}.} \bibinfo{year}{2018}\natexlab{}.
\newblock \showarticletitle{A Survey on Online Judge Systems and Their
  Applications}.
\newblock \bibinfo{journal}{\emph{ACM Comput. Surv.}} \bibinfo{volume}{51},
  \bibinfo{number}{1}, Article \bibinfo{articleno}{3} (\bibinfo{date}{jan}
  \bibinfo{year}{2018}), \bibinfo{numpages}{34}~pages.
\newblock
\showISSN{0360-0300}
\urldef\tempurl%
\url{https://doi.org/10.1145/3143560}
\showDOI{\tempurl}


\bibitem[Watanobe(2018)]%
        {s_ref42}
\bibfield{author}{\bibinfo{person}{Yutaka Watanobe}.}
  \bibinfo{year}{2018}\natexlab{}.
\newblock \bibinfo{title}{Aizu Online Judge}.
\newblock
\newblock
\newblock
\shownote{Available: \url{https://onlinejudge.u-aizu.ac.jp/ }}.


\bibitem[Watanobe et~al\mbox{.}(2022a)]%
        {s_ref131}
\bibfield{author}{\bibinfo{person}{Y. Watanobe}, \bibinfo{person}{Md~Mostafizer
  Rahman}, \bibinfo{person}{R. Kabir}, {and} \bibinfo{person}{Md~Faizul~Ibne
  Amin}.} \bibinfo{year}{2022}\natexlab{a}.
\newblock \showarticletitle{Identifying Algorithm in Program Code Based on
  Structural Features Using CNN Classification Model}.
\newblock \bibinfo{journal}{\emph{Applied Intelligence}}
  (\bibinfo{year}{2022}).
\newblock


\bibitem[Watanobe et~al\mbox{.}(2022b)]%
        {s_ref135}
\bibfield{author}{\bibinfo{person}{Yutaka Watanobe},
  \bibinfo{person}{Md.~Mostafizer Rahman}, \bibinfo{person}{Taku Matsumoto},
  \bibinfo{person}{Uday~Kiran Rage}, {and} \bibinfo{person}{Penugonda
  Ravikumar}.} \bibinfo{year}{2022}\natexlab{b}.
\newblock \showarticletitle{Online Judge System: Requirements, Architecture,
  and Experiences}.
\newblock \bibinfo{journal}{\emph{International Journal of Software Engineering
  and Knowledge Engineering}} \bibinfo{volume}{32}, \bibinfo{number}{06}
  (\bibinfo{year}{2022}), \bibinfo{pages}{917--946}.
\newblock
\urldef\tempurl%
\url{https://doi.org/10.1142/S0218194022500346}
\showDOI{\tempurl}


\bibitem[Wei et~al\mbox{.}(2019)]%
        {s_ref124}
\bibfield{author}{\bibinfo{person}{Bolin Wei}, \bibinfo{person}{Ge Li},
  \bibinfo{person}{Xin Xia}, \bibinfo{person}{Zhiyi Fu}, {and}
  \bibinfo{person}{Zhi Jin}.} \bibinfo{year}{2019}\natexlab{}.
\newblock \bibinfo{booktitle}{\emph{Code Generation as a Dual Task of Code
  Summarization}}.
\newblock \bibinfo{publisher}{Curran Associates Inc.}, \bibinfo{address}{Red
  Hook, NY, USA}.
\newblock


\bibitem[Wen-xin and Wei(2005)]%
        {wen2005peking}
\bibfield{author}{\bibinfo{person}{Li Wen-xin} {and} \bibinfo{person}{Guo
  Wei}.} \bibinfo{year}{2005}\natexlab{}.
\newblock \showarticletitle{Peking university oneline judge and its
  applications [j]}.
\newblock \bibinfo{journal}{\emph{Journal of Changchun Post and
  Telecommunication Institute S}}  \bibinfo{volume}{2} (\bibinfo{year}{2005}),
  \bibinfo{pages}{23}.
\newblock


\bibitem[White et~al\mbox{.}(2016a)]%
        {s_ref95}
\bibfield{author}{\bibinfo{person}{Martin White}, \bibinfo{person}{Michele
  Tufano}, \bibinfo{person}{Christopher Vendome}, {and} \bibinfo{person}{Denys
  Poshyvanyk}.} \bibinfo{year}{2016}\natexlab{a}.
\newblock \showarticletitle{Deep learning code fragments for code clone
  detection}. In \bibinfo{booktitle}{\emph{2016 31st IEEE/ACM International
  Conference on Automated Software Engineering (ASE)}}.
  \bibinfo{pages}{87--98}.
\newblock


\bibitem[White et~al\mbox{.}(2016b)]%
        {nref68}
\bibfield{author}{\bibinfo{person}{Martin White}, \bibinfo{person}{Michele
  Tufano}, \bibinfo{person}{Christopher Vendome}, {and} \bibinfo{person}{Denys
  Poshyvanyk}.} \bibinfo{year}{2016}\natexlab{b}.
\newblock \showarticletitle{Deep Learning Code Fragments for Code Clone
  Detection} \emph{(\bibinfo{series}{ASE '16})}.
  \bibinfo{publisher}{Association for Computing Machinery},
  \bibinfo{address}{New York, NY, USA}, \bibinfo{pages}{87–98}.
\newblock
\showISBNx{9781450338455}
\urldef\tempurl%
\url{https://doi.org/10.1145/2970276.2970326}
\showDOI{\tempurl}


\bibitem[Xavier et~al\mbox{.}(2011)]%
        {nref11}
\bibfield{author}{\bibinfo{person}{Jo{\~a}o Cristov{\~a}o Afonso~Sampaio
  Xavier} {et~al\mbox{.}}} \bibinfo{year}{2011}\natexlab{}.
\newblock \showarticletitle{Computer-based assessment system for e-learning
  applied to programming education}.
\newblock  (\bibinfo{year}{2011}).
\newblock


\bibitem[Xiaomeng et~al\mbox{.}(2018)]%
        {nref61}
\bibfield{author}{\bibinfo{person}{Wang Xiaomeng}, \bibinfo{person}{Zhang Tao},
  \bibinfo{person}{Xin Wei}, {and} \bibinfo{person}{Hou Changyu}.}
  \bibinfo{year}{2018}\natexlab{}.
\newblock \showarticletitle{A Survey on Source Code Review Using Machine
  Learning}. In \bibinfo{booktitle}{\emph{2018 3rd International Conference on
  Information Systems Engineering (ICISE)}}. \bibinfo{pages}{56--60}.
\newblock
\urldef\tempurl%
\url{https://doi.org/10.1109/ICISE.2018.00018}
\showDOI{\tempurl}


\bibitem[Yahav(2018)]%
        {s_ref34}
\bibfield{author}{\bibinfo{person}{Eran Yahav}.}
  \bibinfo{year}{2018}\natexlab{}.
\newblock \showarticletitle{From programs to interpretable deep models and
  back}. In \bibinfo{booktitle}{\emph{Computer Aided Verification: 30th
  International Conference, CAV 2018, Held as Part of the Federated Logic
  Conference, FloC 2018, Oxford, UK, July 14-17, 2018, Proceedings, Part I
  30}}. Springer, \bibinfo{pages}{27--37}.
\newblock


\bibitem[Yao et~al\mbox{.}(2018)]%
        {nref34}
\bibfield{author}{\bibinfo{person}{Ziyu Yao}, \bibinfo{person}{Daniel~S Weld},
  \bibinfo{person}{Wei-Peng Chen}, {and} \bibinfo{person}{Huan Sun}.}
  \bibinfo{year}{2018}\natexlab{}.
\newblock \showarticletitle{Staqc: A systematically mined question-code dataset
  from stack overflow}. In \bibinfo{booktitle}{\emph{Proceedings of the 2018
  World Wide Web Conference}}. \bibinfo{pages}{1693--1703}.
\newblock


\bibitem[Ye et~al\mbox{.}(2020)]%
        {s_ref97}
\bibfield{author}{\bibinfo{person}{Fangke Ye}, \bibinfo{person}{Shengtian
  Zhou}, \bibinfo{person}{Anand Venkat}, \bibinfo{person}{Ryan Marucs},
  \bibinfo{person}{Nesime Tatbul}, \bibinfo{person}{Jesmin~Jahan Tithi},
  \bibinfo{person}{Paul Petersen}, \bibinfo{person}{Timothy Mattson},
  \bibinfo{person}{Tim Kraska}, \bibinfo{person}{Pradeep Dubey},
  {et~al\mbox{.}}} \bibinfo{year}{2020}\natexlab{}.
\newblock \showarticletitle{Misim: An end-to-end neural code similarity
  system}.
\newblock \bibinfo{journal}{\emph{arXiv preprint arXiv:2006.05265}}
  (\bibinfo{year}{2020}).
\newblock


\bibitem[Yi et~al\mbox{.}(2014)]%
        {s_ref5}
\bibfield{author}{\bibinfo{person}{Chao Yi}, \bibinfo{person}{Su Feng}, {and}
  \bibinfo{person}{Zhi Gong}.} \bibinfo{year}{2014}\natexlab{}.
\newblock \showarticletitle{A Comparison of Sandbox Technologies Used in Online
  Judge Systems}. In \bibinfo{booktitle}{\emph{Mechanical Design and Power
  Engineering}} \emph{(\bibinfo{series}{Applied Mechanics and Materials},
  Vol.~\bibinfo{volume}{490})}. \bibinfo{pages}{1201--1204}.
\newblock
\urldef\tempurl%
\url{https://doi.org/10.4028/www.scientific.net/AMM.490-491.1201}
\showDOI{\tempurl}


\bibitem[Yin et~al\mbox{.}(2018a)]%
        {yin2018conala}
\bibfield{author}{\bibinfo{person}{Pengcheng Yin}, \bibinfo{person}{Bowen
  Deng}, \bibinfo{person}{Edgar Chen}, \bibinfo{person}{Bogdan Vasilescu},
  {and} \bibinfo{person}{Graham Neubig}.} \bibinfo{year}{2018}\natexlab{a}.
\newblock \showarticletitle{Learning to Mine Aligned Code and Natural Language
  Pairs from Stack Overflow}. In \bibinfo{booktitle}{\emph{Proceedings of the
  15th International Conference on Mining Software Repositories}} (Gothenburg,
  Sweden) \emph{(\bibinfo{series}{MSR '18})}. \bibinfo{publisher}{Association
  for Computing Machinery}, \bibinfo{address}{New York, NY, USA},
  \bibinfo{pages}{476–486}.
\newblock
\showISBNx{9781450357166}
\urldef\tempurl%
\url{https://doi.org/10.1145/3196398.3196408}
\showDOI{\tempurl}


\bibitem[Yin et~al\mbox{.}(2018b)]%
        {s_ref129}
\bibfield{author}{\bibinfo{person}{Pengcheng Yin}, \bibinfo{person}{Bowen
  Deng}, \bibinfo{person}{Edgar Chen}, \bibinfo{person}{Bogdan Vasilescu},
  {and} \bibinfo{person}{Graham Neubig}.} \bibinfo{year}{2018}\natexlab{b}.
\newblock \showarticletitle{Learning to Mine Aligned Code and Natural Language
  Pairs from Stack Overflow}. In \bibinfo{booktitle}{\emph{Proceedings of the
  15th International Conference on Mining Software Repositories}} (Gothenburg,
  Sweden) \emph{(\bibinfo{series}{MSR '18})}. \bibinfo{publisher}{Association
  for Computing Machinery}, \bibinfo{address}{New York, NY, USA},
  \bibinfo{pages}{476–486}.
\newblock
\showISBNx{9781450357166}
\urldef\tempurl%
\url{https://doi.org/10.1145/3196398.3196408}
\showDOI{\tempurl}


\bibitem[Yin and Neubig(2017)]%
        {s_ref130}
\bibfield{author}{\bibinfo{person}{Pengcheng Yin} {and} \bibinfo{person}{Graham
  Neubig}.} \bibinfo{year}{2017}\natexlab{}.
\newblock \showarticletitle{A syntactic neural model for general-purpose code
  generation}.
\newblock \bibinfo{journal}{\emph{arXiv preprint arXiv:1704.01696}}
  (\bibinfo{year}{2017}).
\newblock


\bibitem[Yoshizawa and Watanobe(2019)]%
        {nref57}
\bibfield{author}{\bibinfo{person}{Yuto Yoshizawa} {and}
  \bibinfo{person}{Yutaka Watanobe}.} \bibinfo{year}{2019}\natexlab{}.
\newblock \showarticletitle{Logic error detection system based on structure
  pattern and error degree}.
\newblock \bibinfo{journal}{\emph{Advances in Science, Technology and
  Engineering Systems Journal}} \bibinfo{volume}{4}, \bibinfo{number}{5}
  (\bibinfo{year}{2019}), \bibinfo{pages}{1--15}.
\newblock


\bibitem[Yue et~al\mbox{.}(2018)]%
        {nref40}
\bibfield{author}{\bibinfo{person}{Ruru Yue}, \bibinfo{person}{Zhe Gao},
  \bibinfo{person}{Na Meng}, \bibinfo{person}{Yingfei Xiong},
  \bibinfo{person}{Xiaoyin Wang}, {and} \bibinfo{person}{J~David
  Morgenthaler}.} \bibinfo{year}{2018}\natexlab{}.
\newblock \showarticletitle{Automatic clone recommendation for refactoring
  based on the present and the past}. In \bibinfo{booktitle}{\emph{2018 IEEE
  International Conference on Software Maintenance and Evolution (ICSME)}}.
  IEEE, \bibinfo{pages}{115--126}.
\newblock


\bibitem[Zhang and Tsai(2003)]%
        {s_ref29}
\bibfield{author}{\bibinfo{person}{Du Zhang} {and} \bibinfo{person}{Jeffrey~JP
  Tsai}.} \bibinfo{year}{2003}\natexlab{}.
\newblock \showarticletitle{Machine learning and software engineering}.
\newblock \bibinfo{journal}{\emph{Software Quality Journal}}
  \bibinfo{volume}{11} (\bibinfo{year}{2003}), \bibinfo{pages}{87--119}.
\newblock


\bibitem[Zhang and Khoo(2021)]%
        {s_ref92}
\bibfield{author}{\bibinfo{person}{Fanlong Zhang} {and}
  \bibinfo{person}{Siau-Cheng Khoo}.} \bibinfo{year}{2021}\natexlab{}.
\newblock \showarticletitle{An empirical study on clone consistency prediction
  based on machine learning}.
\newblock \bibinfo{journal}{\emph{Information and Software Technology}}
  \bibinfo{volume}{136} (\bibinfo{year}{2021}), \bibinfo{pages}{106573}.
\newblock


\bibitem[Zhang et~al\mbox{.}(2020)]%
        {nref49}
\bibfield{author}{\bibinfo{person}{Jian Zhang}, \bibinfo{person}{Xu Wang},
  \bibinfo{person}{Hongyu Zhang}, \bibinfo{person}{Hailong Sun}, {and}
  \bibinfo{person}{Xudong Liu}.} \bibinfo{year}{2020}\natexlab{}.
\newblock \showarticletitle{Retrieval-based neural source code summarization}.
  In \bibinfo{booktitle}{\emph{Proceedings of the ACM/IEEE 42nd International
  Conference on Software Engineering}}. \bibinfo{pages}{1385--1397}.
\newblock


\bibitem[Zhang et~al\mbox{.}(2022)]%
        {s_ref33}
\bibfield{author}{\bibinfo{person}{Jie~M. Zhang}, \bibinfo{person}{Mark
  Harman}, \bibinfo{person}{Lei Ma}, {and} \bibinfo{person}{Yang Liu}.}
  \bibinfo{year}{2022}\natexlab{}.
\newblock \showarticletitle{Machine Learning Testing: Survey, Landscapes and
  Horizons}.
\newblock \bibinfo{journal}{\emph{IEEE Transactions on Software Engineering}}
  \bibinfo{volume}{48}, \bibinfo{number}{1} (\bibinfo{year}{2022}),
  \bibinfo{pages}{1--36}.
\newblock
\urldef\tempurl%
\url{https://doi.org/10.1109/TSE.2019.2962027}
\showDOI{\tempurl}


\bibitem[Zhao and Huang(2018)]%
        {nref46}
\bibfield{author}{\bibinfo{person}{Gang Zhao} {and} \bibinfo{person}{Jeff
  Huang}.} \bibinfo{year}{2018}\natexlab{}.
\newblock \showarticletitle{Deepsim: deep learning code functional similarity}.
  In \bibinfo{booktitle}{\emph{Proceedings of the 2018 26th ACM Joint Meeting
  on European Software Engineering Conference and Symposium on the Foundations
  of Software Engineering}}. \bibinfo{pages}{141--151}.
\newblock


\bibitem[Zheng et~al\mbox{.}(2015)]%
        {nref10}
\bibfield{author}{\bibinfo{person}{Ninghan Zheng}, \bibinfo{person}{Shuzhen
  Tian}, {and} \bibinfo{person}{Yongqiang Chen}.}
  \bibinfo{year}{2015}\natexlab{}.
\newblock \showarticletitle{Online Learning Management System}. In
  \bibinfo{booktitle}{\emph{2015 International Conference on Computational
  Science and Computational Intelligence (CSCI)}}. \bibinfo{pages}{293--299}.
\newblock
\urldef\tempurl%
\url{https://doi.org/10.1109/CSCI.2015.160}
\showDOI{\tempurl}


\bibitem[Zhigang et~al\mbox{.}(2001)]%
        {nref13}
\bibfield{author}{\bibinfo{person}{Sun Zhigang}, \bibinfo{person}{Su Xiaohong},
  \bibinfo{person}{Zhu Ning}, {and} \bibinfo{person}{Cheng Yanyu}.}
  \bibinfo{year}{2001}\natexlab{}.
\newblock \showarticletitle{Moodle plugins for highly efficient programming
  courses}.
\newblock  (\bibinfo{year}{2001}).
\newblock


\bibitem[Zhou et~al\mbox{.}(2018)]%
        {nref6}
\bibfield{author}{\bibinfo{person}{Wenju Zhou}, \bibinfo{person}{Yigong Pan},
  \bibinfo{person}{Yinghua Zhou}, {and} \bibinfo{person}{Guangzhong Sun}.}
  \bibinfo{year}{2018}\natexlab{}.
\newblock \showarticletitle{The framework of a new online judge system for
  programming education}. In \bibinfo{booktitle}{\emph{Proceedings of ACM
  turing celebration conference-China}}. \bibinfo{pages}{9--14}.
\newblock


\bibitem[Zhou and Jiang(2012)]%
        {nref29}
\bibfield{author}{\bibinfo{person}{Yajin Zhou} {and} \bibinfo{person}{Xuxian
  Jiang}.} \bibinfo{year}{2012}\natexlab{}.
\newblock \showarticletitle{Dissecting android malware: Characterization and
  evolution}. In \bibinfo{booktitle}{\emph{2012 IEEE symposium on security and
  privacy}}. IEEE, \bibinfo{pages}{95--109}.
\newblock


\bibitem[Zhou et~al\mbox{.}(2019)]%
        {s_ref73}
\bibfield{author}{\bibinfo{person}{Yaqin Zhou}, \bibinfo{person}{Shangqing
  Liu}, \bibinfo{person}{Jingkai Siow}, \bibinfo{person}{Xiaoning Du}, {and}
  \bibinfo{person}{Yang Liu}.} \bibinfo{year}{2019}\natexlab{}.
\newblock \bibinfo{booktitle}{\emph{Devign: Effective Vulnerability
  Identification by Learning Comprehensive Program Semantics via Graph Neural
  Networks}}.
\newblock \bibinfo{publisher}{Curran Associates Inc.}, \bibinfo{address}{Red
  Hook, NY, USA}.
\newblock


\end{thebibliography}
